\documentclass[reprint,aip,longbibliography,superscriptaddress,preprintnumbers,amsmath,amssymb,floatfix]{revtex4-1}
\usepackage{graphicx}
\usepackage{multirow}
\usepackage{booktabs}
\usepackage{amsmath}
\usepackage{amssymb}
\usepackage{mathbbol}

\graphicspath{{./figures/}}
\usepackage[usenames,dvipsnames]{color}

\def \be {\begin{equation}}
\def \ee {\end{equation}}
\def \ben {\begin{eqnarray}}
\def \een {\end{eqnarray}}
\def\re#1{{\color{black} {#1}}}
\def\rc#1{{\color{black} {#1}}}

\begin{document}
\title{Nonadiabatic derivative couplings through multiple Franck-Condon modes dictate the energy gap law for near and short-wave infrared dye molecules}

\author{Pablo Ramos}
\affiliation{Department of Chemistry and Biochemistry, Queens College, City University of New York (CUNY), 65-30 Kissena Boulevard, Queens, New York 11367, USA \\ \& Chemistry and Physics PhD programs, CUNY Graduate Center}

\author{Hannah Friedman }
\affiliation{Department of Chemistry and Biochemistry, University of California, Los Angeles, Los Angeles, California 90095, United States}

\author{Barry Y. Li}
\affiliation{Department of Chemistry and Biochemistry, University of California, Los Angeles, Los Angeles, California 90095, United States}

\author{Cesar Garcia}
\affiliation{Department of Chemistry and Biochemistry, University of California, Los Angeles, Los Angeles, California 90095, United States}

\author{Ellen Sletten}
\affiliation{Department of Chemistry and Biochemistry, University of California, Los Angeles, Los Angeles, California 90095, United States}

\author{Justin R. Caram}
\affiliation{Department of Chemistry and Biochemistry, University of California, Los Angeles, Los Angeles, California 90095, United States}

\author{Seogjoo J. Jang}
\email{seogjoo.jang@qc.cuny.edu}
\affiliation{Department of Chemistry and Biochemistry, Queens College, City University of New York (CUNY), 65-30 Kissena Boulevard, Queens, New York 11367, USA \\ \& Chemistry and Physics PhD programs, CUNY Graduate Center}

\date{Published in the {\it Journal of Physical Chemistry Letters} {\bf  15}, 1802-1810 (2024)}



\begin{abstract}
Near infrared (NIR, $700 - 1,000\ {\rm nm}$) and short-wave infrared (SWIR, $1,000 - 2,000\ {\rm nm}$) dye molecules exhibit significant nonradiative decay rates from the first singlet excited state to the ground state.   While these trends can be empirically explained by a simple energy gap law, detailed mechanisms of the nearly universal behavior have remained unsettled for many cases. Theoretical and experimental results for two representative NIR/SWIR dye molecules reported here clarify \re{the key mechanism for the observed energy gap law behavior}. It is shown that the first derivative nonadiabatic coupling terms serve as major coupling pathways for nonadiabatic decay processes \re{from the first excited singlet state to the ground state for these NIR and SWIR dye molecules} and that vibrational modes other than the highest frequency ones also make significant contributions to the rate.  This assessment is  corroborated by further theoretical comparison with possible alternative mechanisms of intersystem crossing to triplet states and also by comparison with experimental data for deuterated molecules.
\end{abstract}
\maketitle

Despite the utility and reliability of the Born-Oppenheimer approximation\cite{born-adp389,jang-qmc} in general, nonadiabatic couplings  between adiabatic potential energy surfaces can be significant in some cases,  playing key roles for important quantum transfer and relaxation processes.\cite{baer-pr358,yonehara-cr112,Jasper2004,Chen2016,Uratani2020} Well established  examples include electron and proton transfer reactions and internal conversion between excited electronic states of comparable energies, for which various theoretical and computational advances\cite{tully-jcp137,yonehara-cr112,Jasper2004,Chen2016,Uratani2020,song-jctc16,zhao-jpcl11,curchod-jcp144,bian-jcp154,prezhdo-acr54,shu-jctc18,alfonso-hernandez-pccp22} have been made. On the other hand, the role of nonadiabatic derivative couplings (NDCs) as direct routes for nonradiative transitions from the first excited singlet state to the ground electronic state, has long remained unsettled.  A widely held view\cite{ishikawa-jcp37,laposa-jcp42} is that such direct nonadiabatic transitions are insignificant due to large energy gaps between excited and ground states.  Thus, many nonradiative decay processes are hypothesized to occur  through dark forbidden states or via an activated crossing to special nonadiabatic coupling regions such as conical intersections.   Indeed, there are many examples with solid experimental evidences for transitions through such indirect\cite{yonehara-cr112,Jasper2004,kathayat-jacs138} or special routes,\cite{yonehara-cr112,Jasper2004,hare-pnas104} for which advanced theoretical and computational approaches have also been developed.\cite{yonehara-cr112,Jasper2004,izmaylov-jcp135} \rc{However,  for near infrared (NIR) dye molecules with wavelengths in the range of $700 - 1,000\ {\rm nm}$ and short-wave infrared (SWIR) dye molecules in the range of $1,000 - 2,000\ {\rm nm}$, direct nonadiabatic transitions through NDC from the first excited state to the ground state can play a significant role due to relatively small energy gaps and thus need to be examined more thoroughly.} 

 Indeed, a recent work\cite{friedman-chem7} by Friedman {\it et al.} reported fairly universal energy gap law\cite{englman-mp18,jang-jcp155-1} behavior of nonradiative decay rates for a broad class of NIR and SWIR dye molecules.  Similar observation was reported more recently for different kinds of NIR dye molecules based on single molecule spectroscopy as well.\cite{erker-jacs144} The observed experimental trends\cite{friedman-chem7,erker-jacs144}  suggest simple direct nonadiabatic transitions through NDC terms as major routes for nonradiative decays, which however have not been demonstrated clearly  by any theoretical calculations yet.   Given the growing importance of NIR and SWIR chromophores in biological imaging applications\cite{Thimsen2017,Xie2022,Pascal2021,Li2020}, \rc{quantitative understanding of the role of NDCs 
 on lifetimes of excited states also has} significant implications and applications. The present letter elucidates this issue through detailed theoretical modeling of experimental lineshapes and nonradiative rates.

\begin{figure}[ht]
\includegraphics[width=.4\textwidth]{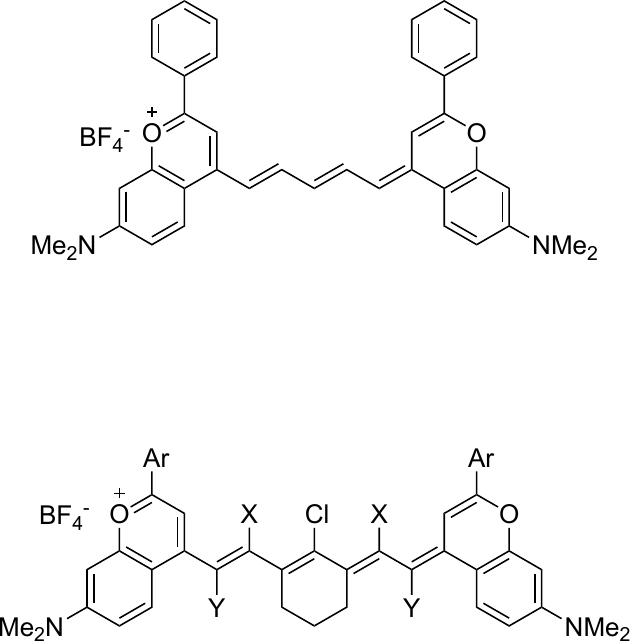}\vspace{.1in}\\
	\caption{Comparison of the two organic dye molecules Flav5 (upper panel) and Flav7 (lower panel).  For the latter, X represent positions of hydrogen atoms being deuterated and Y represents those for which 30\% of hydrogen atoms are deuterated.  Flav7 with all hydrogen atoms in the aryl rings deuterated are also considered.}
	\label{dyes}
\end{figure}

 As representative cases, we here report comparative computational and experimental studies of two similar dye molecules, 7-dimethylamino flavylium pentamethine (Flav5) and 7-dimethylamino flavylium heptamethine (Flav7)  and some deuterated forms of the latter molecule synthesized recently (see SI for experimental details).     Flav5 and Flav7 are NIR/SWIR dye molecules with similar structural motifs (see Fig. \ref{dyes}), but their experimental nonradiative decay rates determined in dichloromethane (DCM) solvent, $3.2\ {\rm ns}^{-1}$ (Flav5) and $14.8\ {\rm ns}^{-1}$ (Flav7), are different by about a factor of five.   
 
 As computational methods for obtaining our theoretical data, we here employ the density functional theory (DFT)\cite{parr1989,dreizler_gross,engel_dreizler} and time dependent DFT (TD-DFT)\cite{parr1989,gros1995,marq2003} methods with CAM-B3LYP functional,\cite{yana2004} which has been tested broadly and is known to produce fairly accurate values of reorganization energies\cite{higashi-jpcb118} and Huang-Rhys (HR) factors.\cite{li-jcc42}  Our choice is also supported by a recent comprehensive study\cite{li-cs13} showing that CAM-B3LYP is one of the best functionals available for calculating excited state properties although it sometimes overestimates excitation energies.  
 
  For each molecule, we conducted optimization of the ground electronic state (${\rm S_0}$) structure,  in the presence of a counter ion, ${\rm BF_4^-}$, employing the DFT method\cite{parr1989,dreizler_gross,engel_dreizler} with the CAM-B3LYP functional\cite{yana2004} and the 6-311+G basis set.    Using the optimized structure of ${\rm S_0}$ as the initial guess, we also optimized the structure of the excited state (${\rm S_1}$) for each molecule using the TD-DFT method\cite{parr1989,gros1995,marq2003} and using the same functional and basis set.
 
 The  theoretical ${\rm S_0 \rightarrow S_1}$ transition energies for the optimized structure of ${\rm S_0}$ in vacuum calculated by the TD-DFT method (with default value of tuning parameter $\omega=0.33\ {\rm bohr^{-1}}$ for CAM-B3LYP functional), are ${\rm 2.2801\ eV}$ for Flav5 and ${\rm 2.0599\ eV}$ for Flav7.  We find that these are significantly larger than  experimental vertical absorption energies (in DCM), which we estimated from lineshape modeling as will be described later and are {\rm 1.440\ eV} for Flav 5 and {\rm 1.207\ eV} for Flav7.   Similarly,  calculated values of vertical ${\rm S_1 \rightarrow S_0}$ transition energies in vacuum for the optimized structure of ${\rm S_1}$ are $2.0935\ {\rm eV}$ for Flav5 and $1.8835\ {\rm eV}$ for Flav7, which are also larger than experimental solution phase values of  $1.378\ {\rm eV}$ for Flav5  and  $1.177\ {\rm eV}$ for Flav7.   For better understanding of the sources of the discrepancies, we also conducted calculations employing other functionals and considering solvation effects  at continuum level.  Supporting Information (SI) provides these values and discusses their implications.  Outcomes of these calculations suggest that SCAN functional\cite{sun-nc8} in fact produces excitation energies in better agreements with experimental ones. However, they are still significantly larger than experimental excitation energies, and the SCAN functional has not yet been tested well enough for calculating FC factors and NDC terms.  In fact, for SCAN, including solvent effect results in blue shift of excitation energies (see Table S2).  Our preliminary calculation of lineshape based on SCAN functional also suggests its performance for  calculating FC factors is not reliable, which is consistent with a recent study in favor of CAM-B3LYP over SCAN.\cite{li-cs13}  Thus, for our theoretical modeling in this work, we employ all computational data obtained from the CAM-B3LYP functional, except for the excitation energies for which we use values obtained from experimental data.  The validity of this approach is supported by good agreements between theoretical and experimental absorption lineshapes as will become clear.

\begin{figure}[ht]
\includegraphics[width=.4\textwidth]{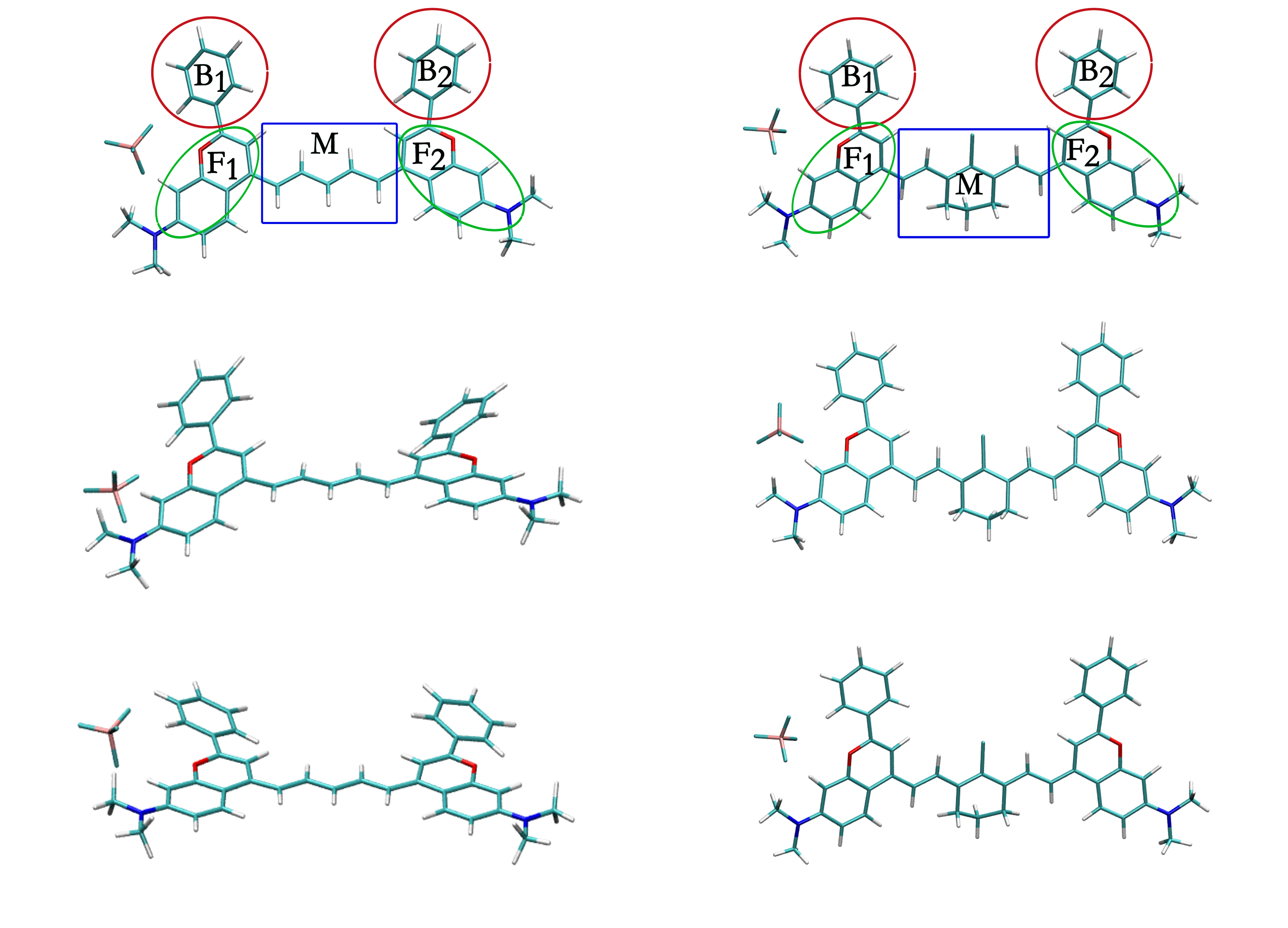}
	\caption{Comparison of the structures of Flav5 (left column) and Flav7 (right column) dye molecules.   The top row provides chemical structures indicating different sections with different labels. The middle row shows optimized structures for the ${\rm S}_0$ state, and the bottom row those for the ${\rm S}_1$ state.}
        \label{optgeo}
\end{figure}

In Fig. \ref{optgeo}, the optimized structures for both dye molecules are shown. Structures shown at the top include labels for different parts of the molecule, which are used for providing a more detailed analyses below.  For each molecule, \re{the structure of ${\rm S_0}$ state  is asymmetric and the F1 flavylium heterocycle is tilted with respect to the molecular plane, which is defined as a plane that contains the M section maximally. On the other hand, the F2 heterocycle is nearly coplanar with the molecular plane.}   The angle between the F1 heterocycle and the \re{molecular plane}  is 18.0${\rm ^o}$ for Flav5 and 26.0${\rm ^o}$ for Flav7. The B1 ring has an angle of 20.4${\rm ^o}$ and 25.0${\rm ^o}$ for Flav5 and Flav7 respectively, and the angle of the B2 phenyl group is about 30.0${\rm ^o}$ w.r.t the molecular plane in both dyes, although they are in opposite directions. Thus, for the ${\rm S_0}$ state, the major structural difference of the two dye molecules lies in detailed orientational features of ring parts.  On the other hand, for the ${\rm S_1}$ states, structures are similar and more symmetric, with B1 and B2 heterocycles at  about 17${\rm ^o}$ w.r.t the molecular plane.   Both F1 and F2 parts are nearly  coplanar with the M section, which is flat for Flav5 but is concave for Flav7 due to the presence of bridging sp$^3$ carbons in the 7-methine M group.

\begin{figure}[ht]
\includegraphics[width=0.4\textwidth]{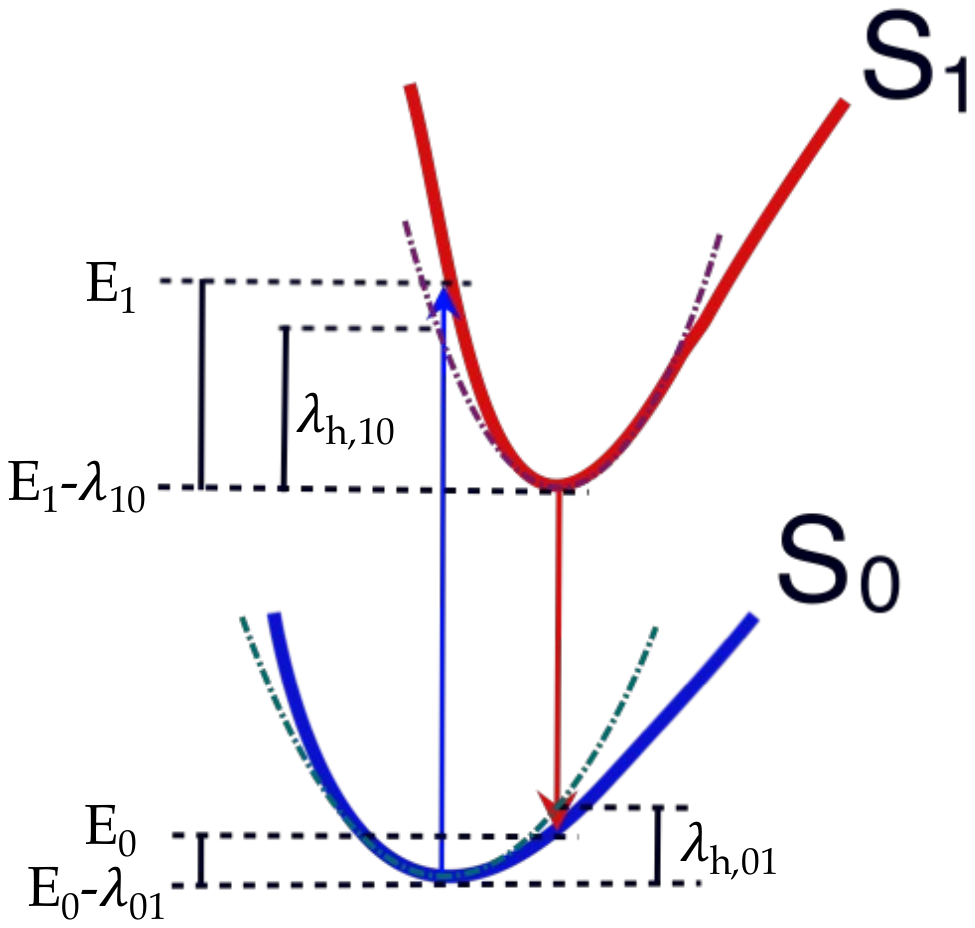}
	\caption{A schematic of adiabatic potential energy surfaces for ${\rm S_0}$ (lower blue line) and ${\rm S_1}$ (upper red line) electronic states, which are in general anharmonic, and relevant energies. $\lambda_{10}$ is the reorganization energy in the ${\rm S}_1$ state from the minimum energy structure of ${\rm S}_0$ to the optimized structure of ${\rm S_1}$, and  $\lambda_{01}$ is the reorganization energy in the ${\rm S}_0$ state from the minimum energy structure of ${\rm S}_1$ to the optimized structure of ${\rm S_0}$.  Note that $\tilde E_0=E_0-\lambda_{01}$ and $\tilde E_1=E_1-\lambda_{10}$.  Reorganization energies within the harmonic approximation, $\lambda_{h,10}$ and $\lambda_{h,01}$ are also shown (see text for more detailed explanation). }
        \label{egap-reorg}
\end{figure}

For theoretical calculation of absorption and emission lineshapes, we evaluated HR factors\cite{reimers2001} for both the vertical ${\rm S_0\rightarrow S_1}$ transition at the optimized structure of ${\rm S_0}$ and also those for ${\rm S_1\rightarrow S_0}$ transition at the optimized structure of ${\rm S_1}$. For each molecule, there were clear differences between the two distributions of HR factors  at ${\rm S_0}$ and ${\rm S_1}$ states, which reflects the presence of significant anharmonic effects.  This is understandable considering the significant differences in the optimized structures for the ${\rm S}_0$ and ${\rm S}_1$ states as can be seen from Fig. \ref{optgeo}.    We account for such anharmonic contributions by introducing  two different baths of harmonic oscillators defined differently depending on the structure of the initial state as described below.   Before providing more detailed description of models, we find it useful to clarify notations we adopt in this work.  ${\rm S_0}$ and ${\rm S_1}$ denote the whole set of the ground and the first excited adiabatic electronic states, respectively, parametrized by nuclear coordinates.   On the other hand, we introduce two diabatic states based on crude adiabatic approximation, $|\tilde {\rm S}_0\rangle$ and $|{\rm S}_1\rangle$, which are determined at the optimized minimum energy nuclear coordinates of the ${\rm S}_0$ surface.  Similarly, we define another two diabatic states,  $|{\rm S}_0\rangle$ and $|\tilde {\rm S}_1\rangle$, which are determined at the optimized minimum energy nuclear coordinates of the ${\rm S}_1$ surface.

We assume that the molecular Hamiltonian near the structure optimized for the ${\rm S_0}$ state of each molecule can be approximated by the following model:
\ben
\lefteqn{\hat H_0 }&&=\tilde E_0|\tilde S_0\rangle\langle \tilde S_0|\nonumber \\
&&+\left \{E_1+\sum_n \hbar\omega_n g_{n,10}(\hat b_n+\hat b_n^\dagger)\right\}|S_1\rangle\langle S_1| \nonumber \\
&&+\sum_n \hbar \omega_n (\hat b_n^\dagger \hat b_n+\frac{1}{2}) , \label{eq:h0-def} 
\een
where $\tilde E_0$ and $E_1$ are electronic energies of ${\rm S_0}$ and ${\rm S_1}$ states respectively for a fully optimized structure of the molecule in the ${\rm S_0}$ state, and  $\hat b_n$ and $\hat b_n^\dagger$ represent lowering and raising operators for each harmonic oscillator normal mode of molecular vibrations calculated for the optimized structure of  the ${\rm S_0}$ state (see Fig. \ref{egap-reorg} for a schematic illustration.).  The coupling constant $g_{n,10}$ represents the extent of displacement of the $n$th normal mode upon excitation, the square of which is the corresponding HR factor for the ${\rm S}_0 \rightarrow {\rm S}_1$ transition near the structure optimized for the ${\rm S_0}$ state.   Note that the tilde symbol in $\tilde E_0$ indicates that it is the energy for the fully relaxed (optimized) structure of the ${\rm S_0}$ state, in contrast to $E_1$.  Thus, $E_1-\tilde E_0$ is the vertical excitation energy for the ${\rm S_0\rightarrow S_1}$ transition at the optimized structure of ${\rm S_0}$.

On the other hand, for the structures optimized at the ${\rm S_1}$ state, we assume that the molecular Hamiltonian can be approximated by the following model:
\ben
\lefteqn{\hat H_1}&&= \left (E_0+\sum_n \hbar\omega_n g_{n,01}(\hat b_n+\hat b_n^\dagger)\right )|S_0\rangle\langle S_0|\nonumber \\
&&+\tilde E_1|\tilde S_1\rangle\langle \tilde S_1|+\sum_n \hbar \omega_n (\hat b_n^\dagger \hat b_n+\frac{1}{2}) , \label{eq:h1-def}
\een 
where $E_0$ and $\tilde E_1$ are electronic energies of ${\rm S_0}$ and ${\rm S_1}$ states respectively for a fully optimized structure of the molecule in the ${\rm S_1}$ state, and  $\hat b_n$ and $\hat b_n^\dagger$ represent lowering and raising operators for each harmonic oscillator normal mode of molecular vibrations calculated for the optimized structure of  the ${\rm S_1}$ state (see Fig. \ref{egap-reorg} for a schematic illustration.).  The coupling constant $g_{n,01}$ represents the extent of displacement of the $n$th normal mode, the square of which is the corresponding HR factor, for the ${\rm S}_1 \rightarrow {\rm S}_0$ transition near the structure optimized for the ${\rm S_1}$ state.

Note that $\tilde E_1=E_1-\lambda_{10}$ and $E_0=\tilde E_0+\lambda_{01}$ (see Fig. \ref{egap-reorg}).  Thus, $\lambda_{10}$ is the reorganization energy in the ${\rm S_1}$ electronic state for the structural change from that of the minimum ${\rm S_0}$ energy to that of ${\rm S_1}$.  On the other hand, $\lambda_{01}$ is the reorganization energy in the ${\rm S_0}$ electronic state for the structural change from that of the minimum ${\rm S_1}$ energy to that of ${\rm S_0}$.    Thus, $\tilde E_1-E_0$ is the vertical emission energy for the transition to ${\rm S_0}$ from the fully relaxed ${\rm S_1}$ state.  \re{Note that  $\lambda_{10}$ and $\lambda_{01}$ include all the anharmonic effects of conformational changes as well as harmonic contributions of the molecular vibrations.  Thus, they are different from those based on the harmonic oscillator bath modes.}  Note also that the oscillator modes  in eqs. \ref{eq:h0-def} and \ref{eq:h1-def} are different but are denoted with the same labels.

\begin{widetext}
  \begin{table}[h]
\caption{Energy gap values best matching peak positions of  experimental lineshapes for Flav 5 and Flav 7.   Theoretical values of reorganization energies are also shown.  Note that $\lambda_{10}=E_1-\tilde E_1$ and $\lambda_{01}=E_0-\tilde E_0$. }
\begin{tabular}{ccccccc}
\hline
\hline 
Dye &${\rm E_1-\tilde E_0\ (eV)^{~}}$ &${\rm \tilde E_1-E_0\ (eV}$)&$\lambda_{10} ({\rm eV})$& $\lambda_{h,10} ({\rm eV})$&$\lambda_{01} ({\rm eV})$&$\lambda_{h,10} ({\rm eV})$  \\
\hline 
	Flav5 & 1.440   & 1.378   &0.817 & 0.455 &0.630 &0.349\\    
\hline              
 	Flav7 & 1.207 & 1.177  &0.701 & 0.631 & 0.524 & 0.283\\
\hline
\hline
        \end{tabular}
        \label{table1}

\end{table}
\end{widetext}

Let us define the following bath spectral densities: 
\be
{\mathcal J}_{ij}(\omega)=\pi\hbar \sum_n \delta (\omega-\omega_n) \omega_n^2 g_{n,ij}^2 , \label{eq:spd-ij}
\ee
where $i,j=1,0$ (for eq. \ref{eq:h0-def}) or $0,1$ (for eq. \ref{eq:h1-def}), and the corresponding real and imaginary parts of the lineshape function: 
 \ben 
G_{R,ij}(t)&=&\frac{1}{\pi\hbar} \int_0^\infty d\omega \frac{{\mathcal J}_{ij}(\omega)}{\omega^2} \coth \left (\frac{\hbar\omega}{2k_BT} \right)\nonumber \\
&&\hspace{1in}\times \left (1-\cos(\omega t)\right) ,  \label{eq:gr-ij}\\
G_{I,ij}(t)&=&\frac{1}{\pi\hbar} \int_0^\infty d\omega \frac{{\mathcal J}_{ij}(\omega)}{\omega^2} \left (\sin(\omega t)-\omega t\right ) \nonumber \\
&=&\frac{1}{\pi\hbar} \int_0^\infty d\omega \frac{{\mathcal J}_{ij}(\omega)}{\omega^2} \sin(\omega t)  -\frac{\lambda_{h,ij}}{\hbar} t,   \label{eq:gi-ij}
\een
with $\lambda_{h,ij}=\hbar\sum_n \omega_n g_{n,ij}^2$.  Note that $\lambda_{h,10}$ and $\lambda_{h,01}$ are harmonic approximations for $\lambda_{10}=E_1-\tilde E_1$ and $\lambda_{01}=E_0-\tilde E_0$, as indicated in Fig. \ref{egap-reorg}.    Thus, $\lambda_{10}-\lambda_{h,10}$ represents contributions of anharmonicity in the ${\rm S}_1$ surface, whereas $\lambda_{01}-\lambda_{h,01}$ represents that in  the ${\rm S}_0$ surface.  In the simplest displaced harmonic oscillator model for which ${\rm S}_1$ and ${\rm S_0}$ are parabolic forms of the same curvature, $\lambda_{01}=\lambda_{10}=\lambda_{h,01}=\lambda_{h,10}$. 

The normalized absorption lineshape for ${\rm S_0}\rightarrow {\rm S_1}$ transition, for the Hamiltonian of eq. \ref{eq:h0-def}, based on the Fermi's golden rule (FGR) for interaction with radiation within the dipole approximation,\cite{jang-exciton} can be expressed as 
\ben
I_{S_0\rightarrow S_1}(\omega)&=&\frac{1}{2\pi}\int_{-\infty}^\infty dt\exp\Bigg \{ i\left ( \omega-\frac{(E_1-\tilde E_0)}{\hbar} \right) t  \nonumber \\
&&\hspace{.4in} -G_{R,10}(t)-iG_{I,10}(t) \Bigg \} .\label{eq:abs-line}
\een
On the other hand, the normalized emission lineshape for ${\rm S_1}\rightarrow {\rm S_0}$ transition, for the Hamiltonian of eq. \ref{eq:h1-def}, can be expressed as\cite{jang-exciton} 
\ben
E_{S_1\rightarrow S_0}(\omega)&=&\frac{1}{2\pi}\int_{-\infty}^\infty dt \exp\Bigg \{ i\left ( \omega-\frac{(\tilde E_1-E_0)}{\hbar} \right) t  \nonumber \\
&&\hspace{.4in} -G_{R,01}(t)+iG_{I,01}(t) \Bigg \} . \label{eq:ems-line}
\een

\begin{figure}[ht]
\includegraphics[width=.4\textwidth]{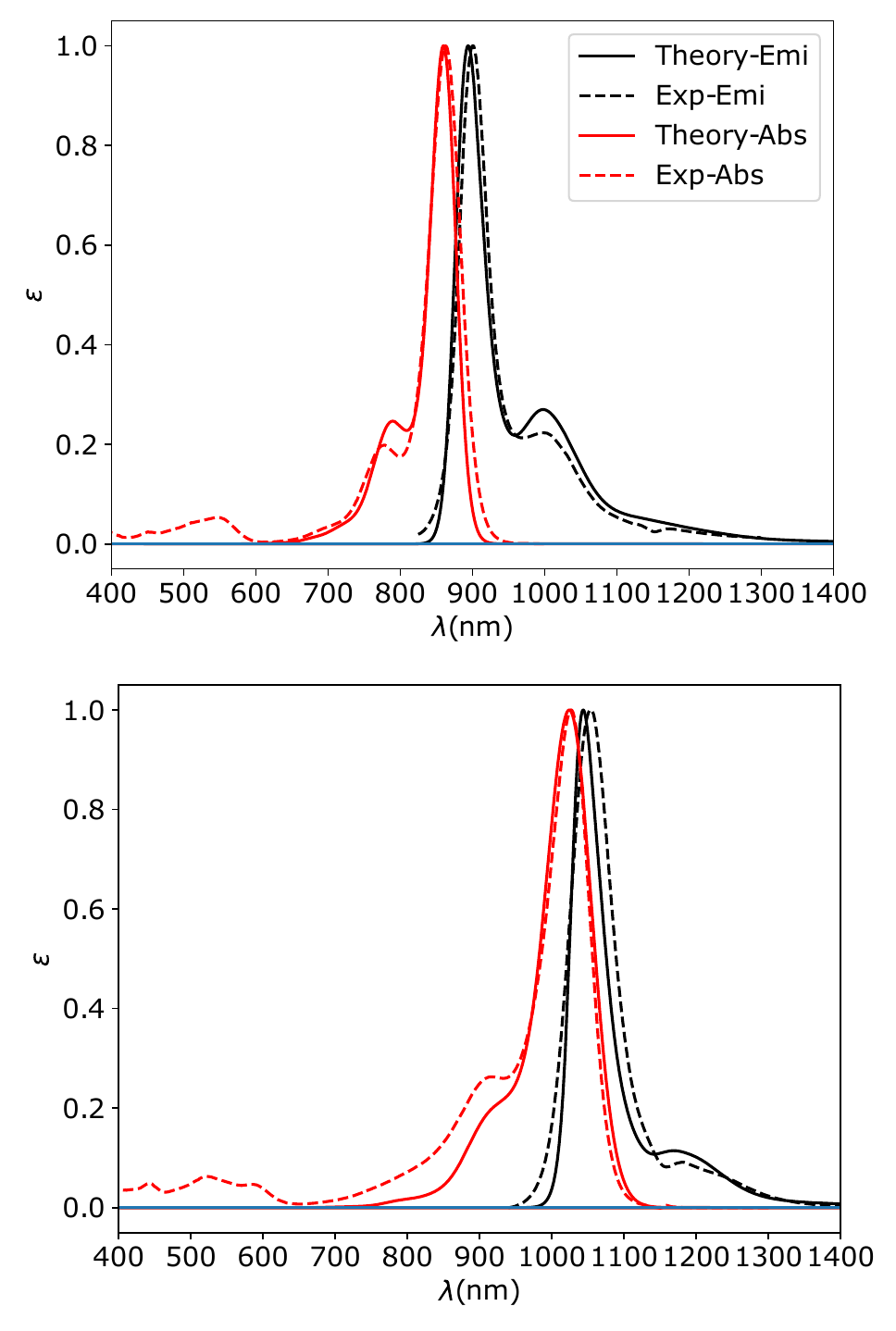}
	\caption{Absorption ($S_0 \rightarrow S_1$) and emission ($S_1 \rightarrow S_0$) lineshapes for Flav5 (upper) and Flav7 (lower). Experimental lineshapes in solution phase are compared with theoretical lineshapes of isolated molecules, which were shifted to match the corresponding maxima of experimental spectra.  For the absorption, the shifts were $1.423\ {\rm eV}$ for Flav5 and $1.191\ {\rm eV}$ for Flav7. For the emission, the shifts were $ 1.405\ {\rm eV}$ for Flav5 and $1.185\ {\rm eV}$  for Flav7.  We also normalized heights with respect to those of their maxima.}
	\label{ls1}
\end{figure}

For actual calculation of lineshapes, we employed a Lorentzian function for the delta function in eq. \ref{eq:spd-ij}.  Thus, the following approximations were used.
\ben
{\mathcal J}_{ij}(\omega)&\approx& \hbar \sum_{n} \frac{\omega_n^2 g_{n,ij}^2 \sigma}{\sigma^2+(\omega-\omega_n)^2} \nonumber \\
&=&2\pi c\hbar \sum_n \frac{\tilde \nu_n^2 g_{n,ij}^2 \tilde \nu_\sigma}{\tilde \nu_\sigma^2+(\tilde \nu_n -\tilde \nu)^2}  ,\label{eq:spd-0-app} 
\een
 where $\tilde \nu=\omega/(2\pi c)$, $\tilde \nu_n=\omega_n/(2\pi c)$, and $\tilde \nu_\sigma =\sigma/(2\pi c)$.   The choice of $\tilde \nu_\sigma$ can be made such that it is small enough to discern all different vibrational frequencies, and we found that  the choice of $\tilde \nu_\sigma =22\ {\rm cm^{-1}}$ reasonable.  \re{The resulting spectral densities calculated by the TD-DFT method with CAM-B3LYP functional are provided in Figs. S1 and S2 in SI.}  

Figure \ref{ls1} compares theoretical absorption and emission lineshapes for Flav5 and Flav7 \re{calculated according to eqs. \ref{eq:abs-line} and \ref{eq:ems-line} }respectively with experimental lineshapes.  The peak maxima of theoretical lineshapes were shifted uniformly to match experimental ones as indicated in the figure caption, but no other corrections were made.   \rc{The resulting values of $E_1-\tilde E_0$ and $\tilde E_1-E_0$ determined in this manner are listed in Table \ref{table1}.  Theoretical values of reorganization energies and \rc{their harmonic approximations} for both the ground and excited state energies are provided as well. }
 Considering the simplicity of underlying models and the fact that solvation effects have not been included, the agreements are excellent.  This suggests that major vibronic interactions are reasonably represented by normal modes of molecular vibrations, modeled as harmonic oscillators, and that contributions of solvents on \rc{details of lineshapes are minor}.    \re{This good agreement also means that HR factors calculated by using CAM-B3LYP functional are quite reliable. } It is worth noting that small peaks around 500 nm in experimental absorption lineshapes correspond to excitations to higher electronic states, which are not included in our calculations.   In addition, while other small disagreements between theoretical and experimental lineshapes could in principle be accounted for by inclusion of solvent effects, simple calculations based on polarizable continuum model turned out to worsen theoretical lineshapes compared to experimental ones (see SI).  Therefore, more accurate solvation model is necessary for better quantitative agreement.

As can be seen from Fig.  \ref{ls1}, the absorption lineshape for each molecule has a prominent sideband followed by broad tail in the blue region.  For Flav5, the major sideband is attributed to two groups of normal modes (see SI for the depiction of major normal modes.).  The first group corresponds to wagging motions of H atoms at 1,026 cm$^{-1}$, which are broadened by those of heavy atoms at around 675 cm$^{-1}$.  The second group consists of the scissoring motion of heavy atoms at $1,237\ {\rm cm^{-1}}$ and H atoms at 1,242 cm$^{-1}$ respectively. The broad tail region is due to the high energy modes and are related to the rocking motion of  H atoms in the structure and the stretching motion of H atoms in the aliphatic chain.  For Flav7, the major sideband is due to the wagging motion of the benzene ring carbons at 825 cm$^{-1}$, scissoring motion of H atoms at 1,252 cm$^{-1}$, and the wagging motion of H atoms at 1,509 cm$^{-1}$. The broad blue tail is also contributed by these wagging modes, with the addition of the stretching motion of H atoms within the central parts of molecules. The stretching of the aliphatic chain at 2,980 cm$^{-1}$ and the stretching of H atoms in aromatic rings at 3,150 cm$^{-1}$ also make some contributions.  Similar features appear in the respective emission lineshapes on the red side of major peaks.  \re{More detailed information on bath spectral densities and major vibrational modes can be found in SI}. 

\begin{figure}
\includegraphics[width=0.4\textwidth]{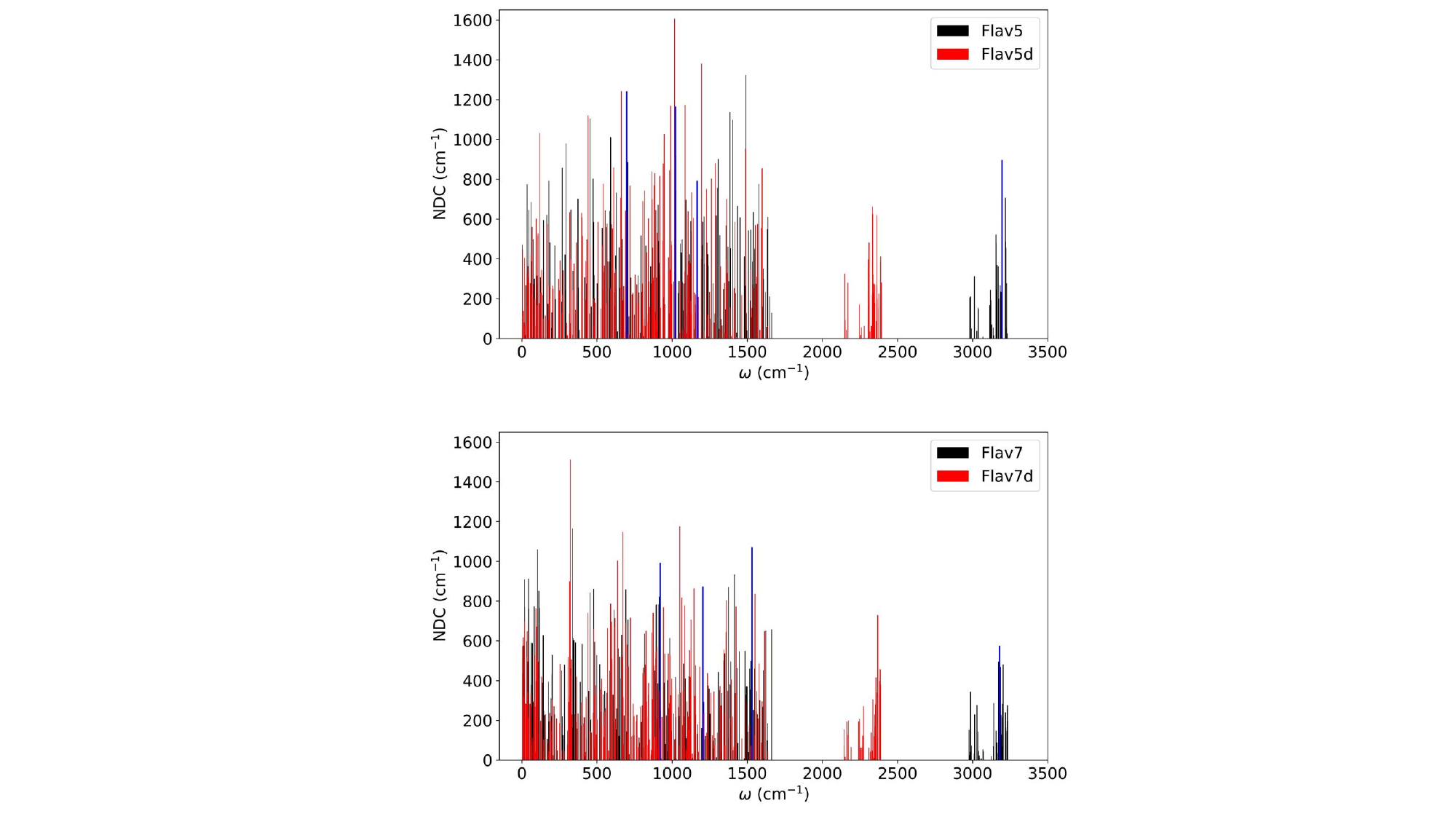}
	\caption{Nonadiabatic couplings projected on to each normal mode for the Flav5 (upper panel) and Flav7 (lower panel).   Black lines represent results for original molecules and red lines represent those for fully deuterated ones.   }
	\label{nacf7}
\end{figure}

For the optimized structure in the ${\rm S_1}$ state of each molecule, we calculated first NDC terms\cite{subot2011,subot2015} in cartesian coordinates, which were then projected onto different mass weighted normal modes of vibration.  The resulting expression is as follows: 
\be
{\rm NDC}_j= \frac{\left | \langle \tilde S_1| \left(\partial \hat H_e({\bf q})/\partial q_{1,j} |_{{\bf q}_1=0}\right) |S_0\rangle\right |}{\tilde E_1-E_0} , \label{eq:ndc_j}
\ee
where $q_{1,j}$ represents the $j$th mass-weighted normal mode in the ${\rm S}_1$ state around its minimum energy structure.  Detailed derivation of the above expression starting from full molecular Hamiltonian in the adiabatic basis\cite{jang-qmc,jang-jcp-nonad} is provided in SI.

In general, nonadiabatic couplings can be very different in different molecular conformations, especially when two energies are near degenerate.  However, for the case of  ${\rm S}_1\rightarrow {\rm S}_0$ transition, where the two electronic states are far from degeneracy and there are no other states of comparable energies  as considered here, it is most likely that the NDC values are fairly insensitive to the displacements from the minimum energy structure of the ${\rm S}_1$ state.    To examine this issue, we conducted additional calculations of NDC terms varying the values of nuclear coordinates, which indeed confirm our assumption and justify using the NDC values for the optimized structures of ${\rm S_1}$ (see SI and Figs. S8-S11).

 Under the same assumption, we also calculated the NDC terms for partially and fully deuterated Flav5 and Flav7 molecules. Figure \ref{nacf7} shows distributions of these projected NDC terms for original and fully deuterated molecules.  Results in Fig. \ref{nacf7} show broad distributions of NDC terms up to about $1,600\ {\rm cm^{-1}}$ and another well separated high frequency bands, which correspond to C-H (or C-D) stretching modes.  It is interesting to note that the contributions of these high frequency modes are rather small compared to those due to broad low frequency modes.  To examine the validity of our simple displaced harmonic oscillator models, we also calculated the Duschinsky rotation matrix\cite{chan2008,ratner2000} from the displacements of the nuclei in the normal modes.  Results provided in SI suggest that Duschinsky effects are not likely to affect the major conclusion of this work.

The transition due to NDC is intrinsically non-Condon type because it involves nuclear momentum operators.  On the other hand, the derivation of the energy gap is based on the FGR rate expression within the Condon approximation. Thus, we here introduce an effective coupling constant that accounts for the non-Condon effect in an average manner by using thermally averaged momentum as follows:
\ben
&&J_{eff} = \hbar \sum_{j=1}^{N_v} {\rm NDC}_{j}  \nonumber \\
&&\times \left (\frac{\sum_{v_j=0}^\infty  \langle p_{v_j}^2\rangle^{1/2} e^{-\hbar \omega_j (v_j+1/2)/(k_BT)}}{\sum_{v_j=0}^\infty e^{-\hbar \omega_j (v_j+1/2)/(k_BT)}} \right)   ,\  \label{eq:j-eff-def}
\een
\re{where ${\rm NDC}_{j}$ is given by Eq. (\ref{eq:ndc_j}) and $\langle p_{v_j}^2\rangle^{1/2}=\sqrt{\hbar \omega_j (v_j + \frac{1}{2})}$,  the root-mean-square momentum of each vibrational mode with vibrational quantum number $v_j$.  More detailed justification of the above expression and description of the calculation method are provided in SI.
The resulting values of $J_{eff}$ for the $|\tilde {\rm S_1}\rangle \rightarrow  |{\rm S_0}\rangle$ of different cases are provided in Table \ref{table2}. Note that in calculating these values we have used experimental values determined from emission spectra for $\tilde E_1-E_0$  whereas all other values were obtained from TD-DFT calculations employing the CAM-B3LYP functional.} 

\re{For the calculation of non-radiative transition rate }due to the effective electronic coupling $J_{eff}$ given by eq. \ref{eq:j-eff-def}, we \re{use an effective two state model coupled to harmonic oscillator baths represented by the following Hamiltonian}: 
\ben
\hat H_{nr} &&=\tilde E_1|\tilde S_1\rangle\langle \tilde S_1|+ \left ( E_0 +\sum_n \hbar\omega_n g_{n,01}(\hat b_n+\hat b_n^\dagger)  \right ) |S_0\rangle \langle S_0|\nonumber \\
&&+J_{eff}(|\tilde S_1\rangle\langle S_0|+|S_0\rangle\langle \tilde S_1|)  +\sum_n \hbar \omega_n\left (\hat b_n^\dagger \hat b_n+\frac{1}{2}\right) .  \label{eq:hamil-nr}
\een
Note that the above model Hamiltonian is the same as that used for the emission lineshape except that the two electronic states are now coupled by $J_{eff}$.

The FGR rate for the above model Hamiltonian is as follows:\cite{jang-exciton} 
\ben
k_{FG} &=& \frac{J_{eff}^2}{\hbar^2}\int^{\infty}_{-\infty} dt  \exp\Bigg \{ \frac{i}{\hbar} (\tilde E_1-E_0)t  \nonumber \\
&&\hspace{1in}  - G_{R,01}(t) -G_{I,01}(t)\Bigg \}  , \label{eq:rate-knr}
\een
where $G_{R,01}(t)$ and $G_{I,01}(t)$ are respectively defined by eqs. \ref{eq:gr-ij} and \ref{eq:gi-ij} with $i=0$ and $j=1$. 

 \begin{figure}
 \ \vspace{.4in}\\
\includegraphics[width=0.45\textwidth]{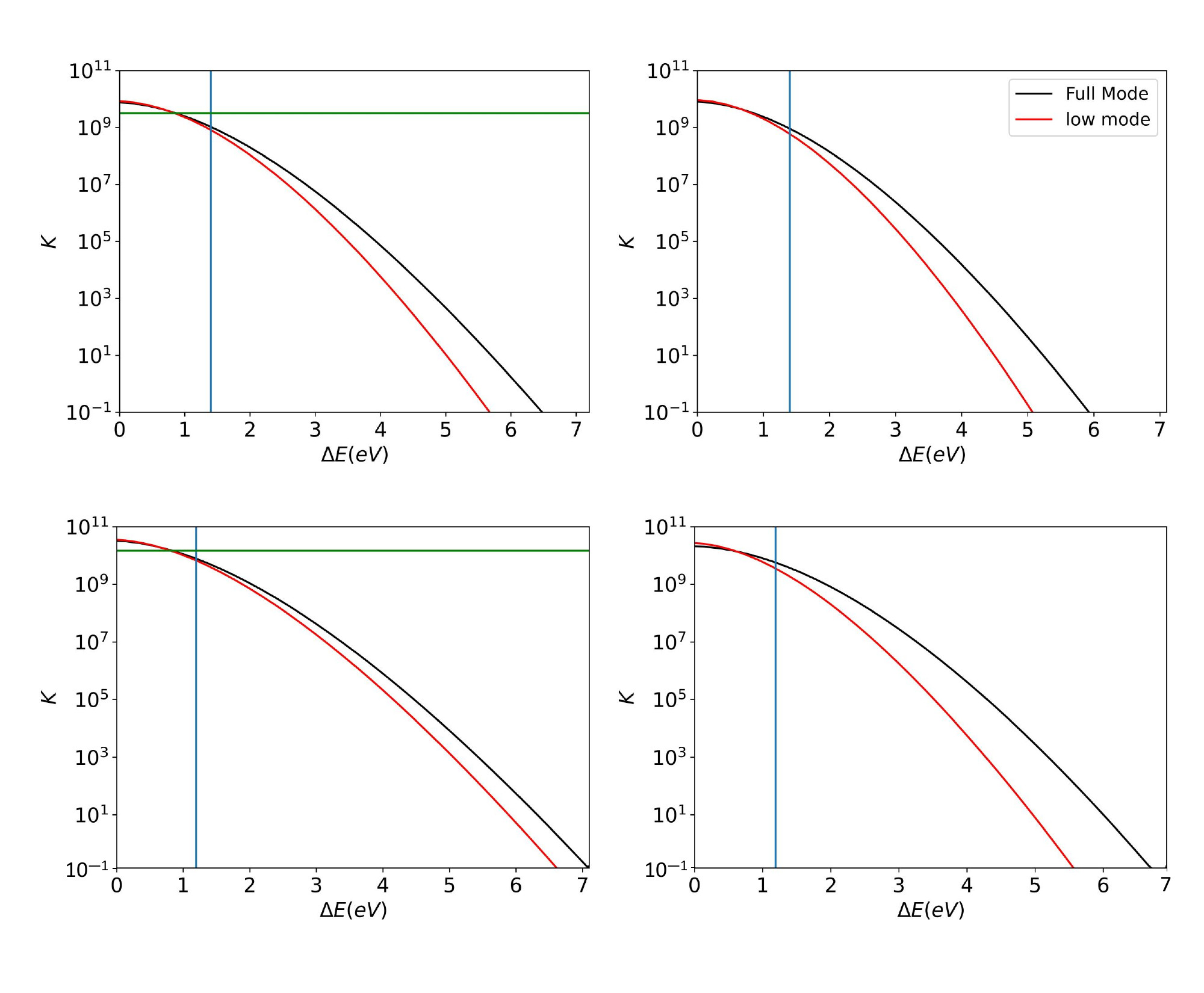} \vspace{.3in}\\
\caption{Nonadiabatic transition rates (in logarithmic scale) versus energy gap for the $S_1 \rightarrow S_0$ electronic transitions for both Flav5 (upper panel) and Flav7 (lower panel) dyes, where the vertical lines represent the experimental energy gap value between the two states, $\tilde E_1-E_0$, which are $1.378\ {\rm eV}$ for Flav5 and $1.177\ {\rm eV}$ for Flav7.  Rates versus the energy gap are shown for the original molecules (left) and deuterated (right) molecules.  The experimental rates for the original molecules are shown as horizontal lines. Red curves represent transition rates calculated with the high energy vibrational modes nullified.}
	\label{ks1s0}
\end{figure}

Figure \ref{ks1s0} shows calculated nonradiative rates for a range of energy gap values.  The actual values of energy gaps, which are shown as vertical lines, are $1.378\ {\rm eV}$ for Flav5 and $1.177\ {\rm eV}$ for Flav7.  These confirm that actual transitions indeed occur within ranges where the energy gap law behavior\cite{englman-mp18,jang-jcp155-1} is expected.    In order to determine the contribution of the high frequency C-H or C-D stretching modes to these rates, we also calculated rates with contributions of those modes to HR factors turned off.  The resulting data are shown as red lines.   Eliminating these high frequency contributions indeed reduces rates, more significantly as the gap increases. 

Table \ref{table2} shows values of theoretical rates determined at the \re{actual experimental energy gap values (vertical lines) }in Fig. \ref{ks1s0}.  Experimental nonradiative rates are also provided for comparison \re{(see SI for details of how these rates were determined)}.   The theoretical values are smaller than experimental ones, which \re{we attribute to contributions of Duschinsky rotation effects\cite{mebel-jpca103,niu-scc51,niu-jpca114,borrelli-jcp129,wang-jcp154} and non-Condon effects that are not accounted for by our model Hamiltonian and rate expression. In particular, we expect that inclusion of explicit consideration of momentum non-Condon effects in NDC, for which closed form rate expression is possible,\cite{jang-rhee-manu2023} can account for the significant portion of the discrepancy.  }  Nonetheless, the theoretical trends agree well with experimental ones.    For both original and fully deuterated molecules, we also determined rates for the hypothetical situation where high frequency C-H or (C-D) stretching modes do not contribute.  The resulting theoretical values, \re{which are shown in parentheses, clarify} that contributions of high frequency modes to the total NDCs and rates are relatively small for these cases.  
  
In addition to fully deuterated cases, we also calculated rates with partial deuteration. The Flav7 molecule denoted with m.D2 and m.D2.12 are those where the deuteration takes place at the aliphatic hydrogen whereas that with Ar.D10 has aromatic hydrogens deuterated. These changes result in modest reduction of effective couplings and rates.  This indicates that \re{deuteration of hydrogen atoms, which affects both HR factors and the effective couplings mainly through frequency changes of normal modes, indeed makes some contributions to the rate but are not a determining factor.  
We find that these computational results are also consistent with experimental data (see Table \ref{table2}).}

  \begin{table}[ht]
\caption{Theoretical data for the  effective electronic coupling $J_{eff}$, eq. \ref{eq:j-eff-def}, and the nonradiative FGR rate, eq. \ref{eq:rate-knr}, for $S_1\rightarrow S_0$ transitions for Flav5, Flav5${\rm_D}$ (fully deuterated), Flav7, and Flav7${\rm_D}$ (fully deuterated), and partially deuterated Flav7 (with different subscripts representing parts of deuteration).  Experimental nonradiative decay rates are also shown.  For the calculation of theoretical rates, experimental values of the energy gap, as listed in Table \ref{table1} were used.  Theoretical data for the hypothetical cases where the high frequency ${\rm C-H}$ (or ${\rm C-D}$) stretching modes are turned off are also shown in parentheses.  }
\begin{tabular}{lccc}
\hline
\hline
\multirow{2}{*}{Dye} &\multicolumn{2}{c}{Theory  (${\rm S_1\rightarrow S_0}$)}&\multirow{2}{*}{$k_{exp} ({\rm ns^{-1}})$} \\
\cline{2-3}
& \makebox[.6in]{$J_{eff}$ (cm$^{-1}$)}&$k_{nr}$(ns$^{-1}$) & \\
\hline
	Flav5 &425.3   & 1.041 & 3.2 \\      
	&(424.8)&(0.812) &\\
\hline
	Flav5$_{\rm D}$  &388.2   & 0.904 &  \\  
	&(388.1)&(0.600)& \\    
\hline              
	Flav7&918.8   &  8.350 & 14.7\\   
	&(917.4)&(7.297) & \\
\hline
	Flav7$_{\rm D}$ &756.8  & 6.081   & \\ 
	&(756.5)&(5.660)&\\
\hline
	Flav7$_{\rm mD2}$ &596.0  & 5.133  & 14.1\\ 
	Flav7$_{\rm mD2.12}$ &654.8  & 3.727  & 13.8\\ 
	Flav7$_{\rm ArD10}$ &482.3 & 1.984   & 13.5\\

\hline
\hline
        \end{tabular}
        \label{table2}

\end{table}

In order to examine the possibility of alternative nonradiative decay processes through triplet states, we also determined first (T$_1$) and second (T$_2$) excited triplet states, and calculated FGR rates for the following four possible transitions: ${\rm T_1\rightarrow S_0}$; ${\rm T_2\rightarrow S_0}$; ${\rm S_1\rightarrow T_1}$; ${\rm S_1\rightarrow T_2}$. \re{For the estimation of these rates, we used  the spectral densities of normal modes determined for the ${\rm S_1}$ state, spin-orbit coupling constants determined as detailed in SI for the effective electronic coupling constant in eq. \ref{eq:rate-knr}.   Figures S10 and S11  provide dependence of these theoretical rates on the value of the energy gap between singlet and triplet states. Table S2 of SI  provides corresponding theoretical values of the energy gap, effective coupling constants, and rates.   These show that all of the singlet-triplet spin-orbit couplings are less than $1\ {\rm cm^{-1}}$ and transition rates are less than $10^5{\rm\ s^{-1}}$.  Although additional approximations were involved in these calculations, it is not likely that more accurate calculations would significantly alter the order of  estimates.  Thus, we believe these results serve as strong evidence that transitions through triplet states do not have appreciable contributions for these dye molecules.  For these molecules, the possibility of going through conical intersections followed by activated processes is not plausible either because activated crossing in the excited state manifold, if a  rate determining step, should be insensitive to the energy gap between excited and ground electronic states and cannot explain the experimental observation.\cite{friedman-chem7}    }

In summary, we have conducted computational modeling and theoretical analyses of non-radiative decay rates from the first singlet excited states for two NIR/SWIR dye molecules, Flav5 and Flav7, which were shown to exhibit the energy gap law behavior.   The trends of our theoretical rates are in  good agreement with those of experimental data.  \re{Theoretical data provided in Table \ref{table2} also elucidate that the difference between the decay rates of Flav5 and Flav7 come from the combination of differences in the values of NDC terms and the energy gap, with the former playing more significant role.} Additional calculations for deuterated dye molecules corroborate our conclusion, and indeed serve as a good evidence that the cumulation of all NDC terms along all Franck-Condon modes in the excited electronic state provide the major route for the nonradiative decay of NIR/SWIR dye molecules exhibiting the energy gap law behavior.  Pending future efforts for broader class of  dye molecules and validation through higher level theoretical/computational studies\cite{mebel-jpca103,niu-scc51,niu-jpca114,borrelli-jcp129,wang-jcp154,alfonso-hernandez-pccp22} that \re{can provide more satisfactory description of}  Duschinsky and non-Condon effects, anharmonic contributions, \re{and more detailed dynamics simulation}, the present work \re{serves as a simple but essential theoretical model capable of providing semiquantitative description of} the nonradiative decays of NIR/SWIR dye molecules \re{exhibiting energy gap law behavior}.

\acknowledgements

S.J.J. acknowledges partial support from the National Science Foundation (CHE-1900170) during the initial stage of this project, and major support from the US Department of Energy, Office of Sciences, Office of Basic Energy Sciences (DE-SC0021413).  S.J.J. also acknowledges support from Korea Institute for Advanced Study through its KIAS Scholar program.  J.R.C., H.C.F., Y.I., C.G. and E.M.S. are supported by NSF Grant CHE-2204263. J.R.C. would also like to acknowledge the Cottrell foundation for support.  We also thank Prof. Daniel Neuhauser (UCLA) for supporting the computational resources for some of the calculations, which also used Expanse CPU at San Diego Supercomputer Center (SDSC) through allocation CHE230099 from the Advanced Cyberinfrastructure Coordination Ecosystem: Services \& Support (ACCESS) program being supported by the National Science Foundation grants (\#2138259, \#2138286, \#2138307, \#2137603, and \#2138296).






\providecommand{\latin}[1]{#1}
\makeatletter
\providecommand{\doi}
  {\begingroup\let\do\@makeother\dospecials
  \catcode`\{=1 \catcode`\}=2 \doi@aux}
\providecommand{\doi@aux}[1]{\endgroup\texttt{#1}}
\makeatother
\providecommand*\mcitethebibliography{\thebibliography}
\csname @ifundefined\endcsname{endmcitethebibliography}
  {\let\endmcitethebibliography\endthebibliography}{}


\ \vspace{5in}\\

\newpage
\newpage

\renewcommand{\theequation}{S\arabic{equation}}
\renewcommand{\thefigure}{S\arabic{figure}}
\renewcommand{\thetable}{S\arabic{table}}
\renewcommand\thepage{S\arabic{page}}
\setcounter{page}{1}
\setcounter{figure}{0}
\setcounter{table}{0}
\setcounter{equation}{0}

\newpage

\begin{widetext}
\begin{center}
{\bf \Large Supporting Information: Nonadiabatic derivative couplings through multiple Franck-Condon modes dictate the energy gap law for near and short-wave infrared dye molecules}
\end{center}
\end{widetext}

\noindent
{\bf S1. Additional details and results of computation}\vspace{.4in}\\
For all the calculations reported in the main text,  geometry optimizations  were  carried out employing the Gaussian 16 program package\cite{g16}, while QCHEM program package\cite{QCHEM} was used for the calculation of electronic excitation energies, Huang-Rhys (HR) factors, spin--orbit couplings and non-adiabatic couplings.     

 We also conducted additional calculations for different functionals in a  similar manner and also employing the ORCA 5.0 program.\cite{neese-jcp152,boerner-orca} For the calculation of the absorption energies, we used different structures optimized for different functionals.  On the other hand, for the calculation of emission energy, we used structure optimized for the CAM-B3LYP functional\cite{yanai-cpl393} since optimization for some functionals in the excited state was difficult to achieve.  Table \ref{table-s1} provides \rc{results of these calculation results, along with those for CAM-B3LYP,}  for the vertical absorption and emission energies in vacuum based on different functionals.

 \begin{table}
\caption{\re{Vertical electronic transition energies from optimized structures,  ${\rm S_0\rightarrow S_1}$ and ${\rm S_1\rightarrow S_0}$,  for Flav5  and Flav7 in vacuum based on different functionals.}}
\begin{tabular}{lccclc}
\hline
\hline
	System &XC--Functional& Vertical abs. (eV)& Vertical em.   (eV)  \\
	\hline
	Flav5 & CAM-B3LYP$^a$&2.2801&2.0935 &  \\ 
& CAM-B3LYP$^b$&2.279&2.091 &  \\
	& B3LYP$^a$ &  2.1038&1.5808&\\
	& B3LYP$^b$ &  2.106&1.581&\\
	&M06-2X$^b$ & 2.262 & 1.955 & \\
	&$\omega$B97X-D3BJ$^b$ & 2.577 & 2.275 & \\
	&SCAN$^b$ & 1.709 & 1.565 & \\
	\hline
	Flav7 & CAM-B3LYP$^a$ & 2.0599& 1.8835&  \\ 
        & CAM-B3LYP$^b$ & 2.061& 1.887&  \\ 
	& B3LYP$^a$ &1.9822  &1.4708 &\\
         & B3LYP$^b$ &1.990 &1.477 &\\
	&M06-2X$^b$ &2.022 & 1.733 & \\ 
	&$\omega$B97X-D3BJ$^b$ & 2.653 &2.026  & \\
	&SCAN$^b$ & 1.653 & 1.469 & \\
\hline
\hline
        \end{tabular}\\
        $^a$: Calculated using QCEM;  $^b$: Calculated using ORCA
        \label{table-s1}

\end{table}
As can be seen from the data in Table \ref{table-s1}, excitation energies calculated by CAM-B3LYP functional are much lower than those calculated by $\omega$B97X-D3BJ,\cite{grimme-jcc27,grimme-jcc32} higher but very close to those for M06-2X,\cite{zhao-tca120} and significantly larger than those for two other functionals. The results for B3LYP functional\cite{lee-prb37} are closer to experimental ones than those for CAM-B3LYP, but this is likely due to cancellation of two opposing errors in the former considering that the latter corrects the self-interaction errors of the former for charge transfer states.   The fact that  the SCAN functional\cite{sun-nc8} produces results  much closer to experimental ones than others, although not yet satisfactory,  may indicate that these are indeed better choices for the two dye molecules.  However,  preliminary calculations of HR factors and the estimation of lineshapes based on SCAN functional resulted in lineshapes that are qualitatively different from experimental ones.   Thus, it is not clear whether other properties of excited states calculated by this functional is  reliable.
 
  \begin{table}[ht]
\caption{\re{Vertical electronic transition energies from optimized structures,  ${\rm S_0\rightarrow S_1}$ and ${\rm S_1\rightarrow S_0}$,  for Flav5  and Flav7 in PCM solvent model of DCM based on different functionals. All calculations were conduced by ORCA.}}
\begin{tabular}{lccccc}
\hline
\hline
	System &XC--Functional& vertical abs (eV) & vertical emi  (eV)   \\
	\hline
	Flav5 & CAM-B3LYP&2.042&1.930 &  \\ 
	& B3LYP &  1.922&1.865&\\
	&M06-2X & 2.010& 1.909 & \\
	&$\omega$B97X-D3BJ & 2.269 & 1.945& \\
	&SCAN & 1.873 & 1.862 & \\
	\hline
	Flav7 & CAM-B3LYP & 1.937& 1.665&  \\ 
	& B3LYP &1.715  &1.674 &\\
	&M06-2X &1.874 & 1.658 & \\ 
	&$\omega$B97X-D3BJ & 2.488 &1.654  & \\
	&SCAN & 1.732 & 1.726& \\
\hline
\hline
        \end{tabular}\\
        \label{table-s2}

\end{table}

We also conducted calculations of excitation energies in the ORCA 5.0 program employing  a quantum-classical interface model called solvation model - density (SMD) model,\cite{marenich-jpcb113} where the full electron density is used to interact with the continuum solvent and is known be better than old continuum models.  It is important to note that the SCAN functional results in higher values of excitation energies when the solvation effect is included, which seems unphysical, whereas all other functionals result in lower values.  The reason for such performance for SCAN functional is unclear, but indicates that it cannot be used  as the best functional for all the properties of excited states we are interested in.     

Although the excitation energies based on the CAM-B3LYP are significantly larger than experimental ones, we believe the discrepancy is likely due to the long range effects of counter ions and correlated effects of explicit solvent molecules which do not affect the local electron-phonon couplings significantly.   This is supported from good agreements of theoretical lineshapes based on this functional with experimental ones.   For this reason, we chose to use CAM-B3LYP functional for the calculation of  the numerator of nonadiabatic derivative coupling term, Eq, (10) in the main text, whereas we used excitation energy values extracted from the experimental data as the denominator.    \vspace{.2in}\\

\begin{figure}[ht]
\includegraphics[width=.45\textwidth]{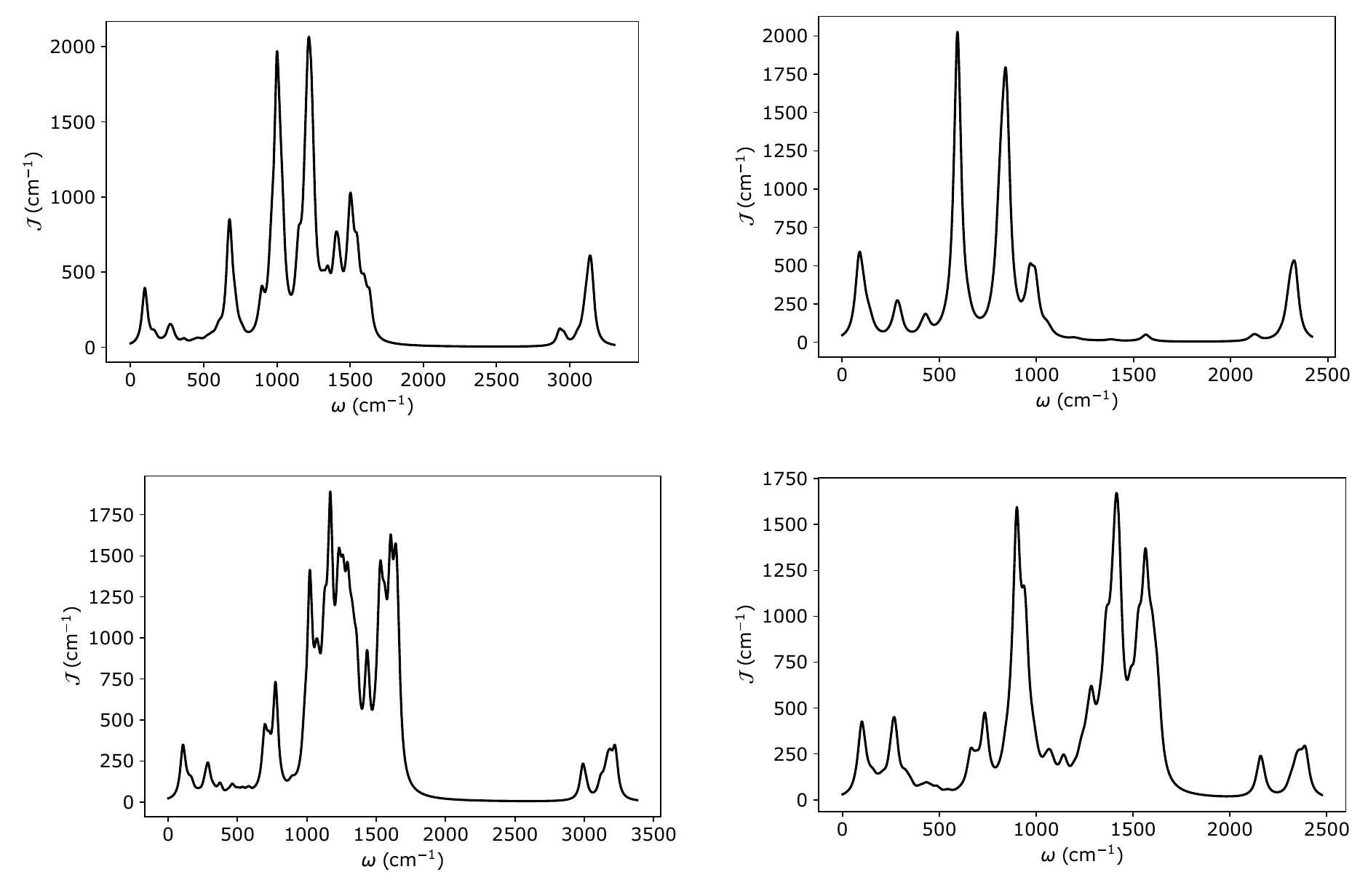}
	\caption{Spectral density ${\mathcal J}_{10}(\omega)$ (upper panels)  and ${\mathcal J}_{01}(\omega)$ (lower panels) for Flav5 for the isolated (left) and deuterated (right) molecules. }
	\label{js5}
\end{figure}

\begin{figure}[ht]
\includegraphics[width=.45\textwidth]{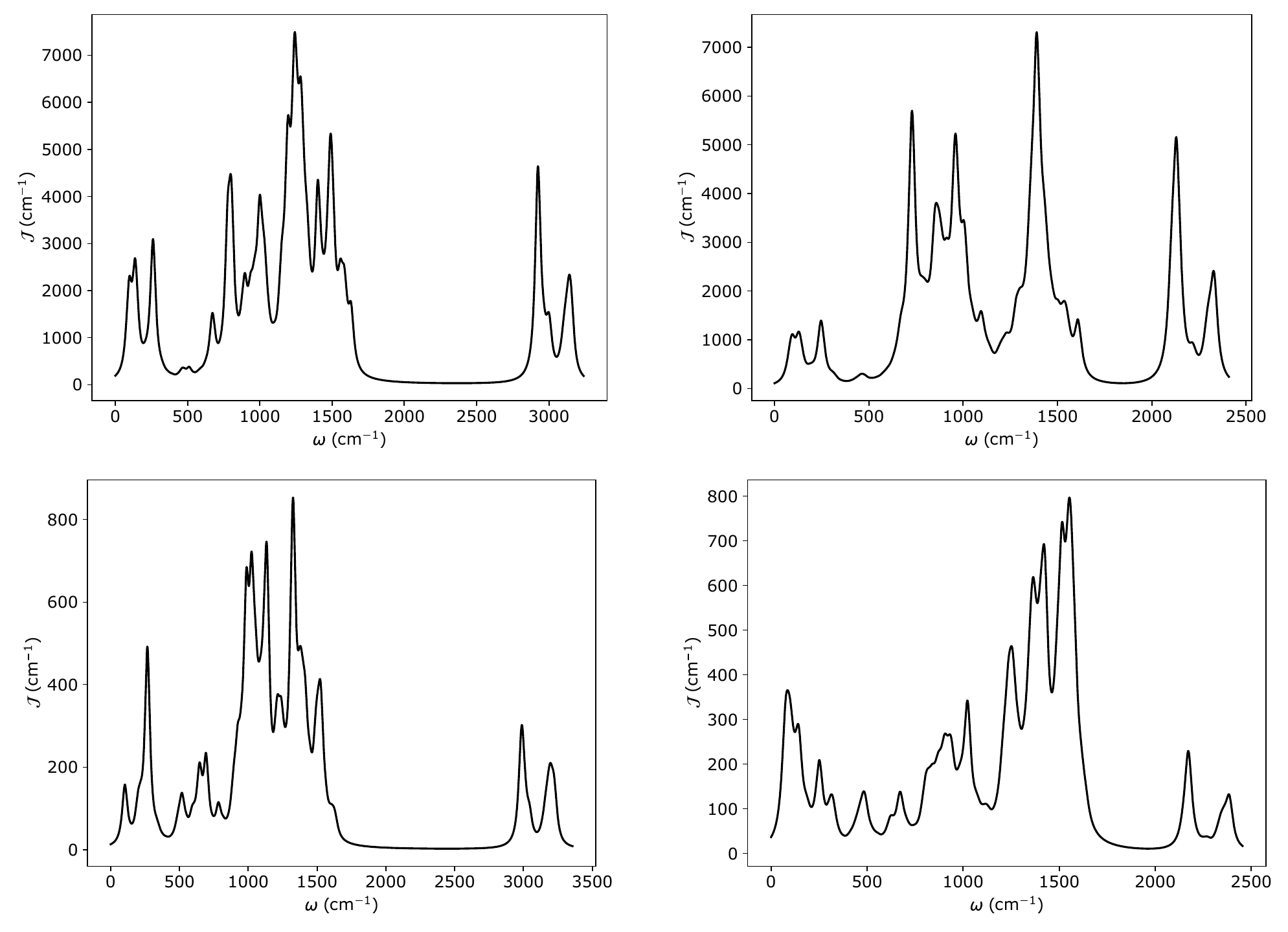}
	\caption{Spectral density ${\mathcal J}_{10}(\omega)$ (upper panels)  and ${\mathcal J}_{01}(\omega)$ (lower panels) for Flav7 for the isolated (left) deuterated (right) molecules.  }
	\label{js7}
\end{figure}

\begin{figure}[ht]
\includegraphics[width=.45\textwidth]{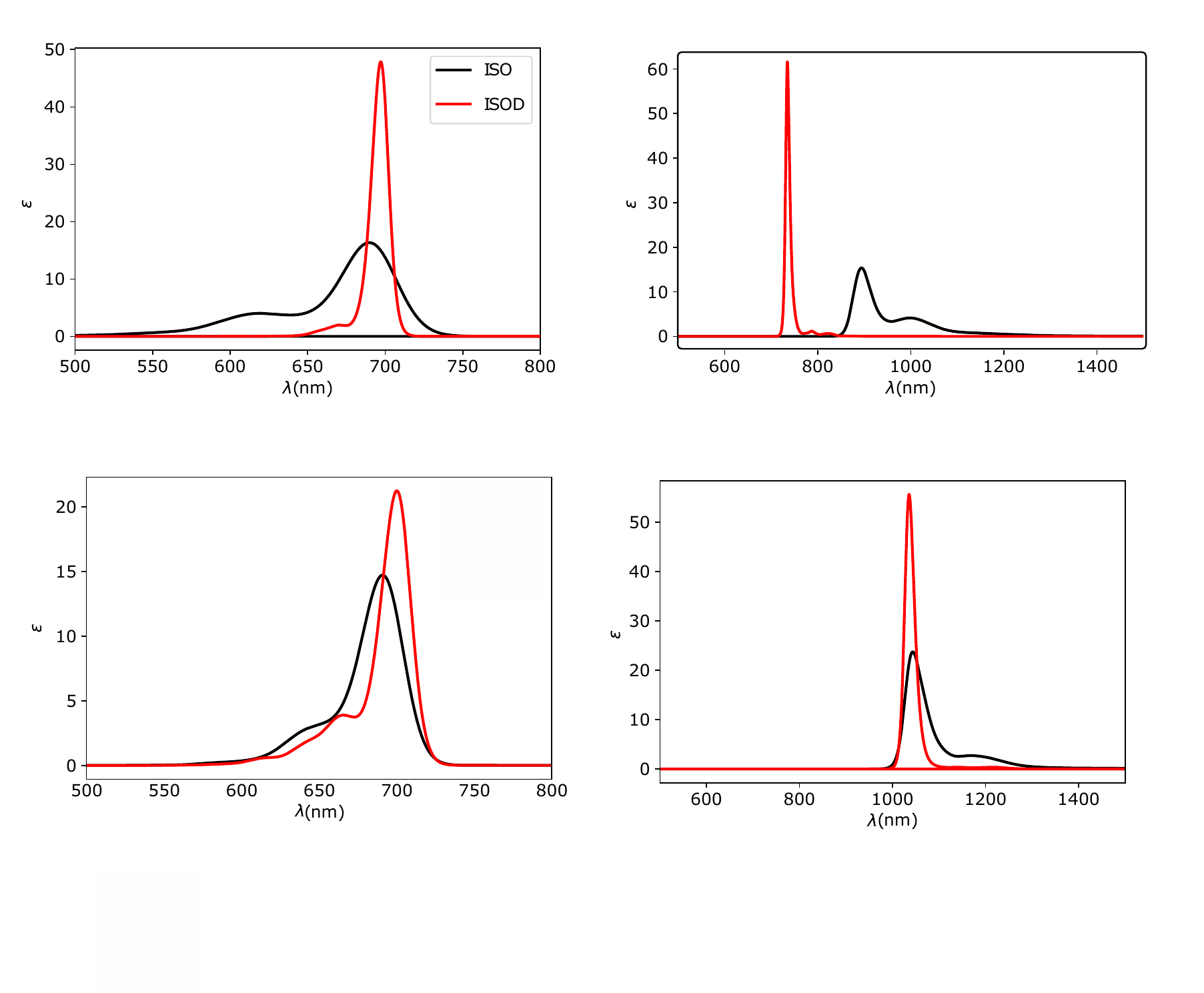}
	\caption{Comparison of the absorption and emission spectra line shape for both isolated and deuterated conditions for the two chromophores Flav5 (upper panels) and Flav7 (lower panels) .   }
	\label{Dls}
\end{figure}

\noindent
{\bf SII. Bath spectral densities}\vspace{.2in}\\
Figures \ref{js5} and \ref{js7} show the bath spectral densities calculated according to Eq. (11) of the main text.  Results for both original isolated molecules and fully deuterated molecules are shown. 
The corresponding absorption and emission lineshapes for fully deuterated molecules are shown in Fig. \ref{Dls}.  The lineshapes for original molecules without deuteration, which were provided in Fig. 4 of the main text, are also shown for comparison.  Note that normalizations of lineshapes in Fig. \ref{Dls} are different from those in Fig. 4 of the main text.

\begin{figure}[ht]
	\includegraphics[width=.45\textwidth]{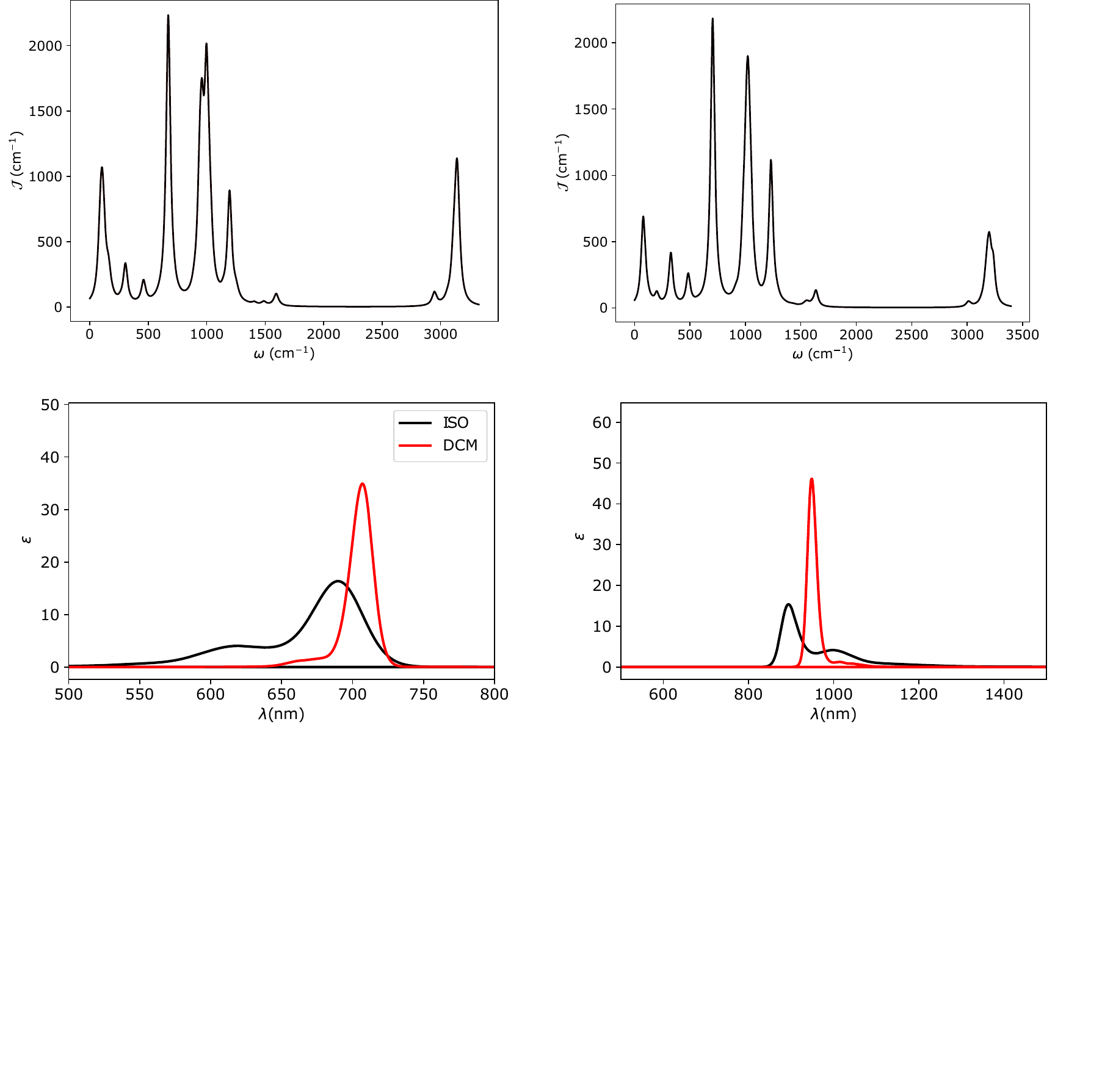}
	\caption{Spectral densities, ${\mathcal J}_{10}(\omega)$ and ${\mathcal J}_{01}(\omega)$ (upper panels), and absorption and emission line shapes (lower panels) of Flav5 molecule in PCM solvent model.  Lineshapes for isolated molecules are also shown for comparison.   }
	\label{ls-flav5}
\end{figure}

\begin{figure}[ht]
\includegraphics[width=.45\textwidth]{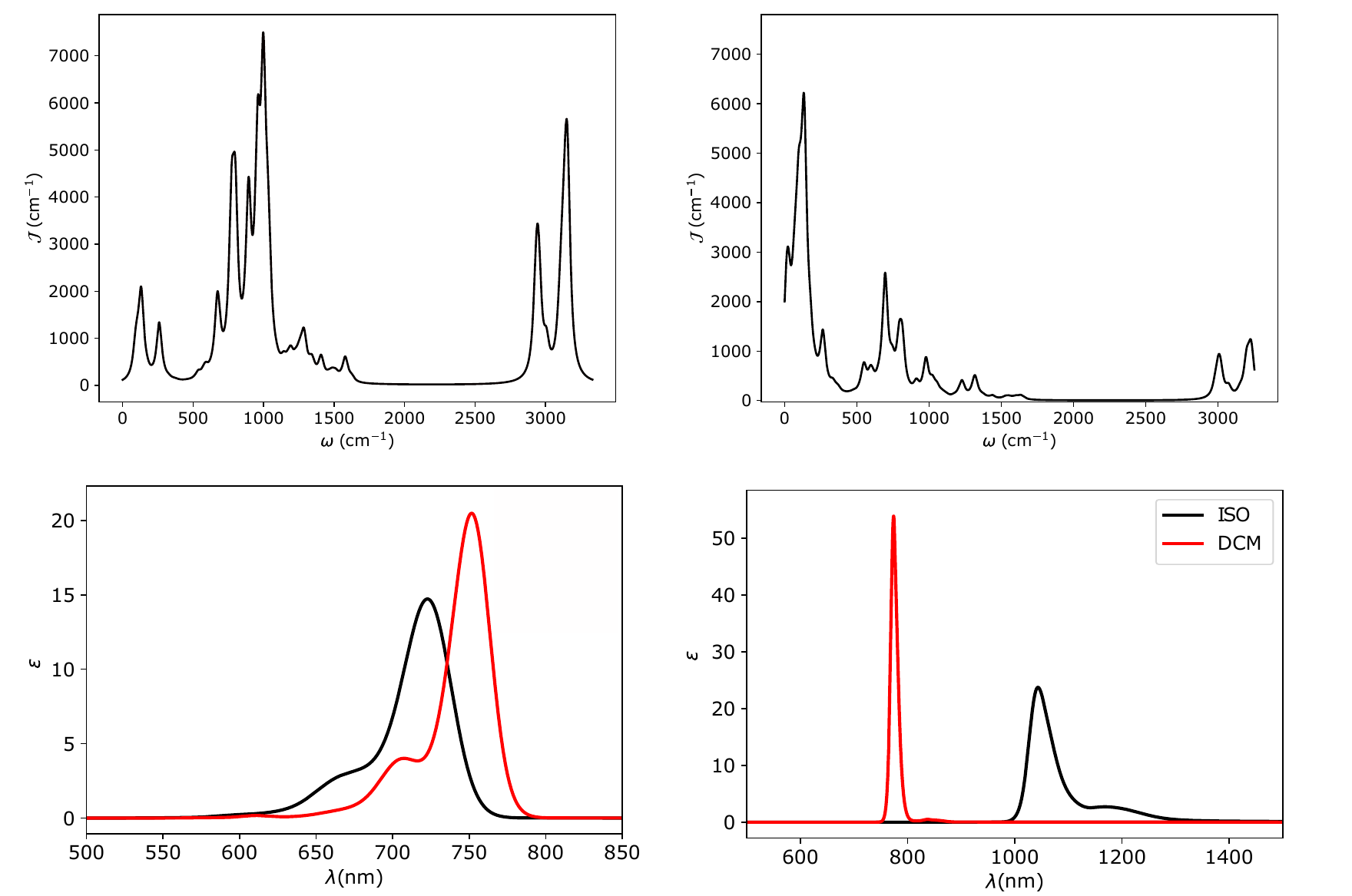}
	\caption{Spectral densities, ${\mathcal J}_{10}(\omega)$ and ${\mathcal J}_{01}(\omega)$ (upper panels), and absorption and emission line shapes (lower panels) of Flav7 molecule in PCM solvent model.  Lineshapes for isolated molecules are also shown for comparison.  }
	\label{ls-flav7}
\end{figure}

In addition to the calculation of gas phase molecules, we also conducted calculations of molecules in dichloromethane (DCM) solvent, modeled at the level of polarizable continuum model  (PCM)\cite{mennucci_book} with the value of the dielectric constant corresponding to that of DCM at room temperature, $\epsilon = 8.93$.    The resulting lineshapes and bath spectral densities are shown in Figs. \ref{ls-flav5} and \ref{ls-flav7}. While inclusion of solvent brings the excitation energy closer to experimental ones, the resulting lineshapes become even worse than isolated ones when compared with experimental data.  We attribute this to that the continuum solvation model restricts the vibrational motion of molecules and makes HR factors become smaller than what they are in actual atomistic solvent environments. \vspace{.2in}\\

\begin{figure}[ht]
	\includegraphics[width=.45\textwidth]{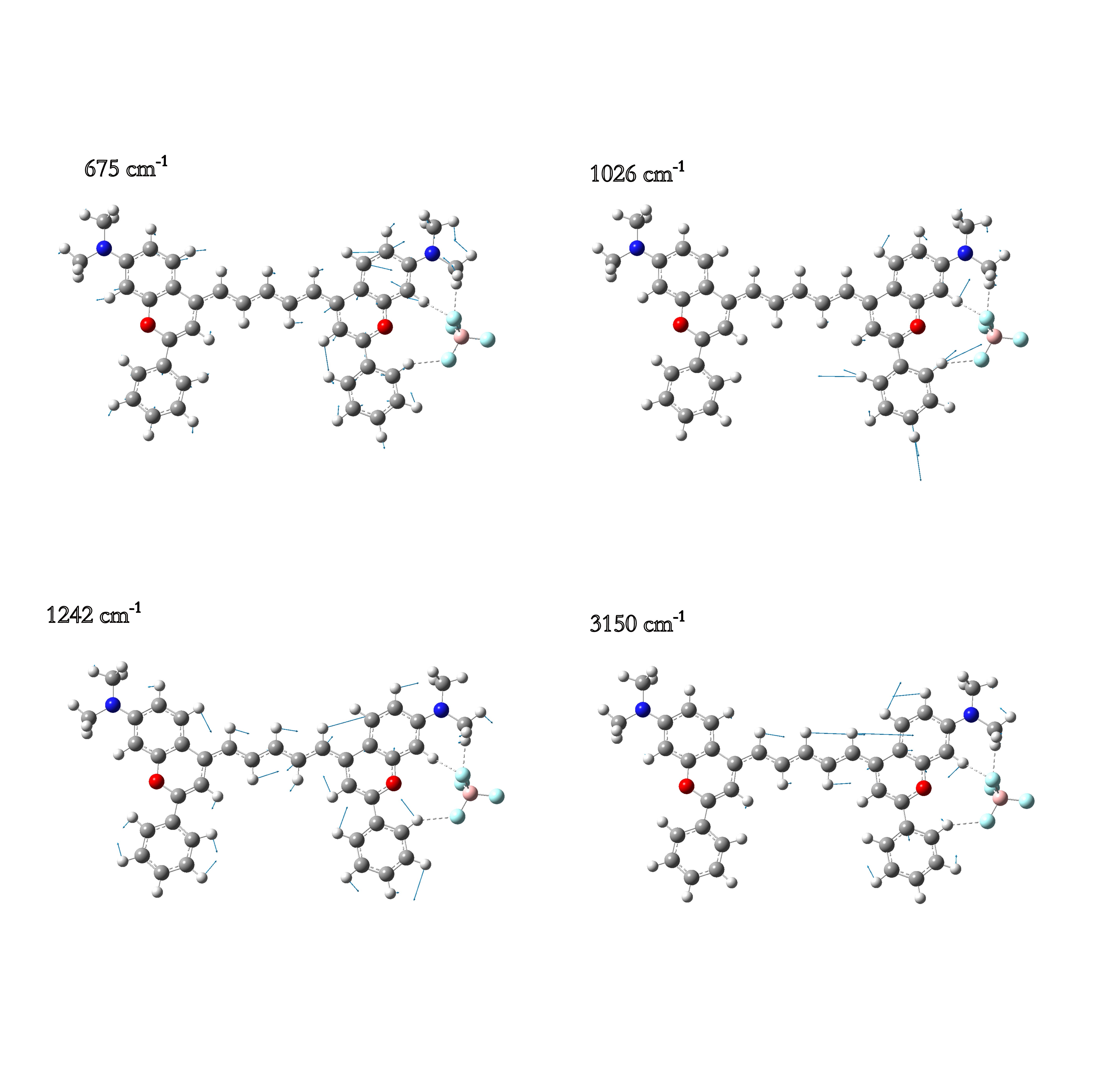}
	\caption{Displacement vectors for the main vibrational modes contributing to the line shape representation of Flav5 molecule.  }
	\label{nm-flav5}
\end{figure}
\begin{figure}[ht]
	\includegraphics[width=.45\textwidth]{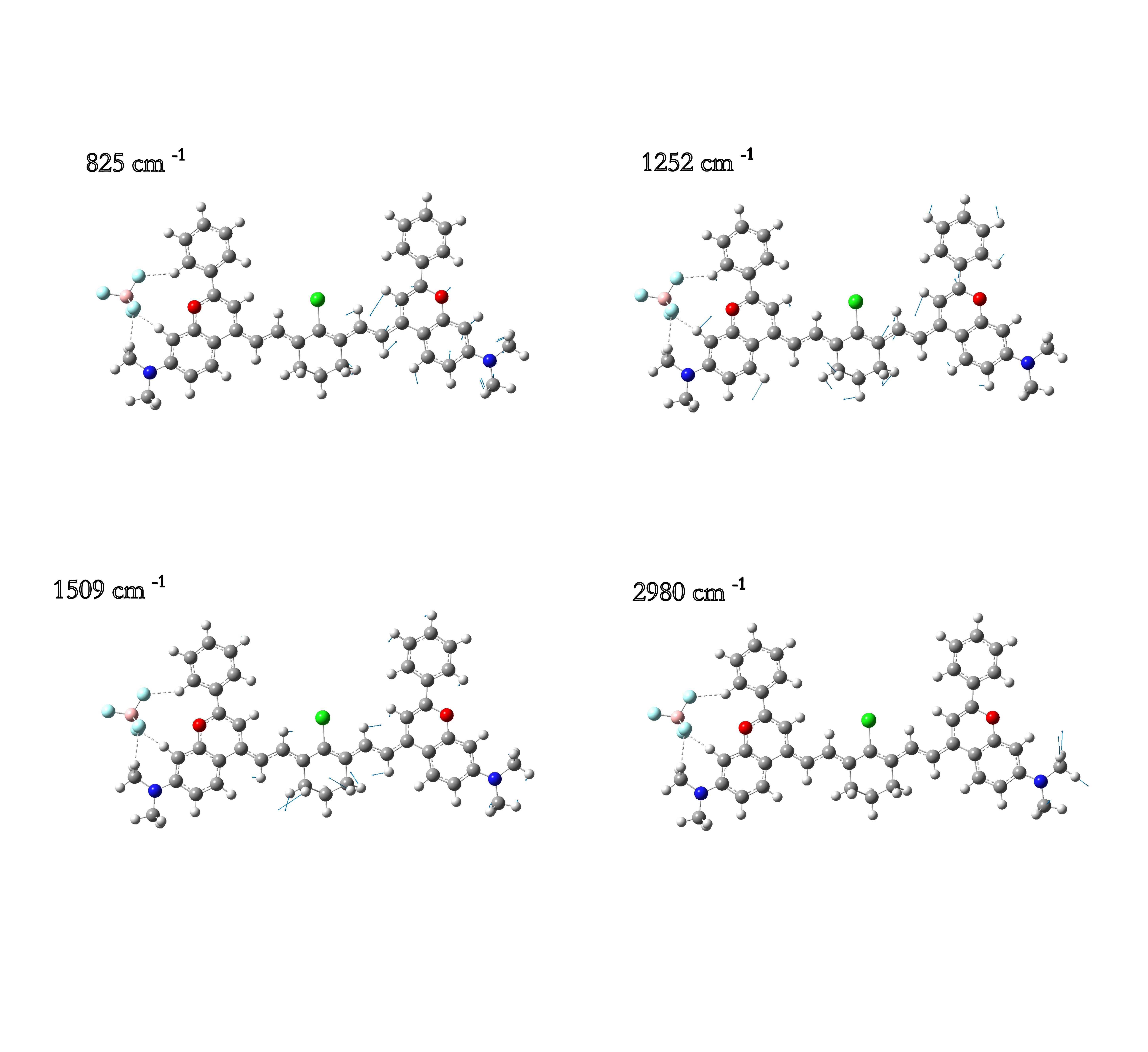}
	\caption{Displacement vectors for the main vibrational modes contributing to the line shape representation of Flav7 molecule.  }
	\label{nm-flav7}
\end{figure}

\noindent
{A. Normal modes}\vspace{.2in}\\
Major normal modes contributing to the lineshapes for Flav5 and Flav7 are depicted in Figs. \ref{nm-flav5} and \ref{nm-flav7}. 

\ \vspace{.4in}\\

{\bf B. Derivative nonadiabatic coupling between ${\rm S_1}$ and ${\rm S_0}$}\vspace{.2in}\\
We here provide a derivation and justification of the effective non-adiabatic coupling, Eq. (12) in the main text.  Consider a molecule consisting of $N_e$ electrons and $N_u$ nuclei.  The corresponding Hamiltonian (in atomic units) can be expressed as\cite{jang-qmc} 
\be
\hat H=\hat H_{en}+\hat H_{nu} ,
\ee
where 
\ben
&&\hat H_{en}=\sum_{\mu=1}^{N_e} \frac{\hat {\bf p}_\mu^2}{2} -\sum_{\mu=1}^{N_e}\sum_{c=1}^{N_u} \frac{Z_c e^2}{|\hat {\bf r}_\mu-\hat {\bf R}_c|^2} \nonumber \\
&&\hspace{.5in}+\frac{1}{2}\sum_{\mu=1}^{N_e}\sum_{\nu\neq \mu} \frac{1}{|\hat {\bf r}_\mu-\hat {\bf r}_\nu|} , \label{eq:h-en}\\
&&\hat H_{nn}=\sum_{c=1}^{N_u} \frac{\hat {\bf P}_c}{2M_c}+\frac{1}{2}\sum_{c=1}^{N_u}\sum_{c'\neq c} \frac{Z_cZ_{c'}}{|\hat {\bf R}_c-\hat {\bf R}_{c'}|}  .
\een
Let us consider the adiabatic electronic Hamiltonian $\hat H_{en}({\bf R})$, which is the same as  Eq. (\ref{eq:h-en}) except that all the nuclear position operator vector $\hat {\bf R}\equiv (\hat {\bf R}_1,\cdots, \hat {\bf R}_{N_u})^T$ in Eq. (\ref{eq:h-en}) is replaced with corresponding vector parameter: ${\bf R}\equiv ( {\bf R}_1,\cdots, {\bf R}_{N_u})^T$.  Then, one can define the following adiabatic electronic states and eigenvalues:
\be
\hat H_{en}({\bf R})|\psi_{e,j}({\bf R})\rangle =E_{e,j}({\bf R})|\psi_{e,j}({\bf R})\rangle , 
\ee
where $j$ denotes collectively the set of all quantum numbers needed to completely define all the electronic states. Thus, the following completeness relation holds in the electronic space:
\be
\hat 1_e=\sum_j |\psi_{e,j}({\bf R})\rangle \langle \psi_{e,j}({\bf R})| .
\ee
It is possible to express $\hat H$ in the basis of the above adiabatic electronic states,\cite{jang-jcp-nonad} which naturally result in the following decomposition of the Hamiltonian into adiabatic and non-adiabatic terms as follows: 
\be
\hat H=\hat H_{ad}+\frac{\hbar}{2} \sum_\alpha \left ( \hat P_\alpha \hat F_\alpha+\hat F_\alpha \hat P_\alpha \right)+\hbar^2 \hat S , \label{eq:h-ad-rep}
\ee
where $\hat H_{ad}$ is the adiabatic approximation given by
\ben
\hat H_{ad}&=&\int d{\bf R} \sum_k |{\bf R}\rangle |\psi_{e,k} ({\bf R})\rangle \Big \{ -\sum_\alpha \frac{\hbar^2}{2M_\alpha} \frac{\partial^2}{\partial R_\alpha^2} \nonumber \\
&&\hspace{1.in} +U_k({\bf R})\Big \}\langle \psi_{e,k} ({\bf R})|\langle {\bf R}| ,
\een
with $U_k({\bf R})=E_{e,k}({\bf R})+\frac{1}{2}\sum_{c=1}^{N_u}\sum_{c'\neq c} \frac{Z_cZ_{c'}}{|{\bf R}_c-{\bf R}_{c'}|}$ and $\alpha$ denoting each one dimensional cartesian direction of a nucleus.  Thus, $\alpha=1,\cdots, 3N_u$. In Eq. (\ref{eq:h-ad-rep}), $\hat P_\alpha$ is the nuclear momentum operator along each direction, $\hat F_\alpha$ is the first derivative nonadiabatic coupling term, and $\hat S$ is the second derivative nonadiabatic coupling term. In more detail, these derivative coupling terms can be expressed as\cite{jang-jcp-nonad}  
\ben
&&\hat F_\alpha=\int d{\bf R} \sum_k\sum_{k'} |{\bf R}\rangle |\psi_{e,k}({\bf R})\rangle F_{\alpha,kk'}({\bf R})\langle \psi_{e,k'} ({\bf R})| \langle {\bf R}| ,  \nonumber \\ \\
&&\hat S=\int d{\bf R} \sum_k\sum_{k'} |{\bf R}\rangle |\psi_{e,k}({\bf R})\rangle S_{kk'}({\bf R})\langle \psi_{e,k'} ({\bf R})| \langle {\bf R}| , \nonumber \\
\een
where
\ben
&&F_{\alpha,kk'} ({\bf R})=-\frac{i}{M_\alpha} \langle \psi_{e,k}({\bf R})| \left (\frac{\partial}{\partial R_\alpha} |\psi_{e,k'}({\bf R})\rangle \right) ,\\
&&S_{kk'}({\bf R})=\frac{1}{2}\sum_\alpha \sum_{k''} M_\alpha F_{\alpha,kk''}({\bf R}) F_{\alpha,k''k'}({\bf R}) .
\een

For the case $k\neq k'$ and $E_{e,k}({\bf R})$ and $E_{e,k'}({\bf R})$ are non-degenerate, the Hellmann-Feynman theorem\cite{hellmann,feynman-pr56} can be used to obtain an alternative expression for $F_{\alpha,kk'}({\bf R})$, which is obtained by taking derivative of the following condition: $\langle \psi_{e,k}({\bf R})|\hat H_e({\bf R})|\psi_{e,k'}({\bf R})\rangle =E_{e,k}({\bf R})\delta_{kk'}$.  The resulting expression is
\be
F_{\alpha,kk'}({\bf R})=\frac{i}{M_\alpha} \frac{\langle \psi_{e,k}({\bf R})|\left (\partial \hat H_e({\bf R}) /\partial R_\alpha\right) |\psi_{e,k'}({\bf R})\rangle}{E_{e,k}({\bf R})-E_{e,k'}({\bf R})} . \label{eq:f-alpha-1}
\ee 

Let us now consider the case where there are two major adiabatic electronic states with significant nonadiabatic couplings, which we denote as $1$ and $2$.  
Then, the adiabatic Hamiltonian including the diagonal components of the second derivative nonadiabatic terms can be expressed as
\begin{widetext}
\ben
&&\hat H_{ad}+\hbar^2 \hat S=\int d{\bf R} |{\bf R}\rangle |\psi_{e,1} ({\bf R})\rangle \Big \{ -\sum_\alpha \frac{\hbar^2}{2M_\alpha} \frac{\partial^2}{\partial R_\alpha^2}  +U_1({\bf R})+S_{11}({\bf R})\Big \}\langle \psi_{e,1} ({\bf R})|\langle {\bf R}|  \nonumber \\
&&\hspace{.2in}+\int d{\bf R} |{\bf R}\rangle |\psi_{e,2} ({\bf R})\rangle \Big \{ -\sum_\alpha \frac{\hbar^2}{2M_\alpha} \frac{\partial^2}{\partial R_\alpha^2}  +U_2({\bf R})+S_{22}({\bf R})\Big \}\langle \psi_{e,2} ({\bf R})|\langle {\bf R}| .
\een
\end{widetext}
Assuming that all $\langle {\bf r}|\psi_{e,k}({\bf R})\rangle$'s are real valued functions, $F_{\alpha,11}({\bf R})=F_{\alpha,22}({\bf R})=0$.  As a result, $S_{12}({\bf R})=S_{21}({\bf R})=0$.  Thus, the only terms that couple the two adiabatic electronic states are the first derivative nonadiabatic terms as follows: 
\be
\hat H_c=\frac{\hbar}{2}\sum_\alpha \left (\hat P_\alpha\hat F_\alpha+\hat F_\alpha\hat P_\alpha \right)  , \label{eq:hc}
\ee
where
\ben
\hat F_\alpha&=&\int d{\bf R}  |{\bf R}\rangle \Big ( F_{\alpha,12}({\bf R}) |\psi_{e,1}({\bf R})\rangle \langle \psi_{e,2} ({\bf R})| \nonumber \\
&& + F_{\alpha,21}({\bf R})|\psi_{e,2}({\bf R})\rangle \langle \psi_{e,1} ({\bf R})|  \Big )\langle {\bf R}|  .
\een
Note that $F_{\alpha,12}({\bf R})=[F_{\alpha,21}({\bf R})]^*$, which makes the above operator Hermitian. 

Let us assume that the initial state is near the minimum energy nuclear coordinates of the $E_{e,1}({\bf R}$), which is denoted as ${\bf R}_1^*$ and that the adiabatic electronic states in this vicinity can be approximated as those for the particular nuclear coordinates, which are respectively denoted as $|\psi_{k,e}\rangle =|\psi_{k,e}({\bf R}_1^*)\rangle$.  Then, the adiabatic Hamiltonian can be approximated as\cite{jang-jcp-nonad} 
\ben
&&\hat H_{ad}+\hbar^2 \hat S\approx  \Big \{ \sum_\alpha \frac{\hat P_\alpha^2}{2M_\alpha}   +U_1(\hat {\bf R})+S_{11}({\bf R}_1^*)\Big \}|\psi_{e,1} \rangle\langle \psi_{e,1}|   \nonumber \\
&&\hspace{.8in}+  \Big \{ \sum_\alpha \frac{\hat P_\alpha^2}{2M_\alpha}   +U_2(\hat {\bf R})+S_{22}({\bf R}_1^*)\Big \}|\psi_{e,2}\rangle \langle \psi_{e,2} | , \nonumber \\ \label{eq:h-ad-1} 
\een
and the first derivative nonadiabatic coupling operator $\hat F_\alpha$ in Eq. (\ref{eq:hc}) can be approximated as
\be
\hat F_\alpha\approx  F_{\alpha,12}({\bf R}_{1}^*) |\psi_{e,1}\rangle \langle \psi_{e,2}|  + F_{\alpha,21}({\bf R}_1^*)|\psi_{e,2}\rangle \langle \psi_{e,1} |    . \label{eq:f-1}
\ee

Let us introduce mass weighted coordinates $\tilde R_\alpha=\sqrt{M_\alpha}R_\alpha$ and the corresponding canonical momentum, $\tilde P_\alpha=P_\alpha/\sqrt{M_\alpha}$.   Then, 
Eqs. (\ref{eq:h-ad-1}) can be expressed as
\ben
&&\hat H_{ad}+\hbar^2 \hat S\approx  \Big \{ \sum_\alpha \frac{\hat {\tilde P}_\alpha^2}{2}   +U_1(\hat {\tilde {\bf R}})+S_{11}(\tilde {\bf R}_1^*)\Big \}|\psi_{e,1} \rangle\langle \psi_{e,1}|   \nonumber \\
&&\hspace{.8in}+  \Big \{ \sum_\alpha \frac{\hat {\tilde P}_\alpha^2}{2}   +U_2(\hat {\tilde {\bf R}})+S_{22}(\tilde {\bf R}_1^*)\Big \}|\psi_{e,2}\rangle \langle \psi_{e,2} | , \nonumber \\ \label{eq:h-ad-2}
\een
Similarly,  $F_{\alpha,12}({\bf R}_{1}^*)$ in Eq. (\ref{eq:f-1}) can be expressed as
\be
F_{\alpha,12}=\frac{i}{\sqrt{M_\alpha}} \frac{\langle \psi_{e,1}|\left (\partial \hat H_e(\tilde {\bf R})/\partial \tilde R_\alpha\right)|_{{\bf R}={\bf R}_1^*}  |\psi_{e,2}\rangle}{E_{e,1}({\bf R}_1^*)-E_{e,2}({\bf R}_1^*)} . \label{eq:f-alpha-2}
\ee 

Expanding $U_1(\hat {\tilde {\bf R}})$ around ${\bf R}_1^*$ up to the second order and diagonalizing the Hessian matrix, we can introduce the following normal coordinates:
\ben
q_{j}&=&\sum_\alpha L_{1,j\alpha} (\tilde R_\alpha -\tilde R_{1,\alpha}^*)\nonumber \\
&=&\sum_\alpha L_{1,j\alpha}\sqrt{M_\alpha} (R_\alpha - R_{1,\alpha}^*) , j=1,\cdots, N_v ,\nonumber \\
\een
with $N_v$ being the total number of normal mode vibrations.  Let us denote the corresponding canonical momentum as $p_{j}$.  Then, 
\be
\sum_\alpha \frac{\hat {\tilde P}_\alpha^2}{2}   +U_1(\hat {\bf R}) \approx \sum_j \left (\frac{1}{2}\hat p_j^2+\frac{\omega_j^2}{2}\hat q_j^2 \right)+U_1({\bf R}_1^*) ,
\ee
\ben
\hat H_c&=&\hbar\sum_\alpha \sum_{j}\sum_{j'} L_{1,j\alpha}L_{1,j'\alpha}  \hat p_j \Big \{ F_{j',12}  |\psi_{e,1}\rangle\langle \psi_{e,2}| \nonumber \\
&&\hspace{1in}+F_{j',21} |\psi_{e,2}\rangle\langle \psi_{e,1}| \Big \} \nonumber \\
&=&\hbar \sum_{j} \hat p_j \Big \{ {F}_{j,12} |\psi_{e,1}\rangle\langle \psi_{e,2}|+F_{j,21} |\psi_{e,2}\rangle\langle \psi_{e,1}| \Big \}  , \nonumber \\ \label{eq:hc-1}
\een
where
\ben
F_{j,12}&=&i \frac{\langle \psi_{e,1}|\left (\partial \hat H_e({\bf R})/\partial  q_j |_{{\bf R}={\bf R}_1^*} \right )  |\psi_{e,2}\rangle}{E_{e,1}({\bf R}_1^*)-E_{e,2}({\bf R}_1^*)} \nonumber \\
&=&i\sum_\alpha \frac{L_{1,j\alpha}}{\sqrt{M_\alpha}}  \frac{\langle \psi_{e,1}|\left (\partial \hat H_e({\bf R})/\partial  R_\alpha |_{{\bf R}={\bf R}_1^*} \right )  |\psi_{e,2}\rangle}{E_{e,1}({\bf R}_1^*)-E_{e,2}({\bf R}_1^*)}  . \nonumber \\ \label{eq:f-alpha-2}
\een 

Equation (\ref{eq:hc-1}) shows that the first derivative non-adiabatic coupling term still includes momentum operator along each normal mode even within the crude adiabatic approximation.  For the purpose of using Fermi's golden rule within the Condon approximation, an average value of the coupling as defined below can be used. 
\be
\hat H_c\approx \hbar \left (\sum_{j} \langle \hat p_{j}^2 \rangle^{1/2} |{F}_{j,12}|\right) \left (|\psi_{e,1}\rangle\langle \psi_{e,2}|+ |\psi_{e,2}\rangle\langle \psi_{e,1}| \right ) . \label{eq:hc-2}
\ee
where $\langle \hat p_{j}^2 \rangle$ is the thermal average over the distribution of vibrational states for the initial electronic state.

For the case of nonradiative decay from ${\rm S}_1$ to ${\rm S}_0$, $|\psi_{e,1}\rangle=|\tilde S_1\rangle$, $|\psi_{e,2}\rangle=|S_0\rangle$, $E_{e,1}({\bf R}_1^*)=\tilde E_1$, and $E_{e,2}({\bf R}_1^*)=E_0$, Eq. (\ref{eq:hc-1}) is expressed as  
\begin{widetext}
\be
\hat H_c\approx \hbar \left (\sum_{j} \langle \hat p_{j}^2 \rangle^{1/2} \frac{\left |\sum_\alpha \frac{L_{1,j\alpha}}{\sqrt{M_\alpha}}\langle \tilde S_1|\left (\partial \hat H_e({\bf R})/\partial  R_\alpha |_{{\bf R}={\bf R}_1^*} \right ) |S_0\rangle \right |}{\tilde E_{1}-E_{0}}\right) \left (|\tilde S_{1}\rangle\langle S_0|+ | S_0\rangle\langle \tilde S_1| \right ) . \label{eq:hc-2}
\ee
\end{widetext}
In this work, the matrix element of $\partial \hat H_e({\bf R})/\partial R_\alpha$ in the above equation is calculated within the adiabatic linear response TD-DFT method.   The detailed ansatz used for deriving the expression can be found in the literature \cite{subot2011,subot2015}. \vspace{.2in}\\

\begin{figure}
\includegraphics[width=.45\textwidth]{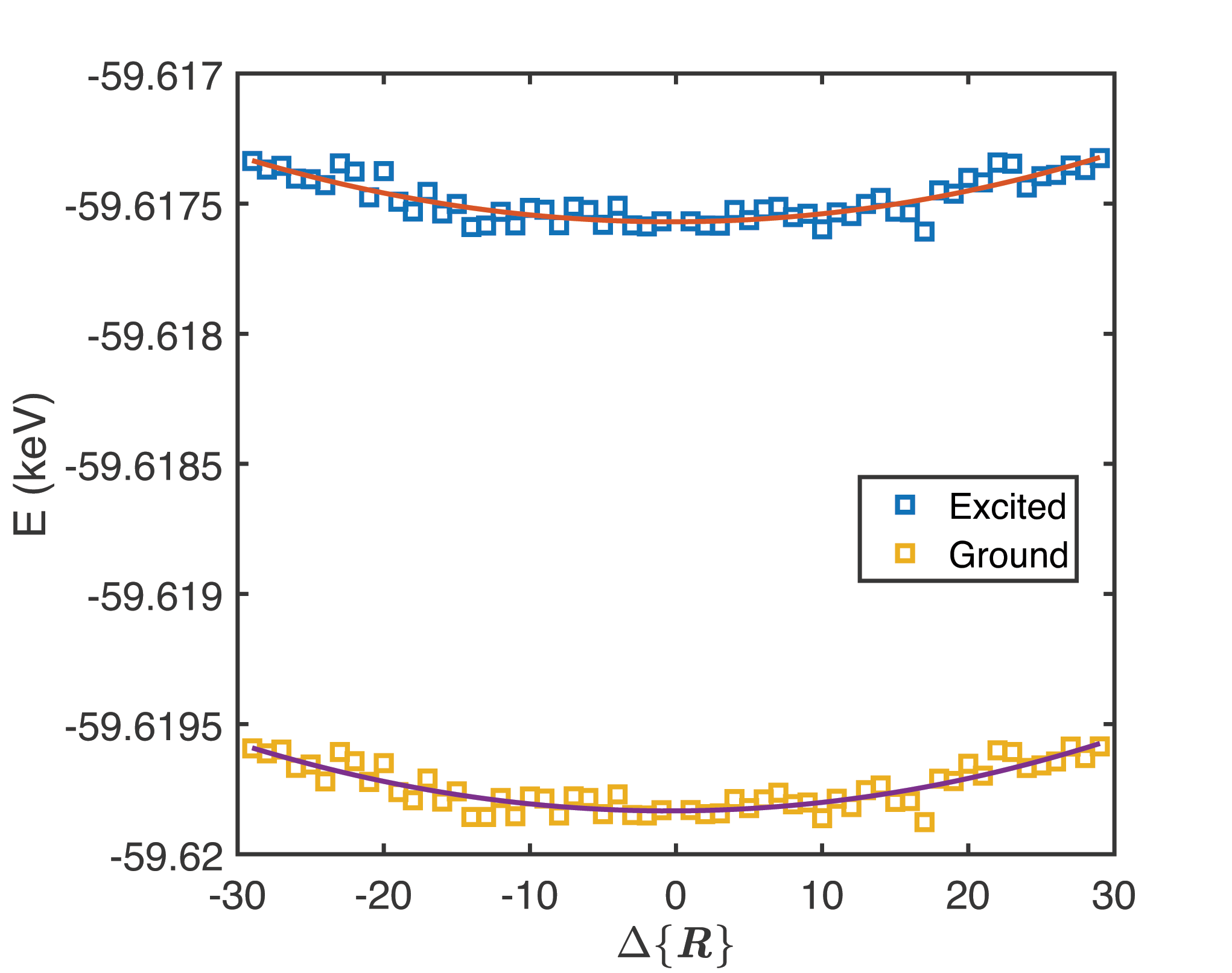} \includegraphics[width=.4\textwidth]{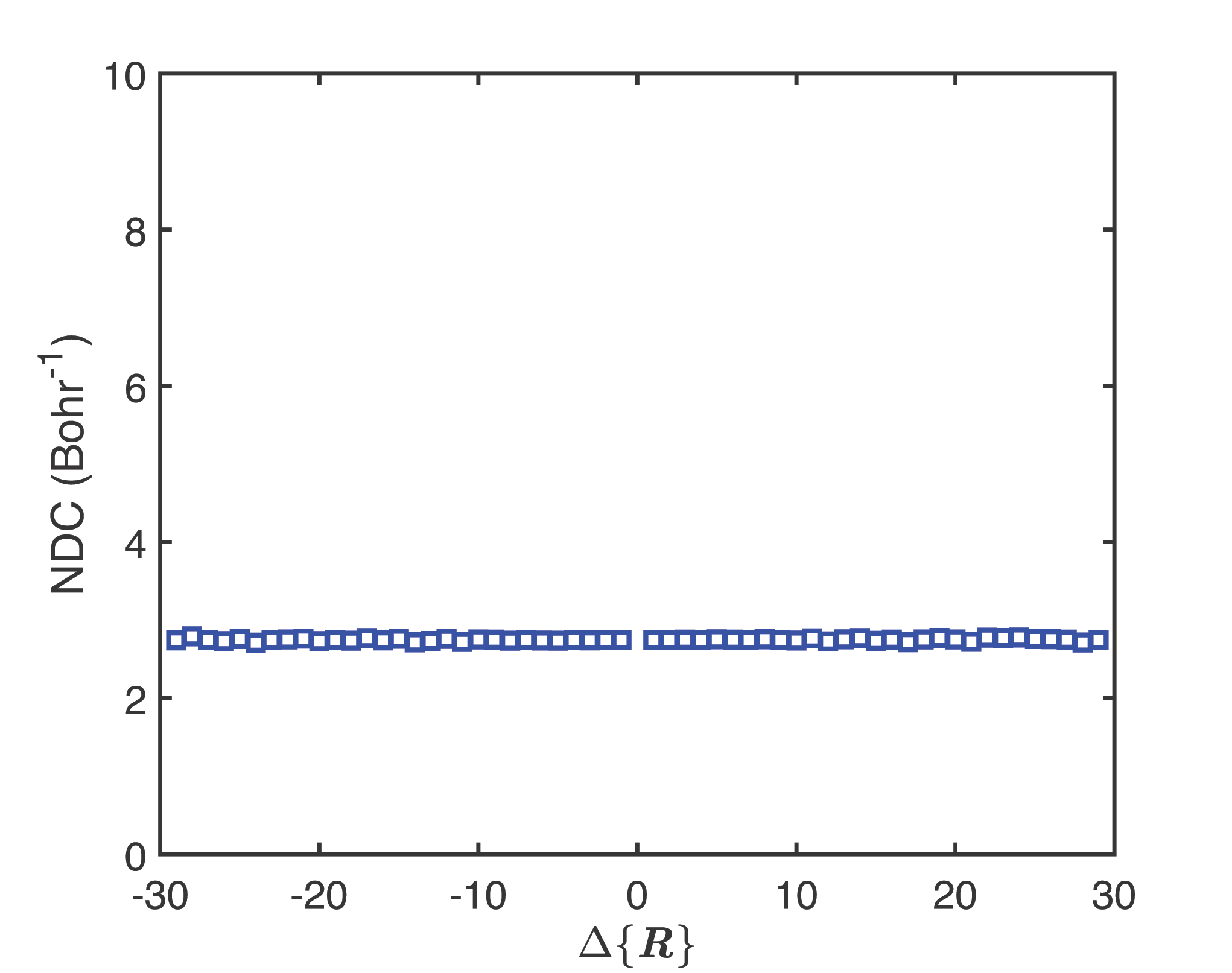} 
	\caption{(a) Variation of potential energy surfaces for ${\rm S_0}$ and ${\rm S_1}$ and (b) theoretical NDC terms, calculated all employing computational data, for Flav5 in vacuum versus displacements of vibrational modes up to those values corresponding  to room-temperature Boltzmann occupation equal to 0.37.  }
	\label{ndc-1}
\end{figure}

\begin{figure}
\includegraphics[width=.45\textwidth]{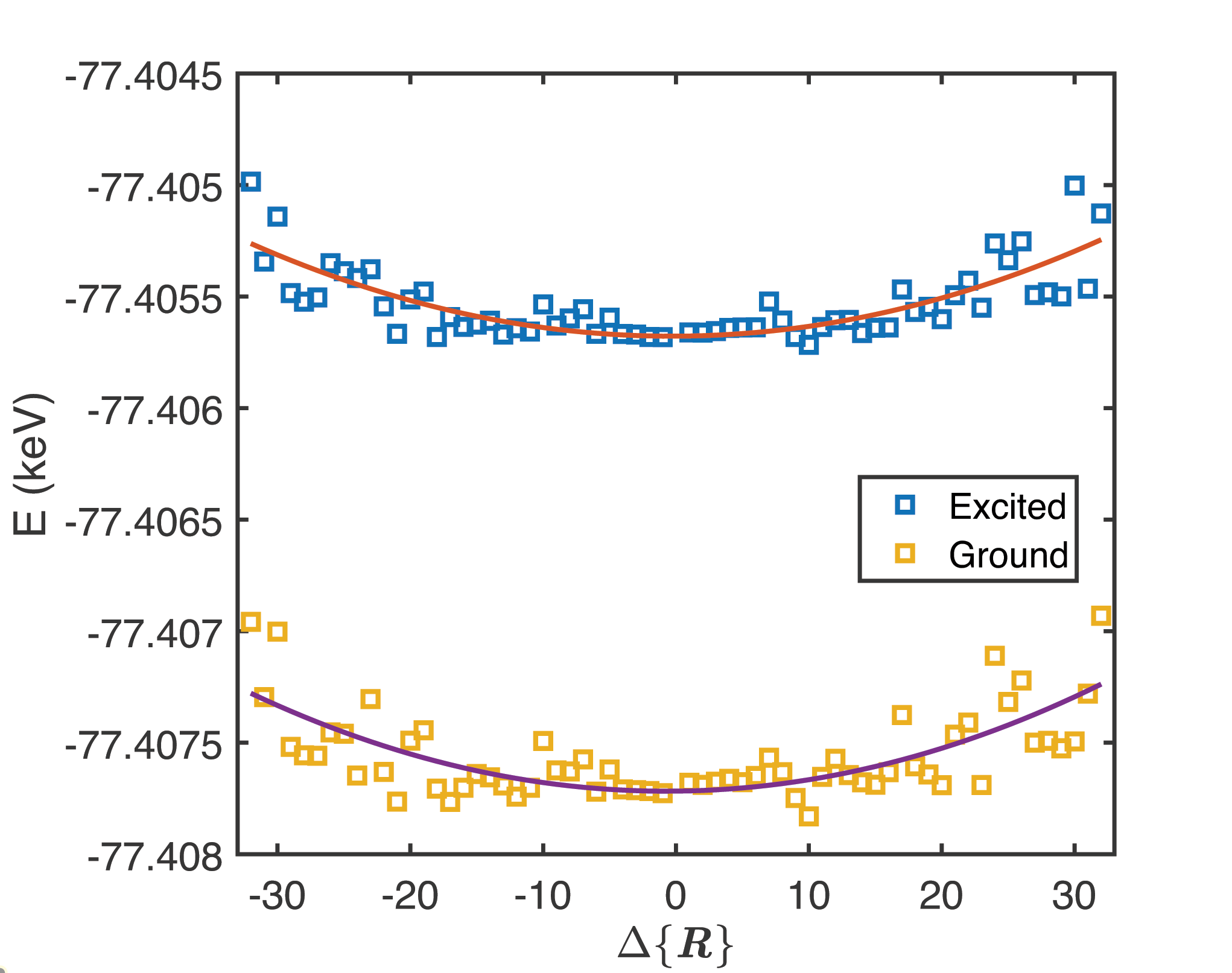} \includegraphics[width=.4\textwidth]{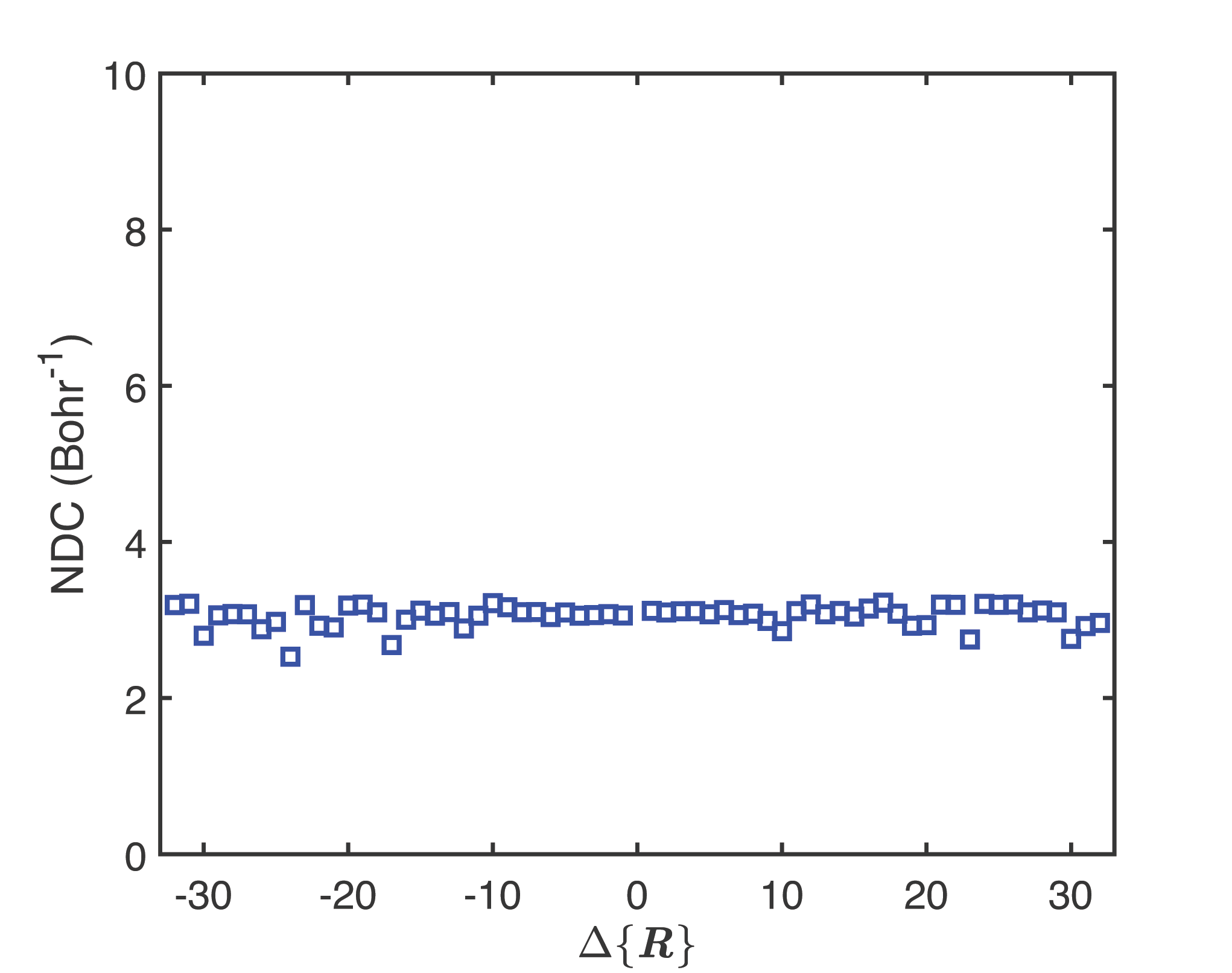} 
	\caption{(a) Variation of potential energy surfaces for ${\rm S_0}$ and ${\rm S_1}$ and (b) theoretical NDC terms, calculated all employing computational data, for Flav7 in vacuum versus displacements of vibrational modes up to those values corresponding  to room-temperature Boltzmann occupation equal to 0.37. }
	\label{ndc-2}
\end{figure}

\begin{figure}
\includegraphics[width=.45\textwidth]{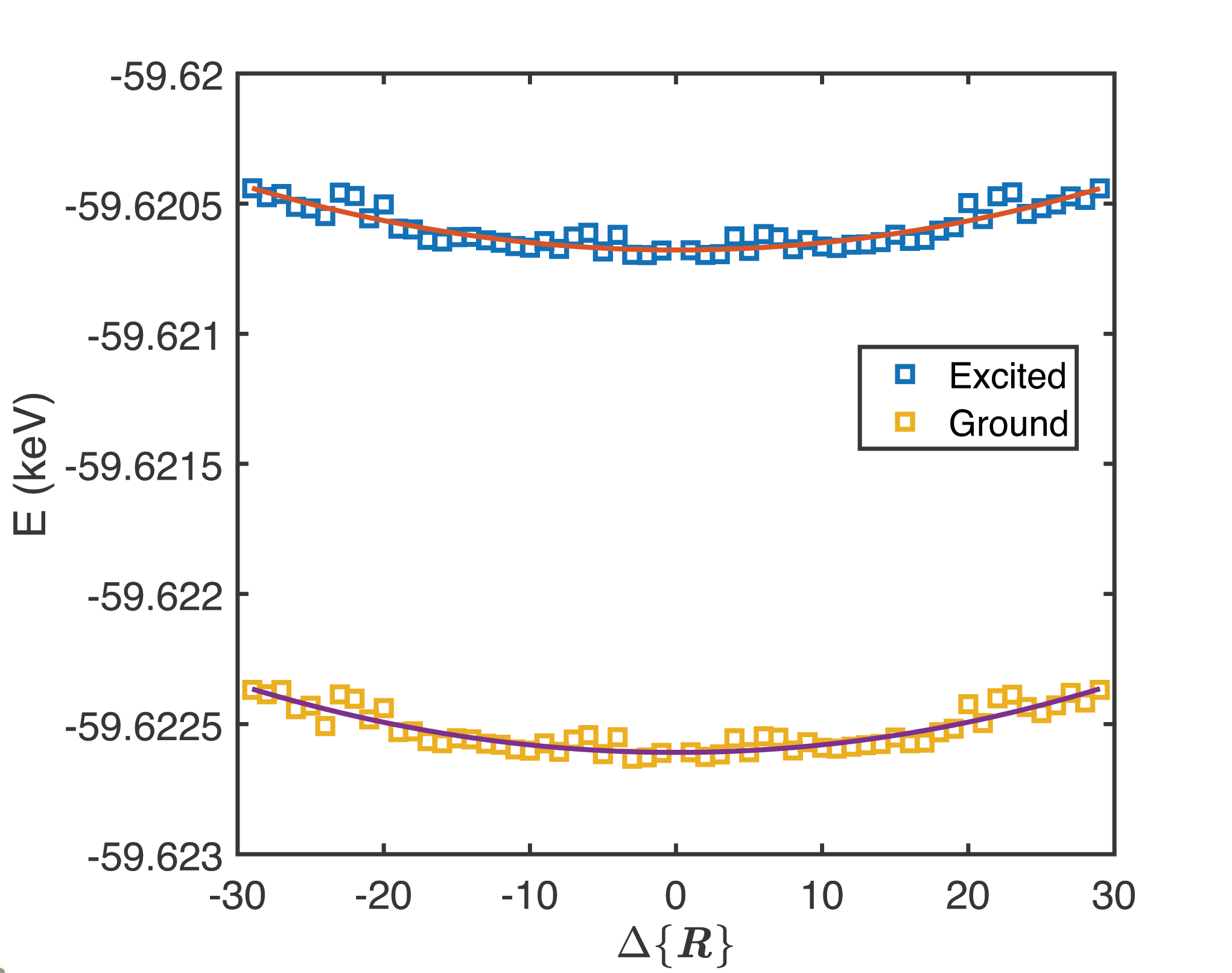} \includegraphics[width=.4\textwidth]{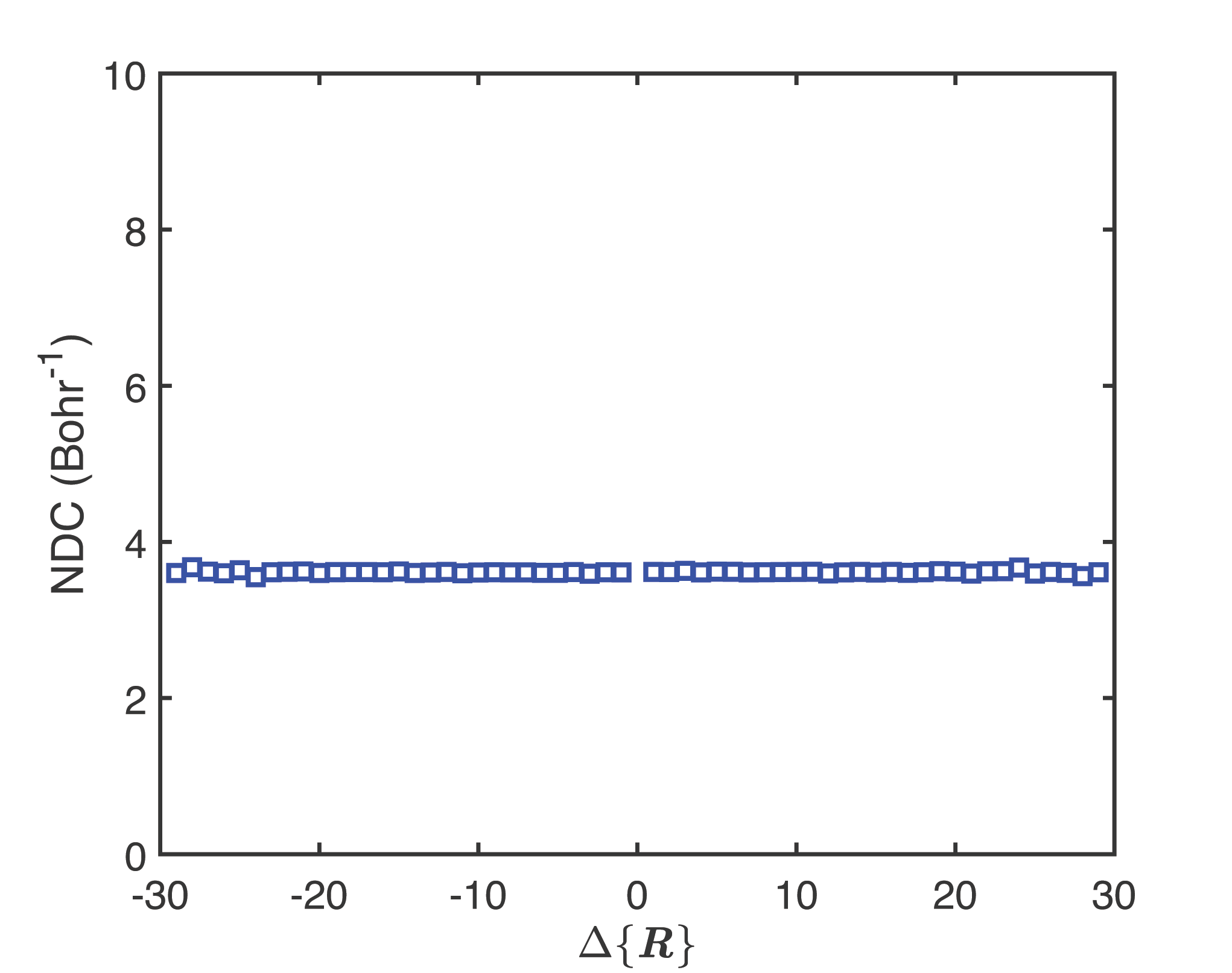} 
	\caption{(a) Variation of potential energy surfaces for ${\rm S_0}$ and ${\rm S_1}$ and (b) theoretical NDC terms, calculated all employing computational data, for Flav5 in DCM solvent versus displacements of vibrational modes up to those values corresponding  to room-temperature Boltzmann occupation equal to 0.37. }
	\label{ndc-3}
\end{figure}

\begin{figure}
\includegraphics[width=.45\textwidth]{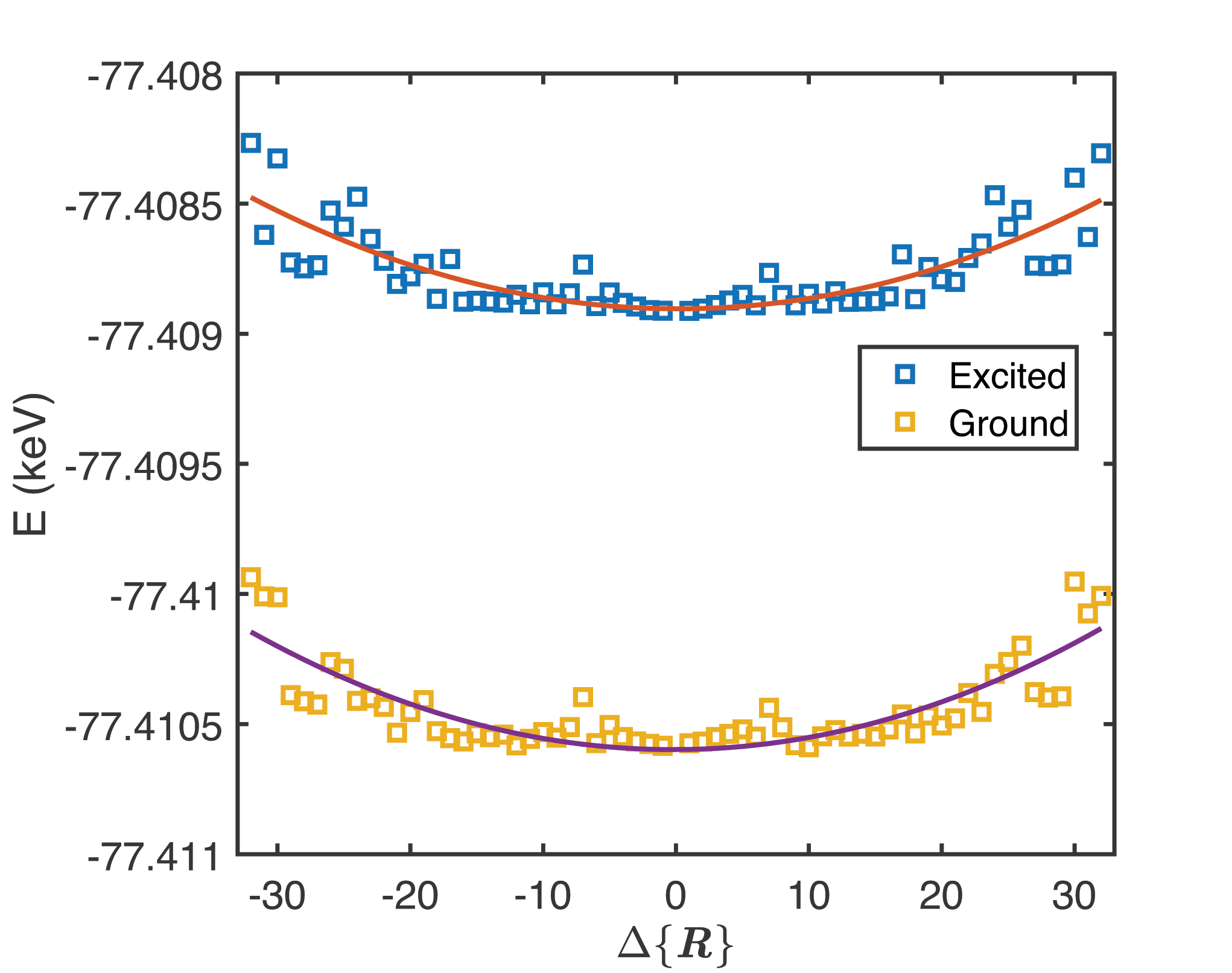} \includegraphics[width=.4\textwidth]{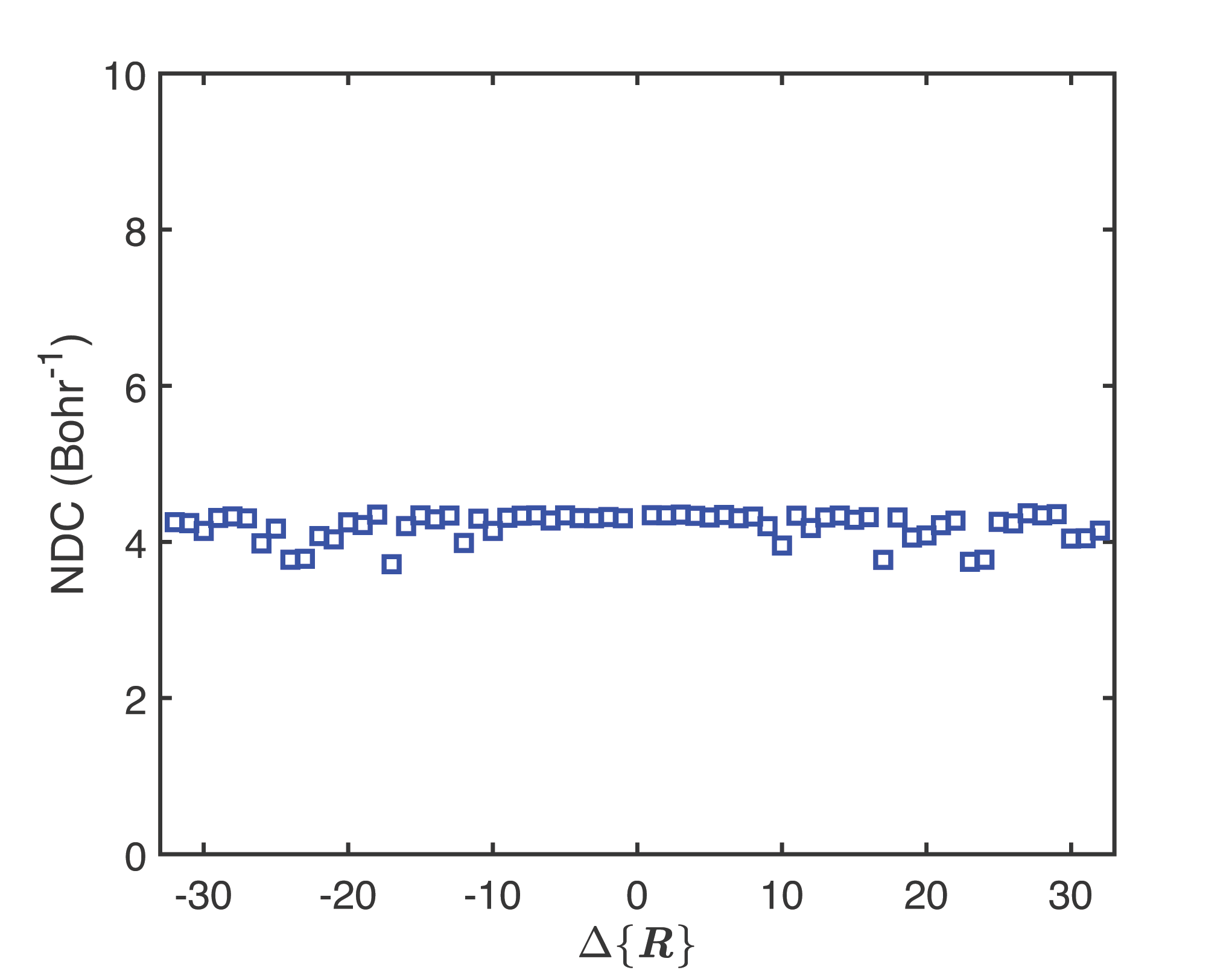} 
	\caption{(a) Variation of potential energy surfaces for ${\rm S_0}$ and ${\rm S_1}$ and (b) theoretical NDC terms, calculated all employing computational data, for Flav7 in DCM solvent versus displacements of vibrational modes up to those values corresponding  to room-temperature Boltzmann occupation equal to 0.37. }
	\label{ndc-4}
\end{figure}

\noindent
{\bf C. Dependence of derivative couplings on nuclear coordinates}\vspace{.2in}\\
We conducted sample calculations NDC terms of Flav5 and Flav7 in S1 electronic states in both vacuum and in solvent employing the TD-DFT method with CAM-B3LYP functional and 6-311+G* basis using the ORCA 5.0 program. First, the vibrational frequency calculations were performed at the optimized S1 geometries of Flav5 and Flav7 with the harmonic approximation. Then, some of the low-energy vibrational modes of the two S1 structures were selected to obtain their near-equilibrium geometries. The mode-selection criterion was that the Boltzmann factor $\exp\left\{-\Delta E_{vib}/k_BT\right \}$  produces a weight greater than or equal to 0.37 at ${\rm 298.15\ K}$, i.e., the mode energy $\leq$ thermal energy at room temperature.  These selected modes are considered thermally accessible, among them, 29 modes of Flav5 and 32 modes for Flav7 meet the criterion. For each mode, the displaced, near-equilibrium geometries are generated by shifting along ±1 scaled mode displacement vectors from the equilibrium S1 geometry, which are labeled as $\Delta \{ {\bf R}\}$.
Figures \ref{ndc-1}-\ref{ndc-4} show the electronic potential energy surface (PES) in the S1 cases and the non-mass-weighted NDC at each point for both molecules (Figures \ref{ndc-1} and \ref{ndc-2} are results in vacuum, and Figs. \ref{ndc-3} and \ref{ndc-4} are results in the solvent of DCM. It can be seen that in these low-energy vibrational modes for two structures, their own near-equilibrium S1 NDCs remain almost the same. Thus, we can conclude that the NDC terms in the S1 state are not affected significantly by thermal fluctuations, which justifies using values at the optimized structures.

\ \vspace{.5in}\\

\begin{figure}
\includegraphics[width=.45\textwidth]{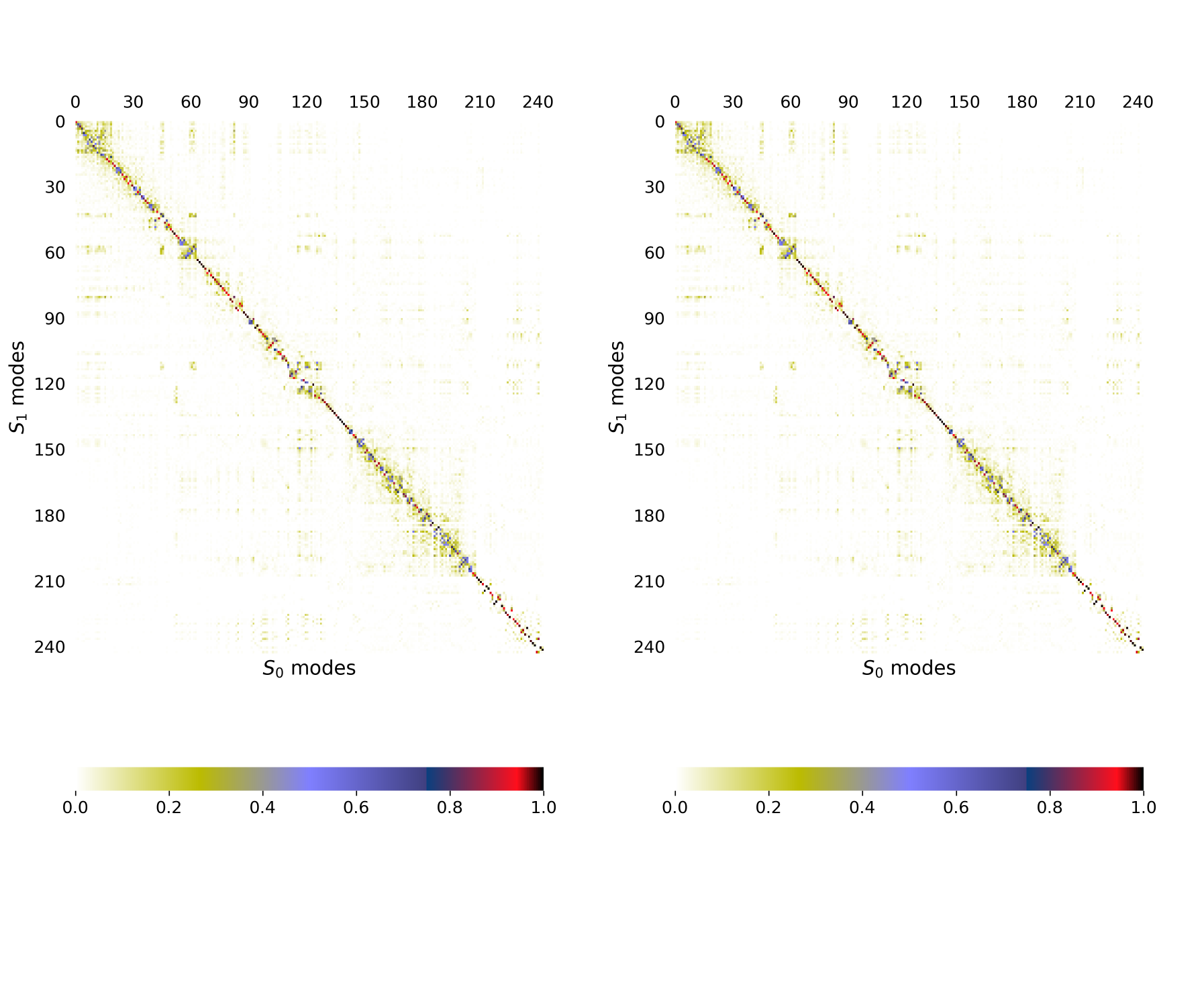}
	\caption{Duschinsky matrix between the  normal modes for the Flav5 dye for the two types of systems, isolated (left), and deuterated (right).}
	\label{dusf5}
\end{figure}

\begin{figure}
\includegraphics[width=.45\textwidth]{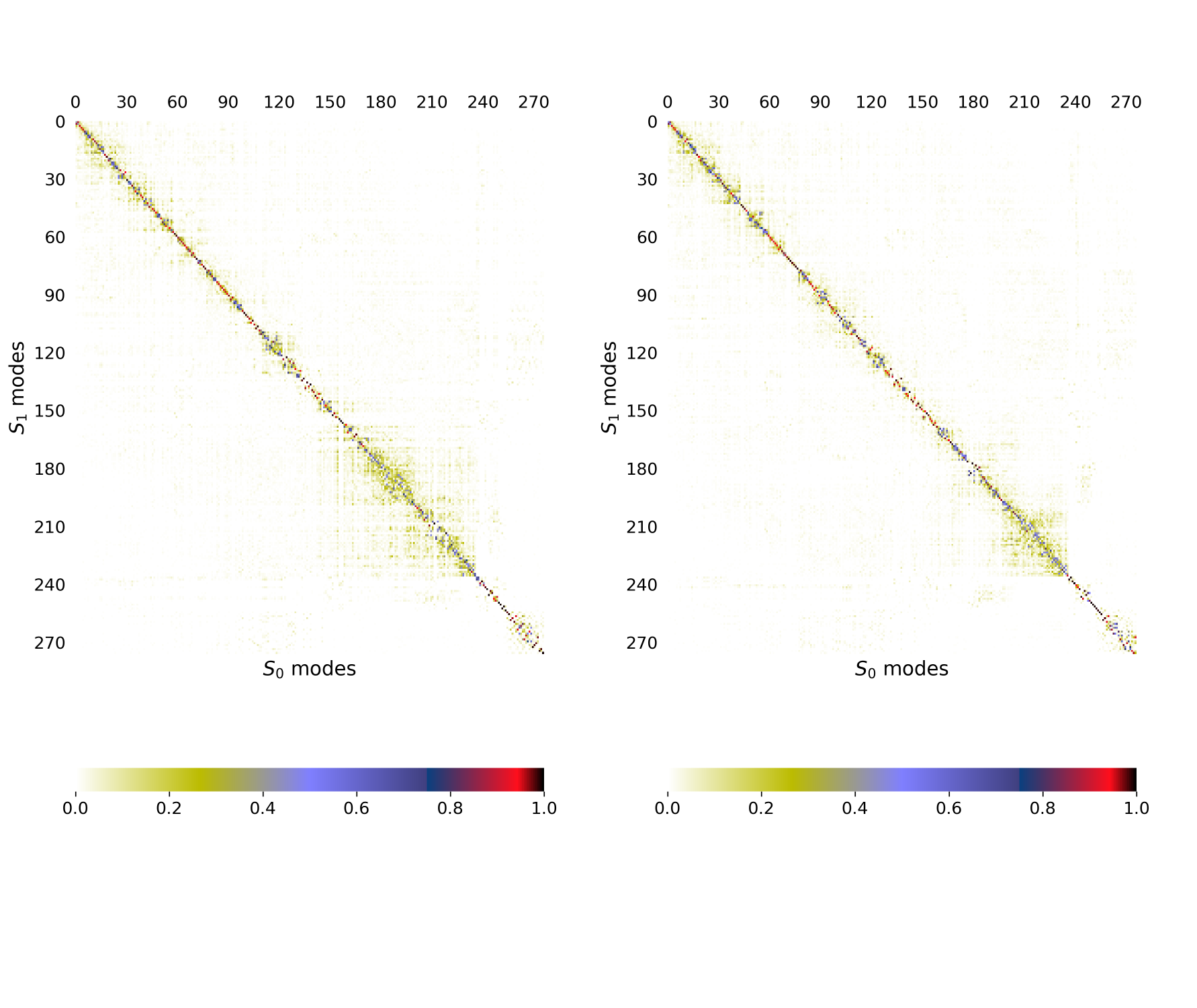}
	\caption{Duschinsky matrix between normal modes for the Flav7 dye for the two types of systems, isolated (left), and deuterated (right).}
	\label{dusf7}
\end{figure}

\noindent
{\bf  D. Duschinsky matrix elements } \vspace{.2in}\\
For further examination of detailed vibronic channels involved in the nonadiabatic transition from $S_1$ to $S_0$, we also calculated the Duschinsky rotation matrix\cite{niu-jpca114,chan2008,ratner2000} from the displacement of the nuclei in the normal modes.   
Let us denote the vector of normal mode vibrational coordinates for ${\rm S}_1$ as ${\bf q}_1$ and that for ${\rm S}_0$ as ${\bf q}_2$, which are assumed to be determined in a common frame of reference that satisfy the Eckart conditions.\cite{niu-jpca114,eckert-pr47,wilson-decius}
Then, 
\ben
&&{\bf q}_1={\bf L}_1 (\tilde {\bf R}-\tilde {\bf R}_{1}^*) , \label{eq:ql1}\\
&&{\bf q}_{0}={\bf L}_{0} (\tilde {\bf R}-\tilde {\bf R}_{0}^*) ,  \label{eq:ql0}
\een
where $\tilde {\bf R}_{1}^*$ and $\tilde {\bf R}_{0}^*$ are mass-weighted nuclear coordinates where the potential energies of ${\rm S_1}$ and ${\rm S_0}$ are minimum respectively.    Employing the relationship, $\tilde {\bf R} ={\bf L}_0^T {\bf q}_0+\tilde {\bf R}_0^*$, which corresponds to the inverse of Eq. (\ref{eq:ql0}), in Eq. (\ref{eq:ql1}), we obtain
\ben
&&{\bf q}_1={\bf L}_1{\bf L}_0^T {\bf q}_0+{\bf L}_1 (\tilde {\bf R}_0^*-\tilde {\bf R}_1^* ) \nonumber \\
&&\hspace{.2in}={\bf J}_{10}{\bf q}_0+{\bf D}_{10} , 
\een
where the first line defines the Duschinsky rotation matrix ${\bf J}_{10}$ and the displacement vector ${\bf D}_{10}$ shown in the second line.  
Figures \ref{dusf5} and \ref{dusf7} show the Duschinsky rotation matrices for Flav5 and Flav7 respectively.  

\ \vspace{.5in}\\

\noindent
{\bf Nonadiabatic rates for transition to triplet states } \vspace{.2in}\\
The major coupling responsible for the transition between singlet and triplet states is  the spin--orbit coupling, which, for TD-DFT states, can be expressed using the one--electron Breit Pauli Hamiltonian\cite{Marian2011,Pellegrini2020} as follows:
\be
\hat{H}_{SO} = - \frac{\alpha_0^2}{2} \sum_{i,c} \frac{Z_c}{r^3_{ic}} (\hat {\bf{r}}_{ic} \times \hat {\bf{p}}_i ) \cdot \hat{s}_i,
\ee
where $\alpha_0 = 137.037^{-1}$ is the fine structure constant,  $\hat{\bf{r}}_{ic}$, $\hat{\bf{p}}_i$  and $\hat{s}_i$ are the position (with respect to nucleus labeled as $c$), momentum, and spin operators of electron $i$, and   $Z_c$ is the nuclear charge.  Employing the second  quantization representation, the Hamiltonian $\hat{H}_{SO}$ takes the following form: 
\ben
&&\hat{H}_{SO_x} = - \frac{\alpha_0^2}{2} \sum_{p,q} \tilde{L}_{xpq}\cdot \frac{\hbar}{2} ( a^\dagger_{p_{\alpha}} a_{q_{\beta}} -  a^\dagger_{p_{\beta}} a_{q_{\alpha}}) ,\\
&&\hat{H}_{SO_y} = - \frac{\alpha_0^2}{2} \sum_{p,q} \tilde{L}_{ypq}\cdot \frac{\hbar}{2} ( a^\dagger_{p_{\alpha}} a_{q_{\beta}} -  a^\dagger_{p_{\beta}} a_{q_{\alpha}}) , \\
&&\hat{H}_{SO_z} = - \frac{\alpha_0^2}{2} \sum_{p,q} \tilde{L}_{zpq}\cdot \frac{\hbar}{2} ( a^\dagger_{p_{\alpha}} a_{q_{\alpha}} -  a^\dagger_{p_{\beta}} a_{q_{\beta}}) ,
\een
where $p$ and $q$ represent molecular orbital indices, $\alpha$ and $\beta$ respectively denote spin up and down,  Finally, $\tilde{L}_{\mu pq}$, with $\mu=x,y$, and $z$, represents  component of the  matrix element (with respect to $p$ and $q$ orbitals) of the operator $\sum_{i,c}Z_c\hat{{\bf L}}_{ic}/r_{ic}^3$, where  $ \hat{\bf L}_{ic} =\hat {\bf{r}}_{ic} \times \hat {\bf{p}}_i$.   On the other hand, the single reference excited states within the Tamm-Dancoff approximation\cite{hira1999b} are expressed as
\ben
&&| \Phi^I_{singlet} \rangle = \sum_{i,a} s^{Ia}_i ( a^\dagger_{a_{\alpha}} a_{i_{\alpha}} -  a^\dagger_{a_{\beta}} a_{i_{\beta}}) | \Phi_{HF}\rangle ,\\
&&| \Phi^{I,m_s=0}_{triplet} \rangle = \sum_{i,a} t^{Ia}_i ( a^\dagger_{a_{\alpha}} a_{i_{\alpha}} -  a^\dagger_{a_{\beta}} a_{i_{\beta}}) | \Phi_{HF}\rangle , \\
&&| \Phi^{I,m_s=1}_{triplet} \rangle = \sum_{i,a} \sqrt{2}t^{Ia}_i  a^\dagger_{a_{\alpha}} a_{i_{\beta}} |\Phi_{HF}\rangle ,\\
&&| \Phi^{I,m_s=-1}_{triplet} \rangle = \sum_{i,a} \sqrt{2}t^{Ia}_i  a^\dagger_{a_{\beta}} a_{i_{\alpha}} |\Phi_{HF}\rangle .
\een
where $s^{Ia}_i$ and $t^{Ia}_i$ are the singlet and triplet excitation coefficients for the $I^{th}$ singlet or triplet excitation.  We use $i,j$ to denote occupied orbitals, and $a,b$ to denote virtual orbitals. These excitations evolve from $| \Phi_{HF}\rangle$, the reference Hartree-Fock ground state. Combining all the terms defined above, we can express the spin--orbit coupling as follows:\cite{bellon2019,zhang2014}  
\ben
&&\langle  \Phi^I_{singlet} |\hat{H}_{SO}|\Phi^{J,m_s=0}_{triplet} \rangle \nonumber \\
&&= \frac{\alpha_0^2 \hbar}{2} \left(\sum_{i,a,b} \tilde{L}_{zab} s^{Ia}_i t^{Jb}_i - \sum_{i,j,a} \tilde{L}_{zij} s^{Ia}_i t^{Ja}_j  \right) ,
\een

\ben
&&\langle  \Phi^I_{singlet} |\hat{H}_{SO}|\Phi^{J,m_s=\pm1}_{triplet} \rangle\nonumber \\
&& = \pm \frac{\alpha_0^2 \hbar}{2\sqrt{2}} \left(\sum_{i,a,b} \tilde{L}_{xab} s^{Ia}_i t^{Jb}_i - \sum_{i,j,a} \tilde{L}_{xij} s^{Ia}_i t^{Ja}_j  \right)\\
&& + \frac{\alpha_0^2 \hbar}{2\sqrt{2i}} \left(\sum_{i,a,b} \tilde{L}_{yab} s^{Ia}_i t^{Jb}_i - \sum_{i,j,a} \tilde{L}_{yij} s^{Ia}_i t^{Ja}_j  \right)\nonumber .
\een
Finally, the net spin--orbit coupling is obtained as the root mean squared of the different contributions as follows:
\ben
J_{so}&=&\langle  \Phi^I_{singlet} |\hat{H}_{SO}|\Phi^{J}_{triplet} \rangle \nonumber \\
&=& \sqrt{\sum_{m_s=0,\pm 1} || \langle  \Phi^I_{singlet} |\hat{H}_{SO}|\Phi^{I,m_s}_{triplet} \rangle||^2 } .
\een

\begin{figure}
\includegraphics[width=0.45\textwidth]{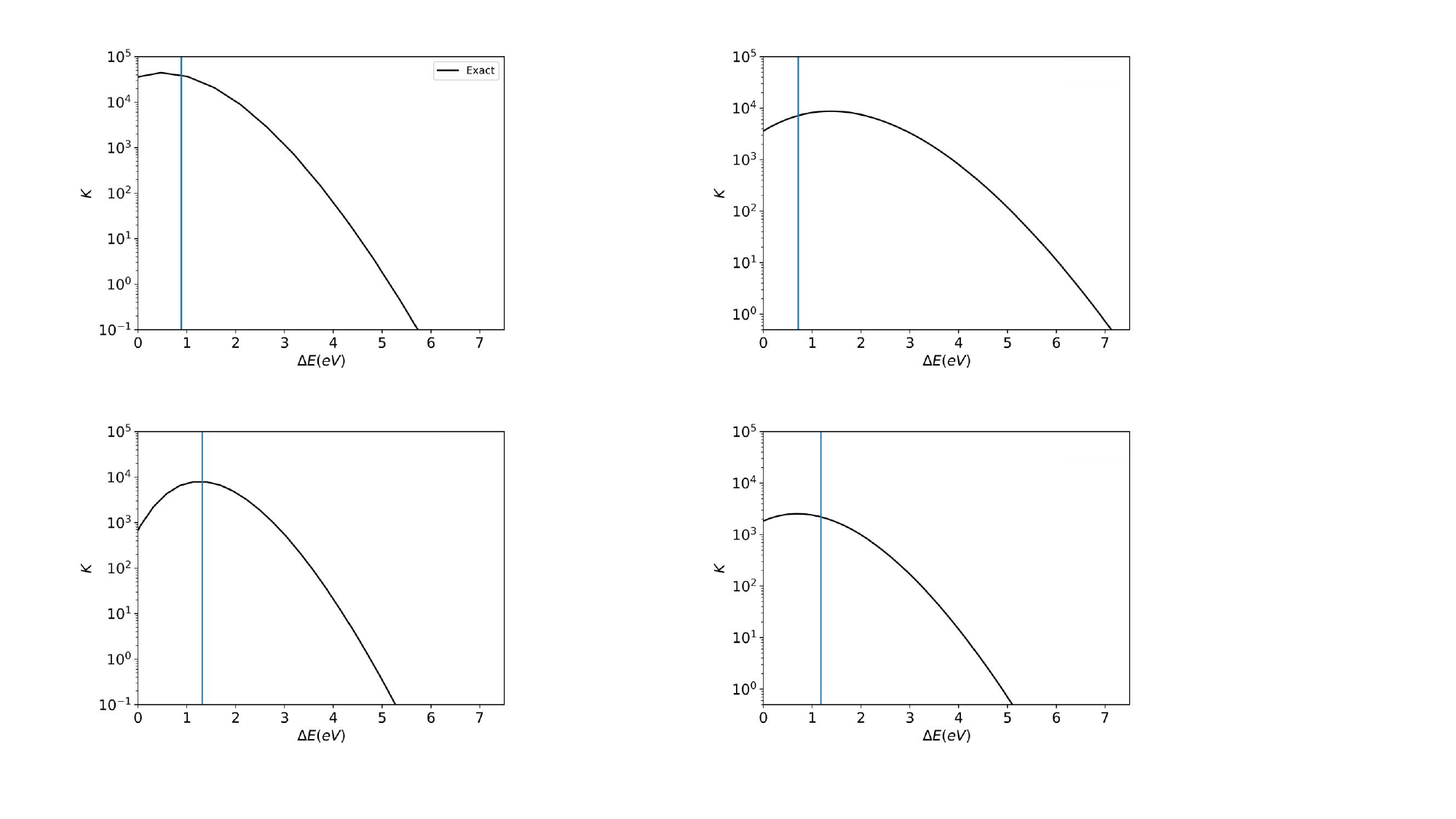}
	\caption{Nonadiabatic transition rates (in logarithmic scale) versus energy gap calculated for the electronic transition in both Flav5 (left column) and Flav7 (right column) dyes. The computed transitions from top to bottom are: ${\rm T_1\rightarrow S_0}$ and ${\rm T_2\rightarrow S_0}$ respectively. The vertical line represents the corresponding energy gap between the coupled states. }
	\label{kf1}
\end{figure}

\begin{figure}
\includegraphics[width=0.45\textwidth]{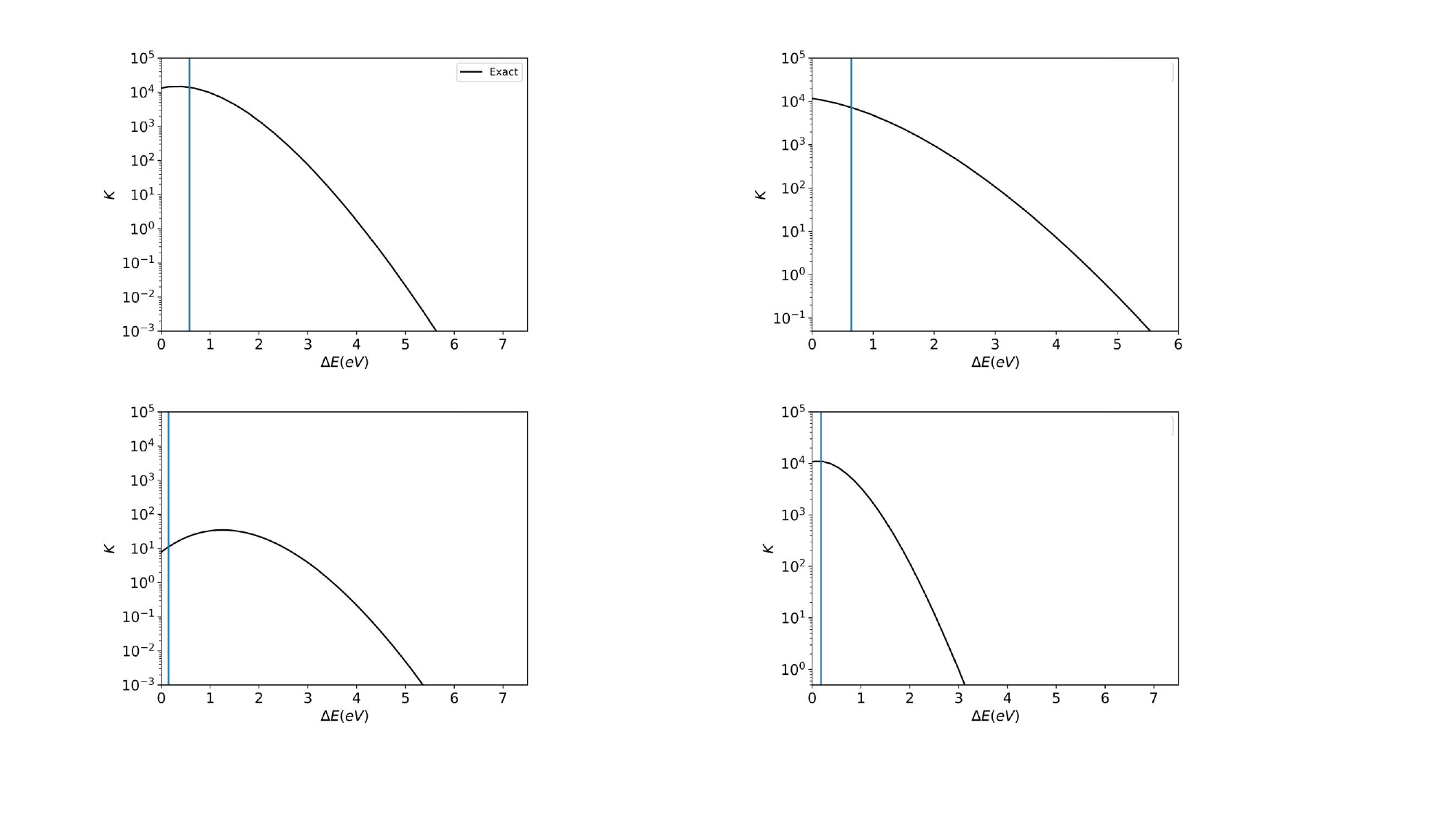}
	\caption{Nonadiabatic transition rates (in logarithmic scale) versus energy gap calculated for the electronic transition in both Flav5 (left column) and Flav7 (right column) dyes. The computed transitions from top to bottom are: ${\rm S_1\rightarrow T_1}$ and ${\rm S_1\rightarrow T_2}$ respectively. The vertical line represents the corresponding energy gap between the coupled states. }
	\label{kf2}
\end{figure}

The resulting singlet-triple transition rates versus the energy gap are shown in Figs. \ref{kf1} and \ref{kf2}. 

 \begin{table}
\caption{\re{Computed energy gap, values of spin-orbit coupling constant $J_{so}$, calculated nonradiative transition rates between singlet and triplet states.}  }
\begin{center}
      \begin{tabular}{lcccc}
\toprule

Dye&transition& $\Delta$E (eV) & $J_{so}$ (cm$^{-1}$) & $k_{nr}$(s$^{-1}$)\\
\hline
\hline
F5 &&&&\\
 & $T_1 \rightarrow S_0$ & 1.0986 & 0.5010& 13567 \\
 & $T_2 \rightarrow S_0$ & 1.3152 & 0.2680& 3926 \\
 & $S_1 \rightarrow T_1$ & 0.5774 & 0.3750& 5139  \\
 & $S_1 \rightarrow T_2$ & 0.1483 & 0.0303& 8.017 \\

\hline
F7 &&&&\\                      
& $T_1 \rightarrow S_0$ & 0.7164 & 0.6610 & 22481 \\
& $T_2 \rightarrow S_0$ & 1.1790 & 0.3213 &6368 \\
& $S_1 \rightarrow T_1$ & 0.6429 & 0.6010 &15931  \\
& $S_1 \rightarrow T_2$ & 0.1803 & 0.5090 & 19.22\\

\bottomrule
\label{tab2}

        \end{tabular}   
        
\end{center}
\end{table}
Table \ref{tab2} provides full theoretical data for the energy gap, spin-orbit coupling, and the resulting nonradiative transition rates calculated by FGR.  \vspace{.2in}\\

\noindent
{S3. Experimental lifetime measurements} \vspace{.2in}\\
The standard procedure for all the lifetime measurements is as follows: We recorded PL lifetimes using a home-built, all-reflective epifluorescence setup.\cite{atallah-jpca123} For all dyes below we used a pulsed 785 nm excitation. Emission was then collected and filtered with a 90:10 beamsplitter and appropriate excitation filters finally reflectively coupled into a single-mode fiber (F-SMF-28-C-10FC, Newport) and detected using an SNSPD (Quantum Opus One).\cite{atallah-jpca123,cosco-jacs143,pengshung-cc56}
Given the short lifetimes of these dyes, lifetimes were fit with a convolution of the instrument response function and an exponential. To determine the lifetime (or decay rate, k) for each TCSPC trace we fit each curve to a convolution of the sum of two Gaussians with a single exponential decay:
\ben
I(t)&=&\frac{I_0}{2} \exp\left \{-k \left( (t-t_0) -\frac{\sigma_1^2 k}{2} \right)\right\} \nonumber \\
&&\times \left \{1+{\rm erf}\left ( \frac{(t-t_0)-\sigma_1^2 k}{\sqrt {2}\sigma_1}\right) \right\} \nonumber \\
&+&\frac{aI_0}{2} \exp\left \{-k \left ( (t-t_0-t_1 )-\frac{\sigma_2^2 k}{2}\right)\right\}\nonumber \\
&&\times \left \{1+{\rm erf}\left ( \frac{(t-t_0-t_1)-\sigma_2^2 k}{\sqrt {2}\sigma_2}\right) \right\} .
\een
The width, $\sigma_1$ and $\sigma_2$ of each Gaussian, the time offset, $t_1$, between the two Gaussians and amplitude scale, $\alpha$, were determined using the instrument response function (IRF) which was measured as the backscatter off of a cuvette with solvent (e.g.,DCM) without the longpass filters (Table \ref{tab3}). The initial peak amplitude, $I_0$, the rate, $k$, and $t_0$ were free fitting parameters, while the time offset, $t_1$, and the IRF widths, $\sigma_1$ and $\sigma_2$ were fixed variables. We use a conservative error of $1\ {\rm ps}$ (the instrument resolution) for our lifetimes.
\begin{center}
\begin{table}
\caption{Parameters used for IRF fit.}
\begin{tabular}{c|c}
\hline 
\hline
IRF fit parameters&Numerical values for the fit \\
\hline
$\sigma_1({\rm ps})$ & $32.0$ \\
\hline
$\sigma_2({\rm ps})$ & $50.6$ \\
\hline
$t_1({\rm ps})$ & $6.6$ \\ 
\hline
$\alpha$ & $1.6$ \\
\hline
\hline
\end{tabular}
\label{tab3}
\end{table}
\end{center}

\ \vspace{.4in}\\

\begin{widetext}
\noindent
{\bf A. Information on materials and general experimental procedures} \vspace{.2in}\\
{\bf 1. Materials} \vspace{.2in}\\
Chemical reagents were purchased from Accela, Acros Organics, Alfa Aesar, Carl Roth, Fisher Scientific, Sigma-Aldrich, or TCI and used without purification. Anhydrous solvents and deoxygenated solvents (toluene, acetonitrile, dimethylformamide, THF) were obtained from a Grubb’s-type Phoenix Solvent Drying System constructed by JC Meyer. Anhydrous solvents (1,4-dioxane, ethanol, n-butanol) were prepared by drying over 4 \AA molecular sieves for at least 3 days. Oxygen was removed from reaction mixtures by three consecutive freeze-pump-thaw cycles in air free glassware. \vspace{.2in}\\

{\bf 2. Instrumentation} \vspace{.2in}\\
Thin layer chromatography was performed using Silica Gel 60 F$_{254}$ (EMD Millipore) plates. Flash chromatography was executed with technical grade silica gel with ${\rm 60\ \AA}$ pores and ${\rm 40 – 63\ \mu m}$ mesh particle size (Sorbtech Technologies). Solvent was removed under reduced pressure with a B\"{u}chi Rotovapor with a Welch self-cleaning dry vacuum pump and further dried with a Welch DuoSeal pump. Bath sonication was performed using a Branson 3800 ultrasonic cleaner. Microwave reactions were performed using a CEM Discover SP microwave synthesis reactor. Reactions were performed in glass 10 mL microwave reactor vials purchased from CEM with silicone/PTFE caps. Flea micro PTFE-coated stir bars were used in the vials with magnetic stirring set to high and 15 seconds of premixing prior to the temperature ramping. All 1H and 13C, spectra are reported in ppm and relative to residual solvent signals (1H and 13C). Spectra were obtained on Bruker AV-300, AV-400, DRX-500, or AV-500 instruments and processed with MestReNova software. Liquid chromatography-mass spectrometry was obtained using an Agilent 6100 series quadrupole LC/MSD. Absorbance spectra were collected on a JASCO V-770 UV-Visible/NIR spectrophotometer with a 2000 nm/min scan rate after blanking with the appropriate solvent. Photoluminescence spectra were obtained on a Horiba Instruments PTI QuantaMaster Series fluorometer. Quartz cuvettes (1 cm) were used for absorbance and photoluminescence measurements. Absorption coefficients in DCM were calculated with serial dilutions with Hamilton syringes in volumetric glassware. Error was taken as the standard deviation of the triplicate experiments. Relative quantum yields were determined in DCM relative to IR-26 in DCM. \vspace{.2in}\\

\noindent
{\bf 3. Abbreviations}\vspace{.3in}\\
DCM = dichloromethane; DMF = dimethylformamide; DMSO = dimethylsulfoxide; EtOH =
ethanol; EtOAc = ethyl acetate; Et2O = diethyl ether; MeOH = methanol, Tol = toluene

\ \vspace{.4in}\\

\noindent
{\bf SIV. Synthetic Schemes, Procedures, and NMR Spectra} \vspace{.4in}\\
\ \vspace{1in}\\ 
\begin{figure}
\includegraphics[width=0.7\textwidth]{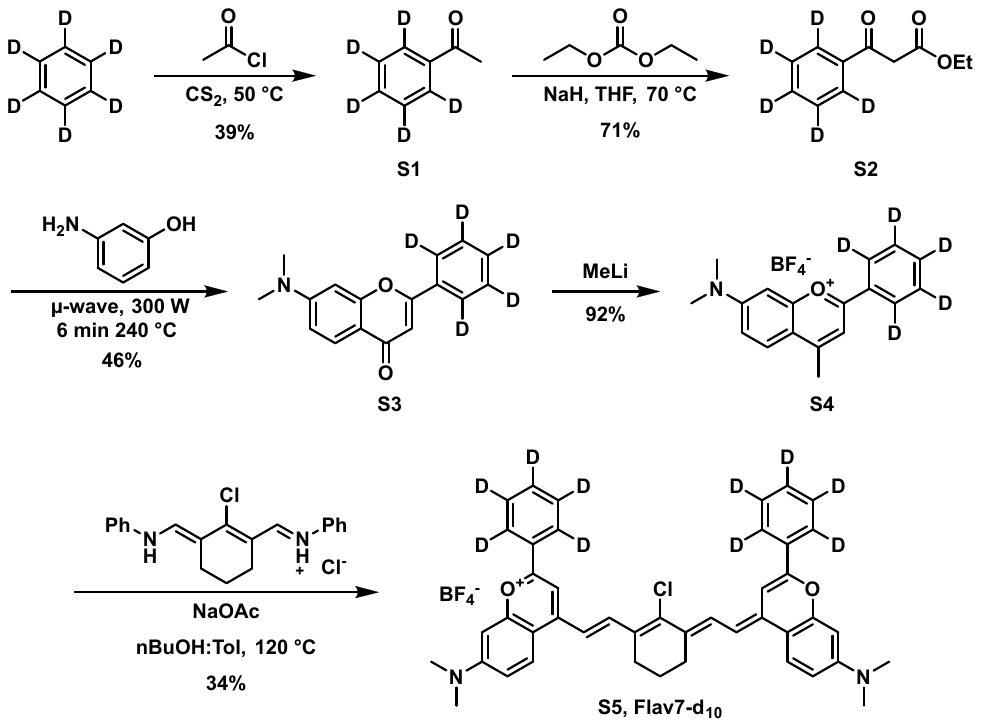}
\caption{Schemes of the overall synthesis}
  \label{syn-schemes}
\end{figure}
\begin{figure}
\includegraphics[width=0.5\textwidth]{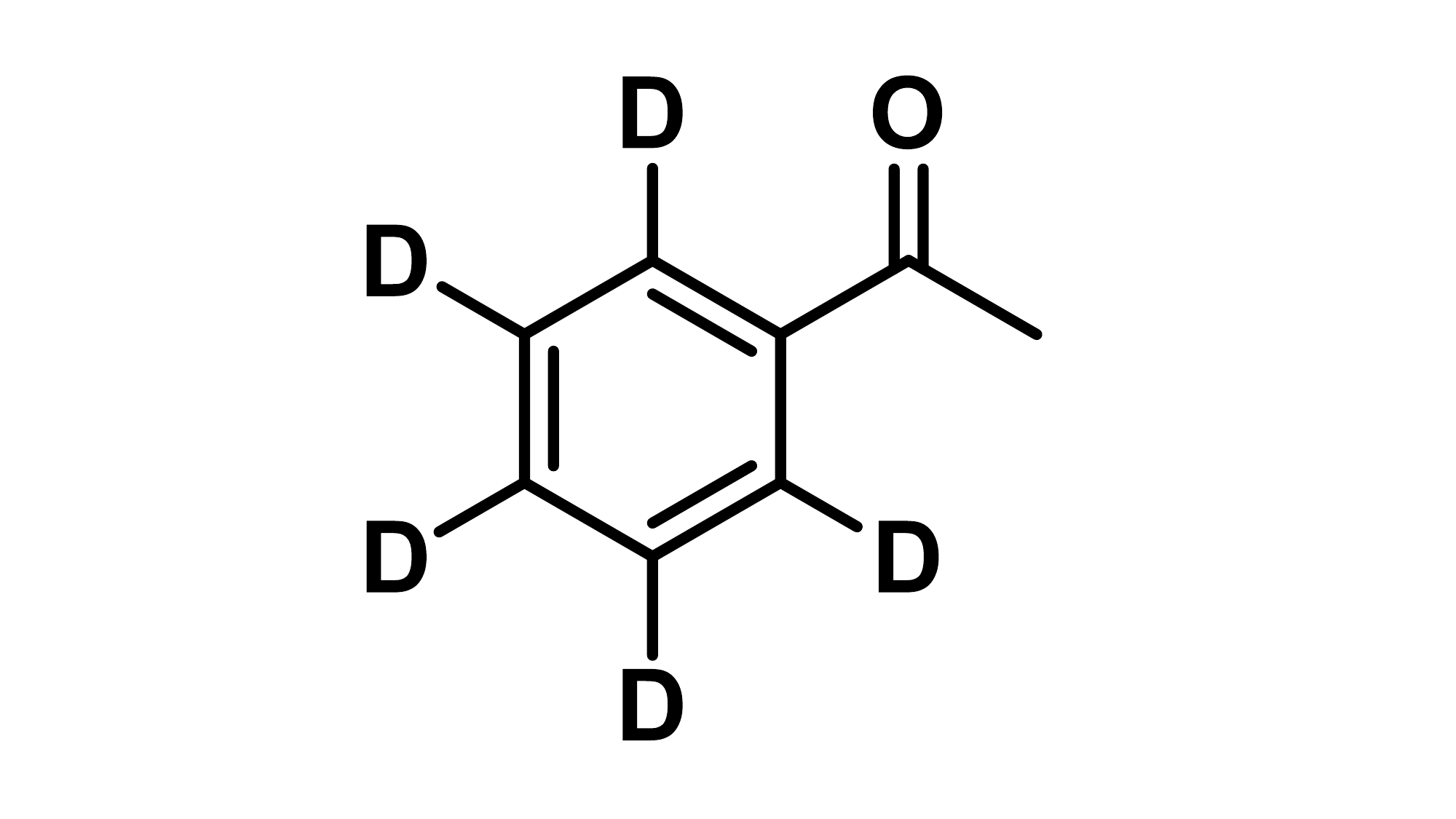}
\caption{{\bf 1-(phenyl-d5)ethan-1-one (S1)}: Benzene-d$_6$ (550 $\mu$L, 5.94 mmol, 1.00 equiv.) and AlCl$_3$ (1.12 g, 7.72 mmol, 1.60 equiv.) were dissolved with 280 $\mu$L CS$_2$ in a dram vial. Acetyl chloride (570 $\mu$L, 7.72 mmol, 1.60 equiv.) dissolved in 500 $\mu$L CS$_2$ was added to the cloudy solution at 0 $^o$C. After stirring for 5 hours at room temperature, the solution was heated to 50 $^o$C for 3 hours. The solution was cooled to 0 $^o$C, quenched via dropwise addition of water and diluted (50 mL), extracted  with DCM (3 x 75 mL), dried over Na$_2$SO$_4$, and concentrated. The pink oil was evaporated onto silica gel and purified via column chromatography with 9:1 Hex/EtOAc. The procedure afforded a pale yellow oil in 64\% yield (468 mg, 3.71 mmol). R$_f = 0.5$ in 9:1 Hex/EtOAc. $^1$H NMR (500 MHz, CDCl$_3$) $\delta$ 2.52 (s, 3H). $^{13}$C NMR (126 MHz, CDCl$_3$) $\delta$ 198.0, 136.9, 132.8 (t, J = 24.5 Hz), 128.4 (q, J = 11.3Hz), 127.9 (t, J = 24.6 Hz), 76.9. HRMS (ESI$^+$) calculated for C$_8$H$_4$D$_5$O$^+$ [M+H]$^+$ : 126.0962; found: 126.0957.}
\end{figure}
\begin{figure}
\includegraphics[width=0.5\textwidth]{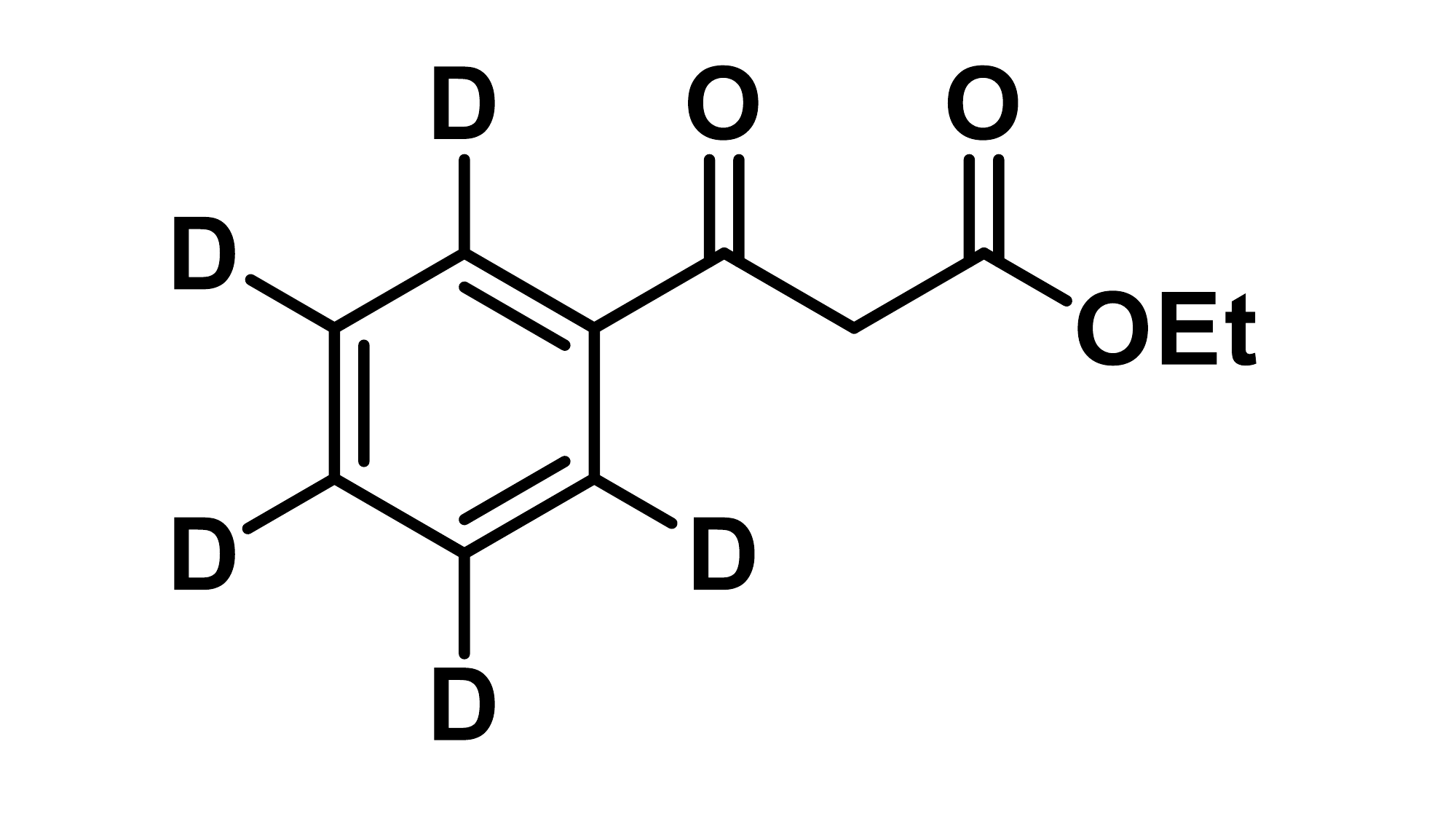}
\caption{{\bf ethyl 3-oxo-3-(phenyl-d5)propanoate (S2): S1} (468 mg, 3.71 mmol, 1.00 equiv.), diethyl carbonate (1.8 mL,  15.3 mmol, 4.00 equiv.), and NaH (264 mg, 11.1 mmol, 3.00 equiv.) were dissolved with THF (4.6 mL, anhydrous) in a sealed synth vial. The reaction mixture was heated to 70 $^o$C and stirred for 3 hours. The reaction was quenched with AcOH (100 $\mu$L), diluted with water, extracted with EtOAc (3 x 50 mL), dried over Na$_2$SO$_4$ and concentrated. The yellow oil was evaporated onto silica gel and purified via column chromatography with 9:1 Hex/EtOAc. The procedure resulted in a pale yellow oil in 71\% yield (523 mg, 2.63 mmol). R$_f = 0.4 in$ 9:1 Hex/EtOAc. $^1$H NMR (500 MHz, Acetonitrile-d$_3$) $\delta$ 4.15 (m, 2H), 4.02 (s, 2H), 1.21 (m, 3H). $^{13}$C NMR (126 MHz, CD$_3$CN) $\delta$ 193.3, 167.7, 135.9, 133.2 (t, J = 24.7 Hz), 128.2 (m), 128.0 (m), 61.0, 45.7, 13.4. HRMS (ESI$^+$) calculated for C$_{11}$H$_8$D$_5$O$_3$$^+$ [M+H]$^+$ : 198.1173; found: 198.1156o.}
\end{figure}
\begin{figure}
\includegraphics[width=0.5\textwidth]{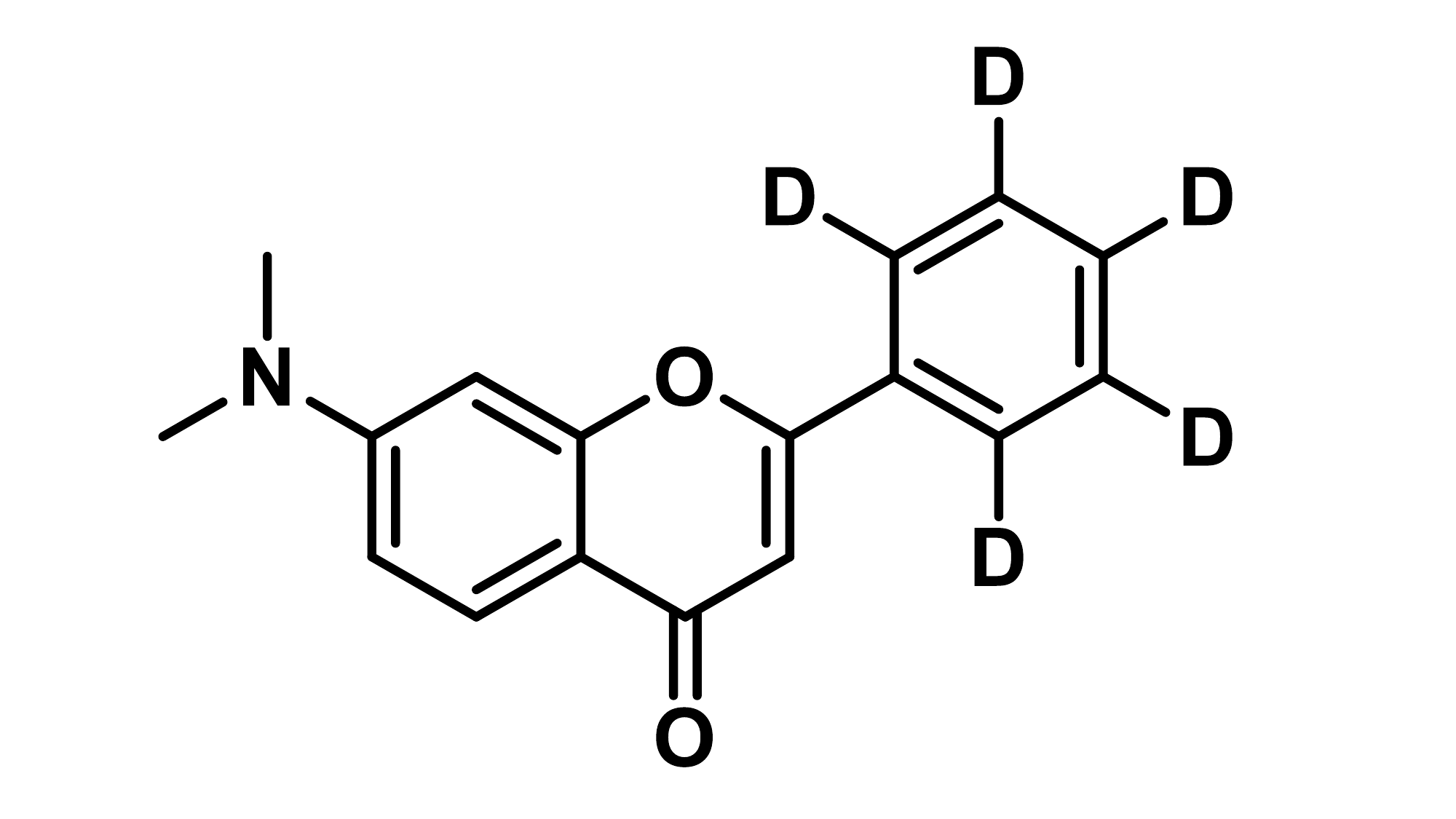}
\caption{{\bf 7-(dimethylamino)-2-(phenyl-d5)-4H-chromen-4-one (S3): S2} (40 mg, 0.20 mmol, 2.0 equiv.) and 3-dimethylamino phenol (14 mg, 0.10 mmol, 1.0 equiv.) were added to a microwave reaction vial and heated for 6 minutes at 240 $^o$C in a microwave reactor to produce a dark red oil. The crude mixture was evaporated onto silica gel and purified via column chromatography with a 9:1 to 0:1 gradient of Hex/EtOAc. The procedure yielded a dark brown solid in 46\% yield (10 mg, 0.039 mmol). R$_f = 0.3$ in 1:2 Hex/EtOAc. $^1$H NMR (500 MHz, Chloroform-d) $\delta$ 8.01 (d, J = 9.0 Hz, 1H), 6.73 (dd, J = 9.0, 2.5 Hz, 1H), 6.67 (s, 1H), 6.55 (d, J = 2.5 Hz, 1H), 3.07 (s, 6H). $^{13}$C NMR (126 MHz, CDCl$_3$) $\delta$ 177.8, 162.1, 158.3, 154.1, 132.1, 130.5 (t, J = 24.2 Hz), 128.3 (t, J = 24.4 Hz), 126.5, 125.6 (t, J = 24.9 Hz), 113.6, 110.7, 107.2, 97.1, 40.1. HRMS (ESI$^+$) calculated for C$_{17}$H$_{11}$D$_5$NO$_2$$^+$ [M+H]$^+$ : 271.1489; found: 271.1507.}
\end{figure}
\begin{figure}
\includegraphics[width=0.5\textwidth]{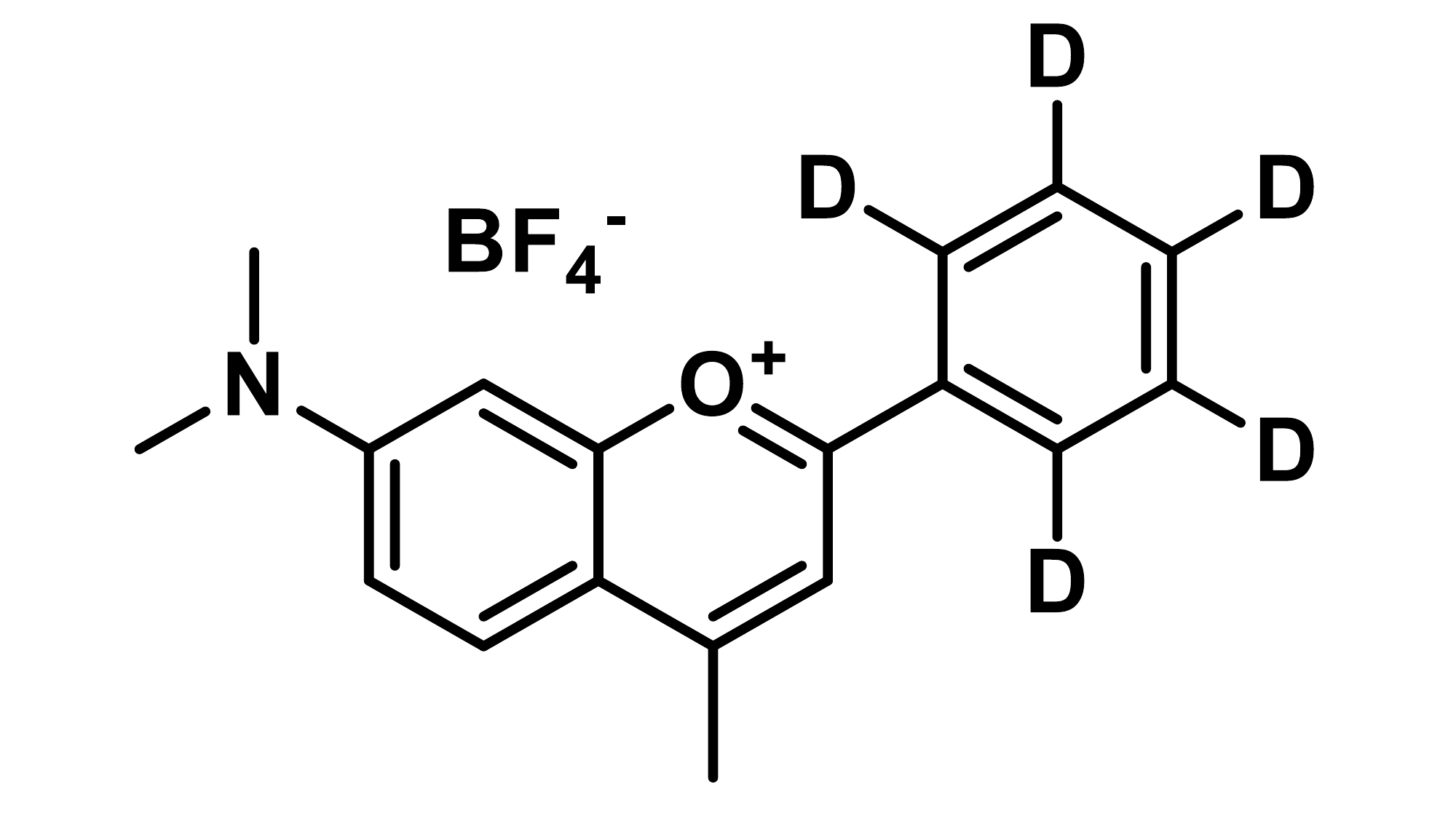}
\caption{{\bf 7-(dimethylamino)-4-methyl-2-(phenyl-d5)chromenylium (S4)}. Compound {\bf S3} (150 mg, 0.55 mmol, 1.0 equiv.) was added to a flame dried round bottom flask, dissolved with THF and cooled to 0 $^o$C under N$_2$ atmosphere. MeLi was added dropwise to the brown solution (690 $\mu$L of 1.6 M solution in ether, 1.11 mmol, 2.00 equiv). After 30 minutes the solution was quenched with 50\% HBF$_4$$^-$ aqueous solution turning the solution red. The resulting solution was diluted with more $50 \%$ ${\rm HBF_4^-}$ aqueous solution (${\rm 75\ mL}$), extracted with DCM (${\rm 3 \times 100 mL}$), dried over ${\rm Na_2SO_4}$ and concentrated. The red solid was then washed with boiling EtOAc (200 mL). The procedure yielded a dark red solid in 95\% yield (141 mg, .520 mmol). $^1$H NMR (500 MHz, CD$_2$Cl$_2$) $\delta$ 8.06 (d, J = 9.6 Hz, 1H), 7.70 (s, 1H), 7.37 (dd, J = 9.6, 2.5 Hz, 1H), 7.03 (d, J = 2.5 Hz, 1H), 3.39 (s, 6H), 2.91 (s, 3H). $^{13}$C NMR (126 MHz, CD$_2$Cl$_2$) $\delta$ 165.8, 163.6, 158.7, 157.9, 134.0 (t, J = 25.3 Hz), 129.3 (t, J = 24.9 Hz), 129.0, 128.5, 127.4 (t, J = 24.0 Hz), 118.2, 117.9, 111.8, 96.1, 41.0, 19.9. HRMS (ESI$^+$) calculated for C$_{18}$H$_{13}$D$_5$NO$^+$ [M]$^+$: 269.1617; found: 269.1677.}
\end{figure}
\begin{figure}
\includegraphics[width=0.5\textwidth]{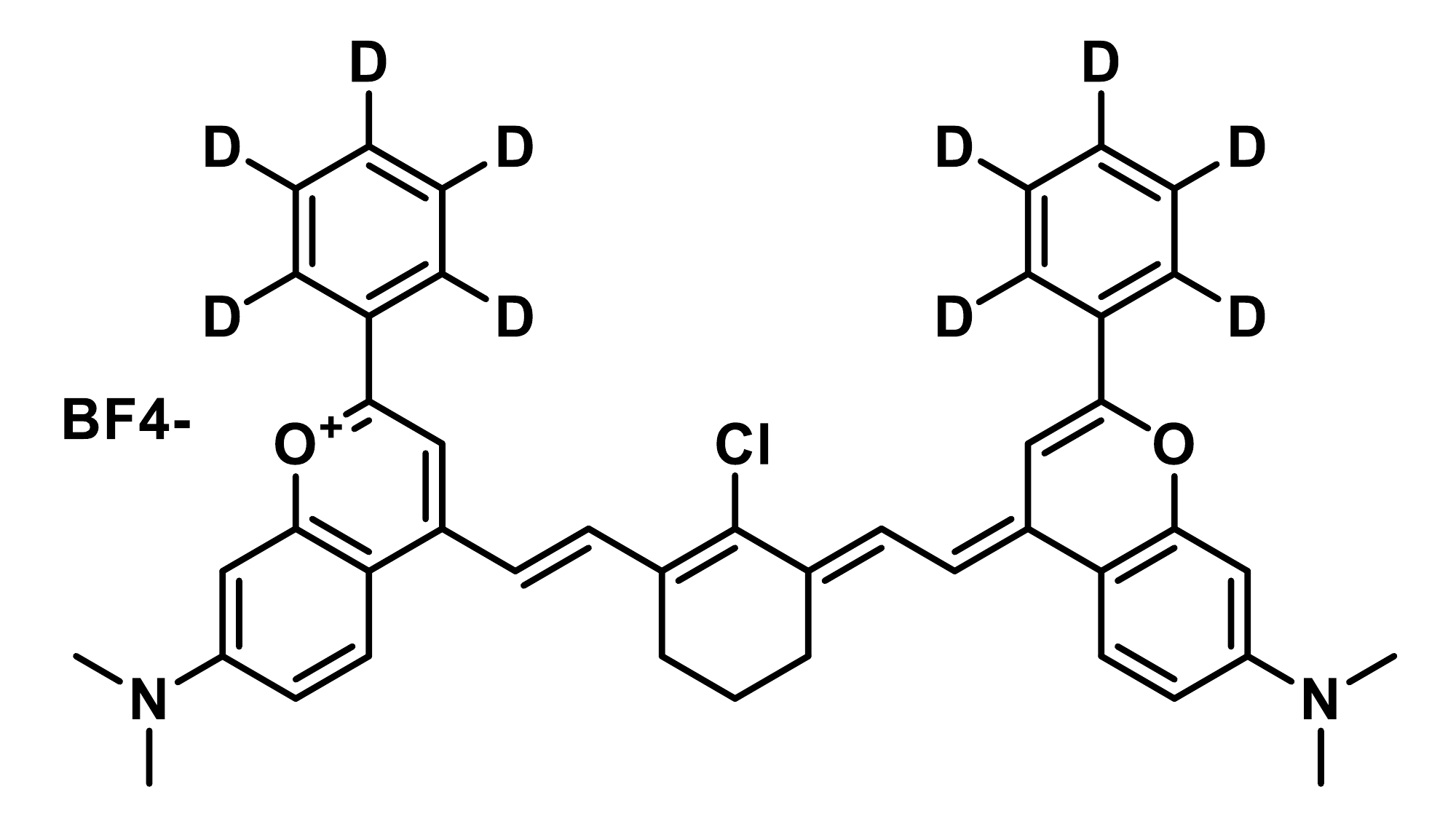}
\caption{{\bf 4-((E)-2-((E)-2-chloro-3-(2-((E)-7-(dimethylamino)-2-(phenyl-d5)-4H-chromen-4-ylidene)ethylidene)cyclohex-1-en-1-yl)vinyl)-7-(dimethylamino)-2-\\-(phenyl-d$_5$) chromenylium (S5, Flav7-d10\#)}. Compound {\bf S4} (80 mg, 0.30 mmol, 2.0 equiv.), N-[(3-(anilinomethylene)2-chloro-1-cyclohexen-1-yl)methylene]aniline hydrochloride, (48 mg, 0.15 mmol, 1.0 equiv.), and sodium acetate (74 mg, .90 mmol, 6.0 equiv.) were dissolved in n-butanol (1.2 mL) and toluene (0.61 mL) in a flame dried Schleck flask under N$_2$ atmosphere. The solution was freeze–pumped–thawed three times and heated to 100 $^o$C for 45 min. The dark brown crude mixture was evaporated, loaded onto silica gel, and purified by column chromatography using a 0.5 to 10\% DCM:EtOH gradient. The procedure yielded a dark burgundy solid in 34\% yield (38 mg, 0.044 mmol). R$_f = 0.5$ in 9:1 DCM/EtOH.  $^1$H NMR (500 MHz, DMSO-d$_6$) $\delta$ 8.18 (d, J = 13.8 Hz, 2H), 8.12 (d, J = 8.9 Hz, 2H), 7.62 (s, 2H), 7.05 (d, J = 13.8 Hz, 2H), 6.96 (d, J = 8.9 Hz, 2H), 6.81 (s, 2H), 3.12 (s, 12H), 2.80 (s, 4H), 1.86 (s, 2H). HRMS (ESI$^+$) calculated for C$_{44}$H$_{30}$D$_{10}$ClN$_2$O$_2$$^+$ [M]$^+$: 673.3401; found: 673.3181. Absorbance/emission (CH$_2$Cl$_2$): 1026, 1054 nm.}
\end{figure}

\begin{figure}
\includegraphics[width=0.7\textwidth]{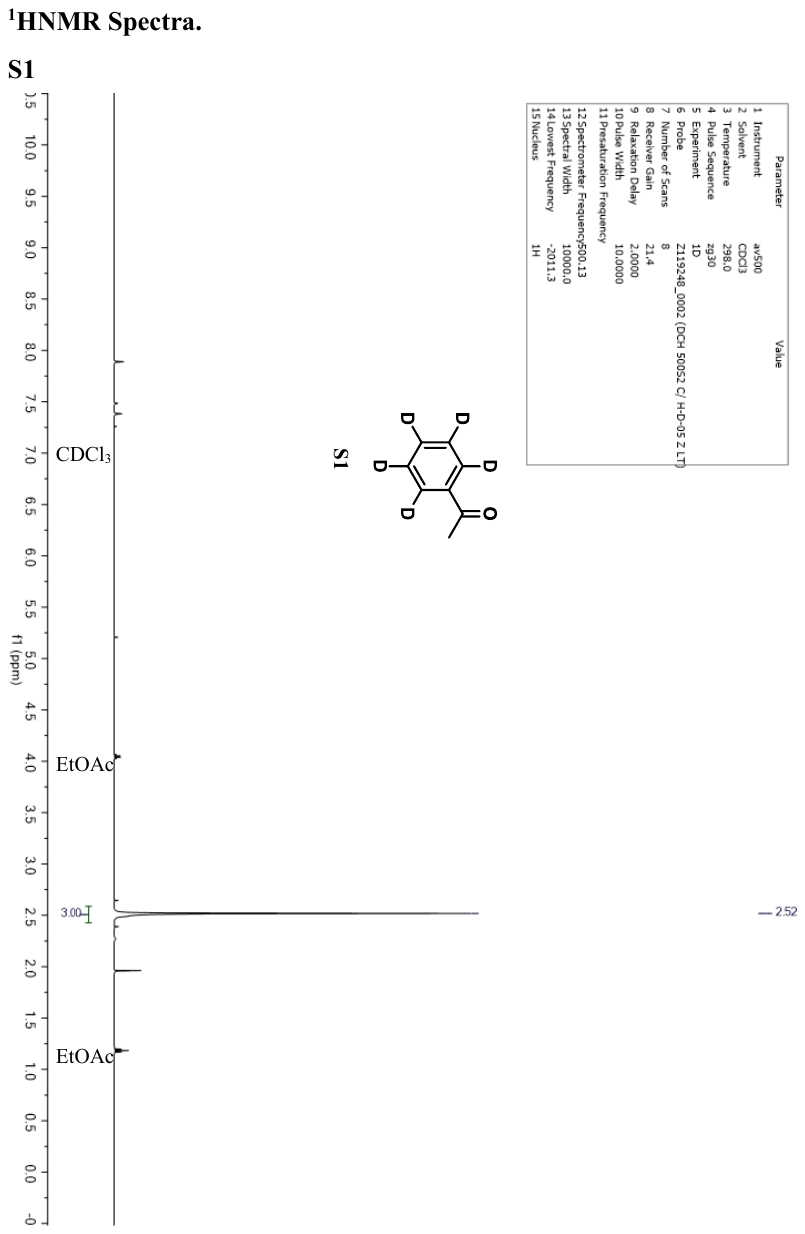}
\caption{$^1$HNMR spectrum of  SH-1}
\end{figure}
\begin{figure}
\includegraphics[width=0.7\textwidth]{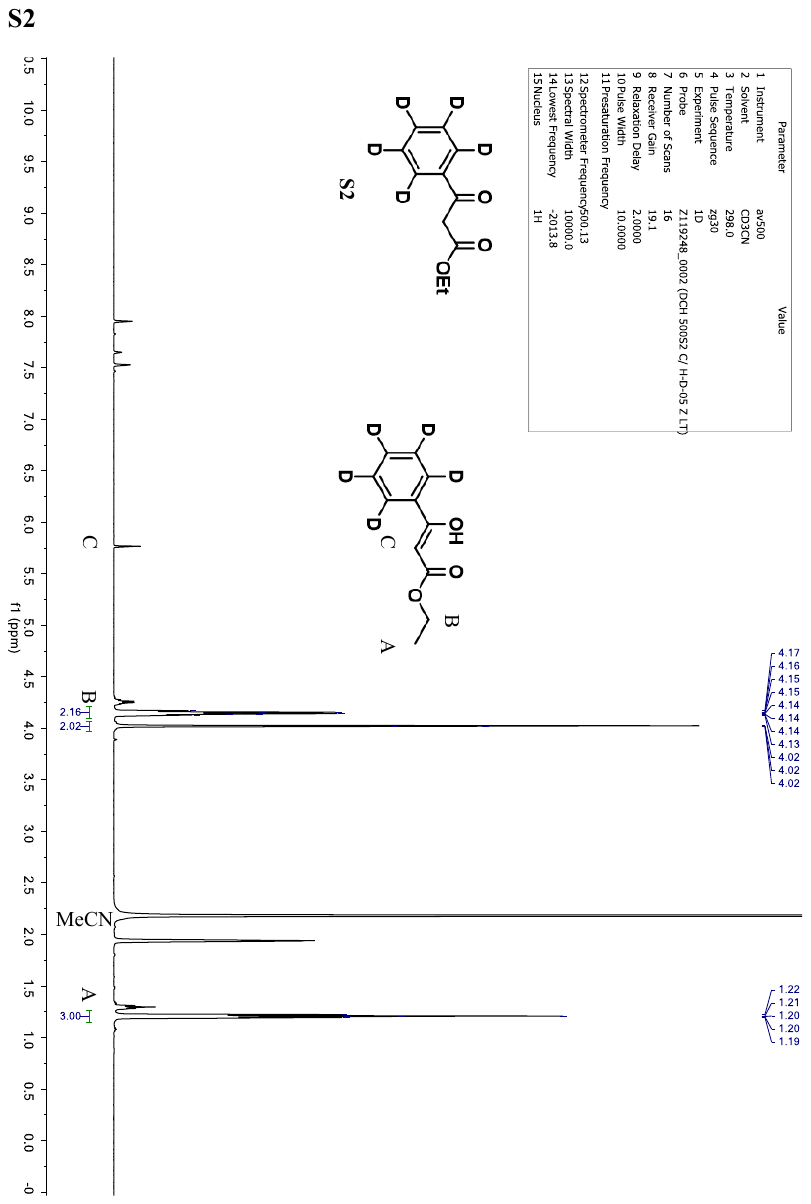}
\caption{$^1$HNMR spectrum of  SH-2}
\end{figure}
\begin{figure}
\includegraphics[width=0.7\textwidth]{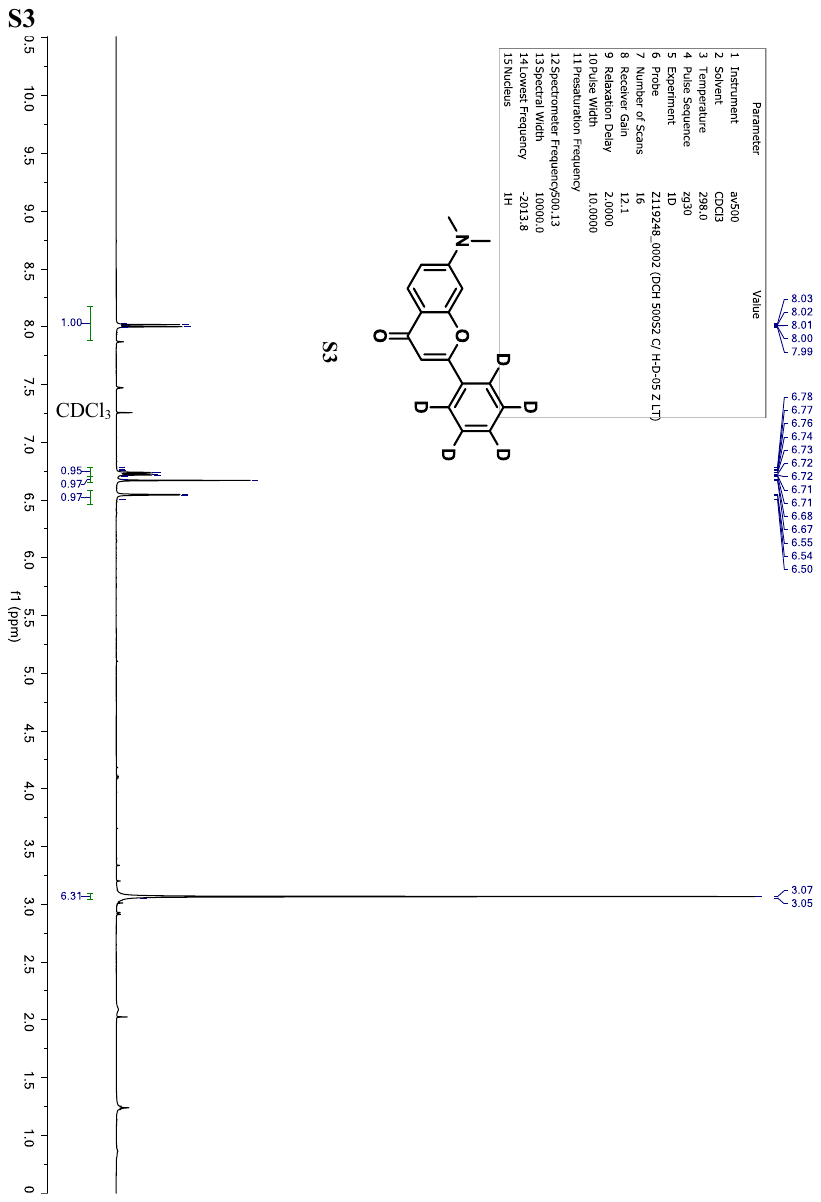}
\caption{$^1$HNMR spectrum of  SH-3}
\end{figure}
\begin{figure}
\includegraphics[width=0.7\textwidth]{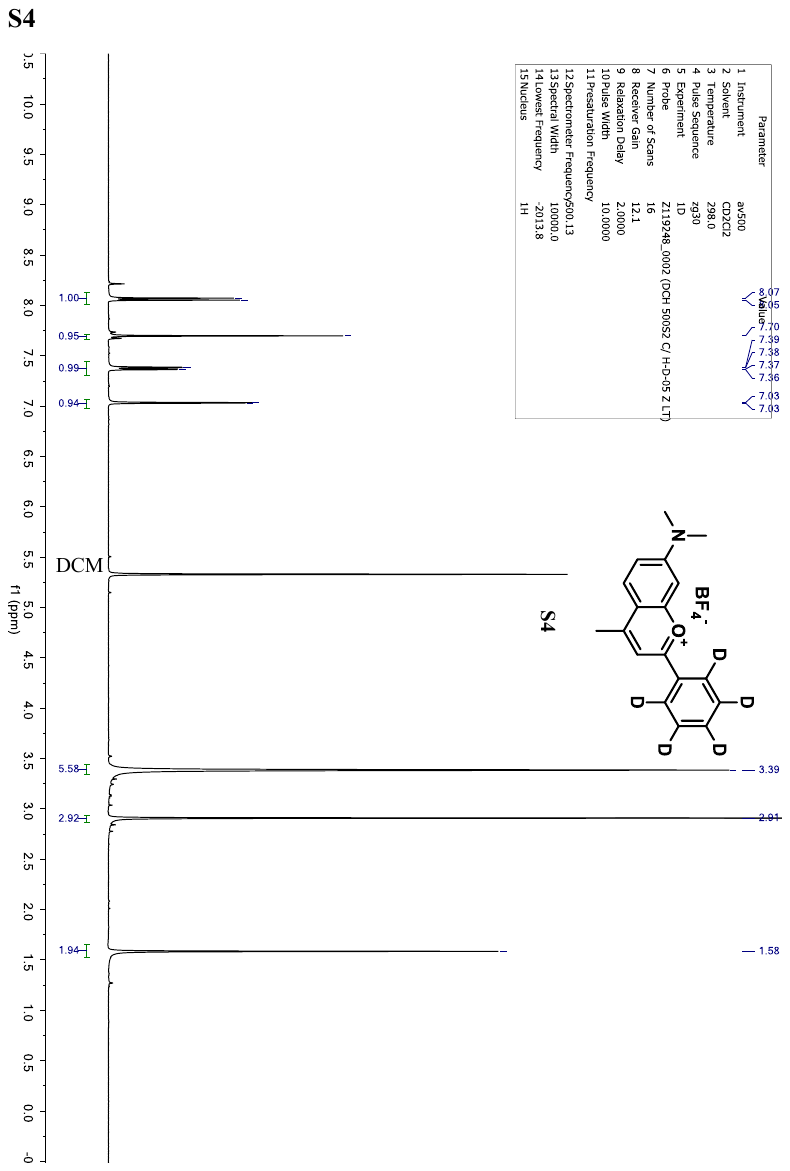}
\caption{$^1$HNMR spectrum of  SH-4}
\end{figure}
\begin{figure}
\includegraphics[width=0.7\textwidth]{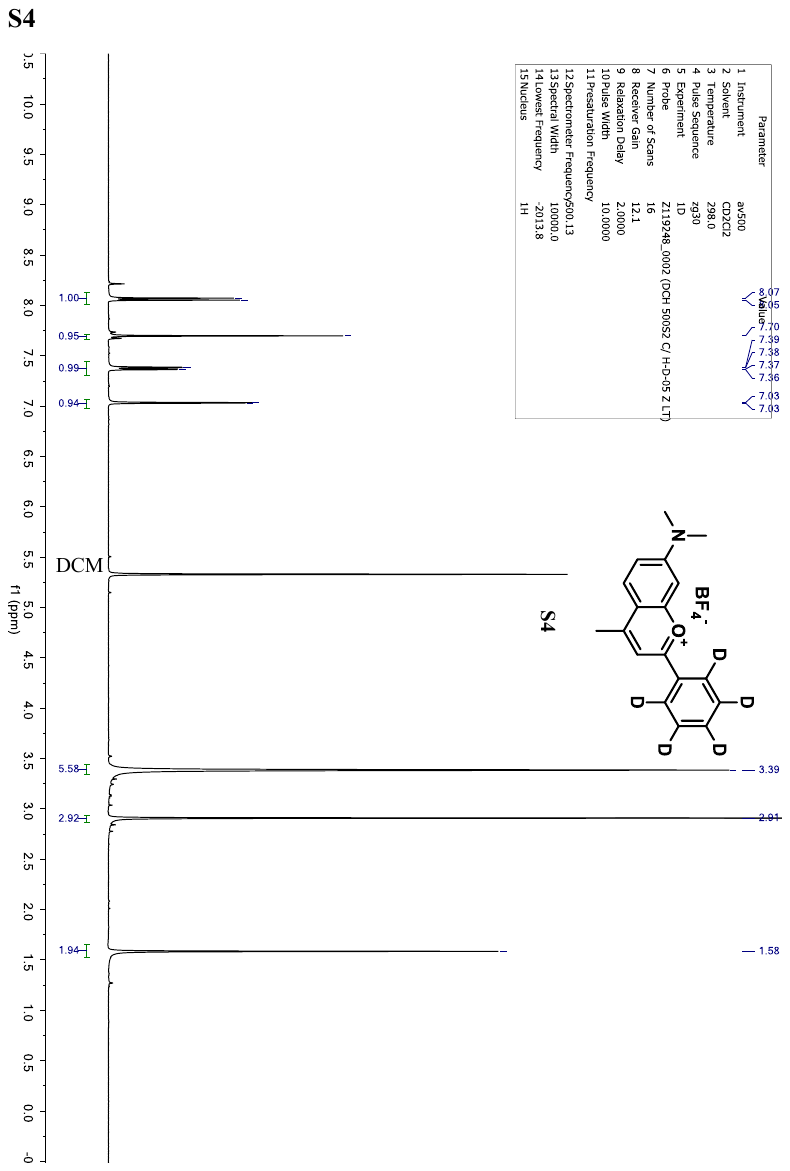}
\caption{$^1$HNMR spectrum of  SH-5}
\end{figure}
\begin{figure}
\includegraphics[width=0.7\textwidth]{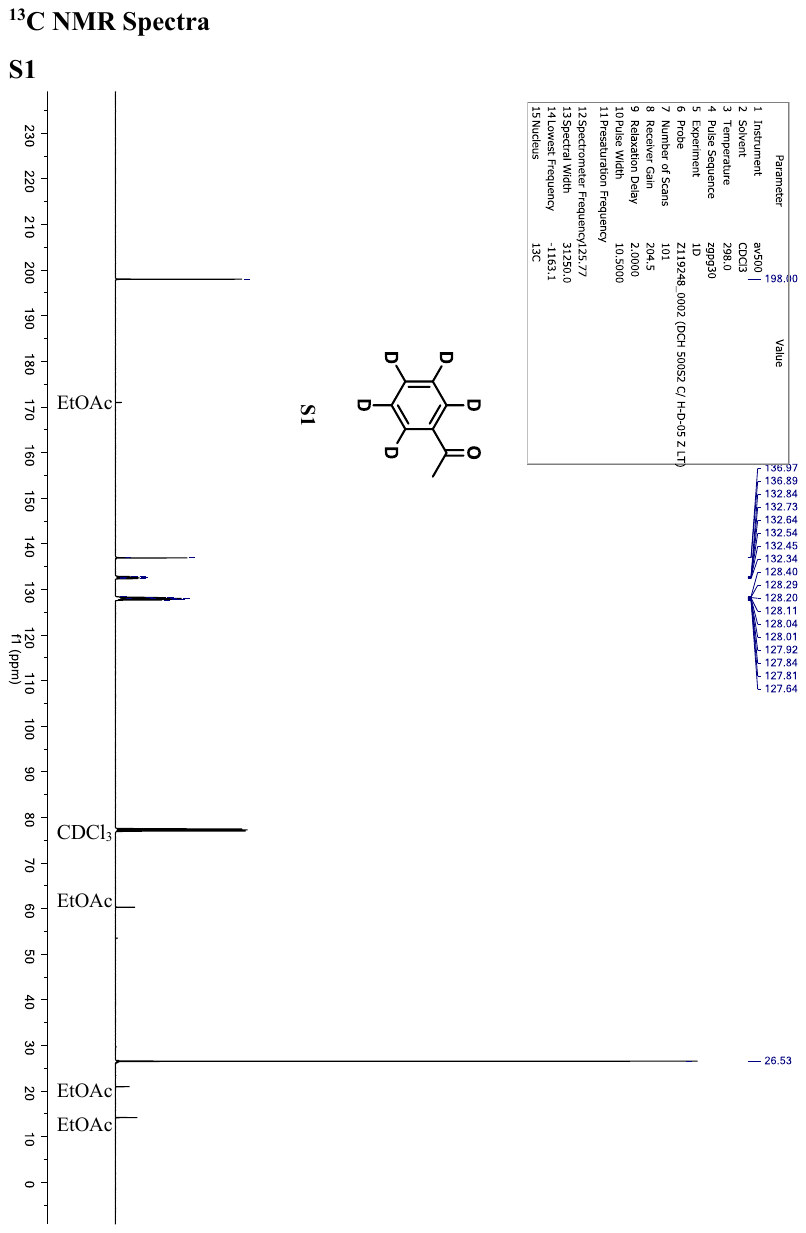}
\caption{$^{13}$CNMR Spectrum of SC-1}
\end{figure}
\begin{figure}
\includegraphics[width=0.7\textwidth]{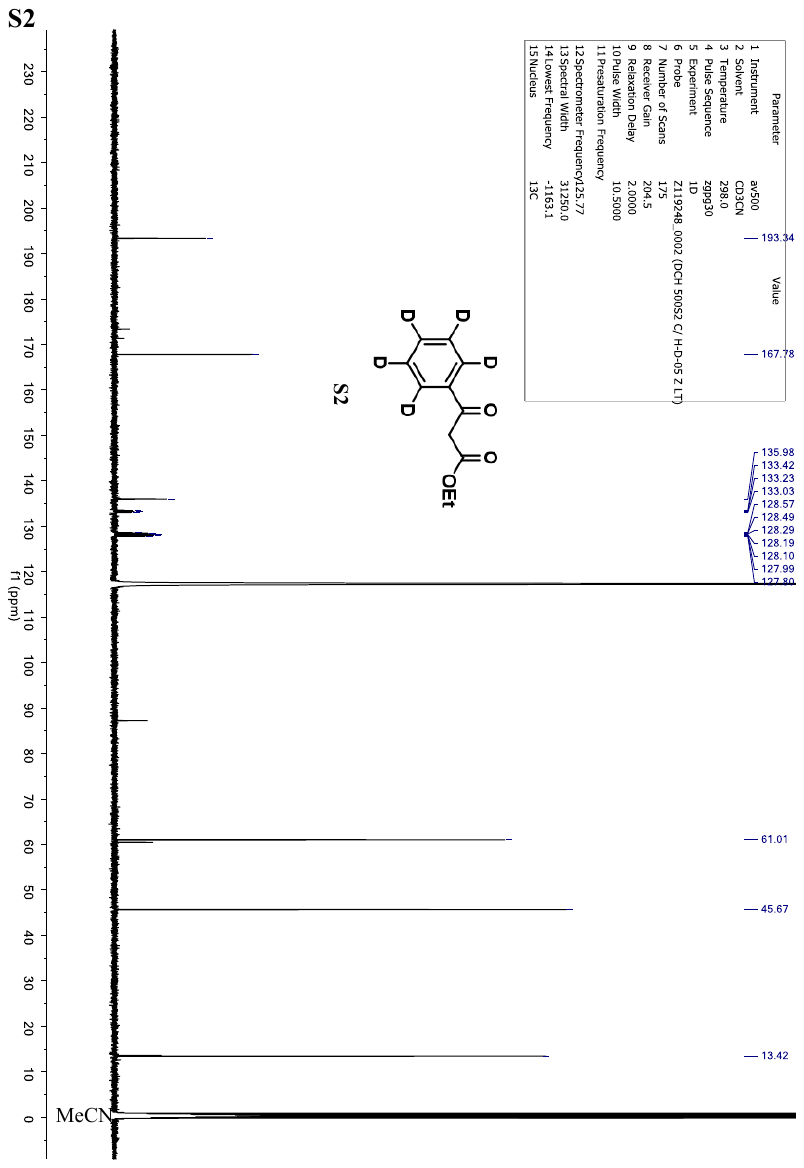}
\caption{$^{13}$CNMR Spectrum of SC-2}
\end{figure}
\begin{figure}
\includegraphics[width=0.7\textwidth]{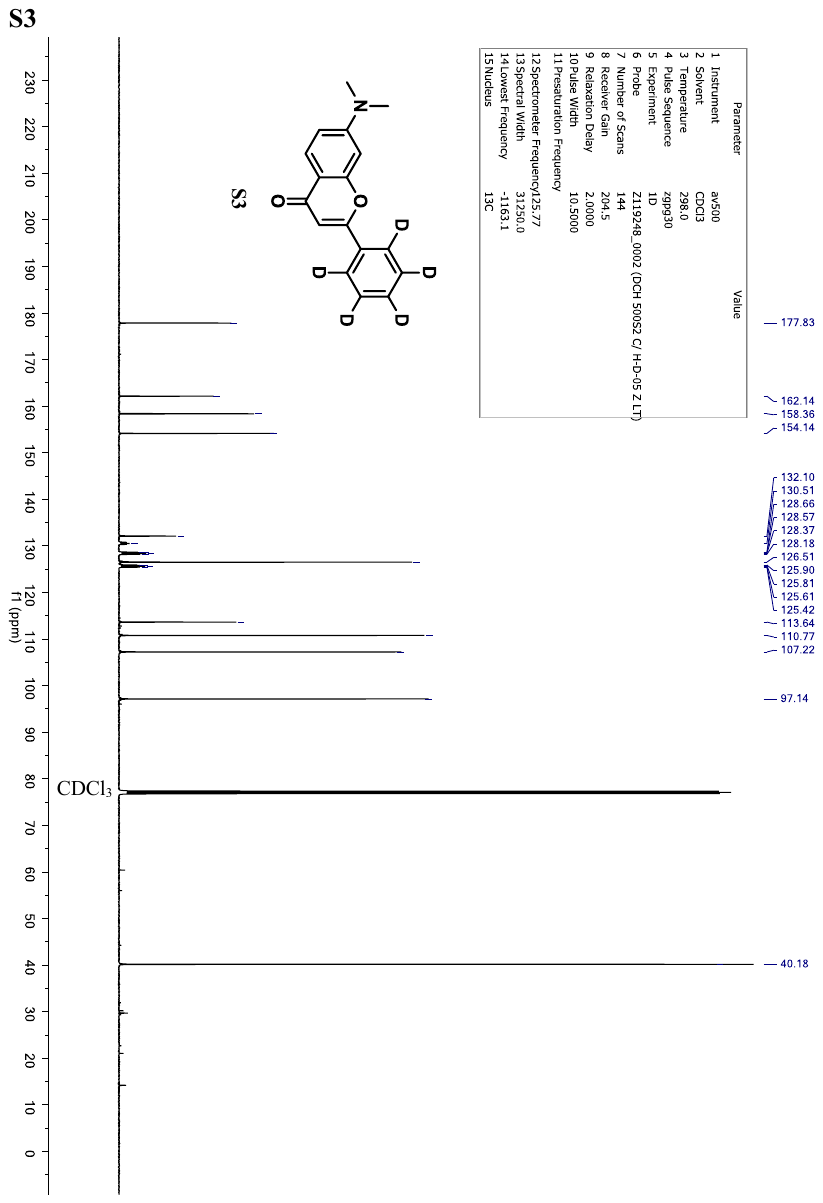}
\caption{$^{13}$CNMR Spectrum of SC-3}
\end{figure}
\begin{figure}
\includegraphics[width=0.7\textwidth]{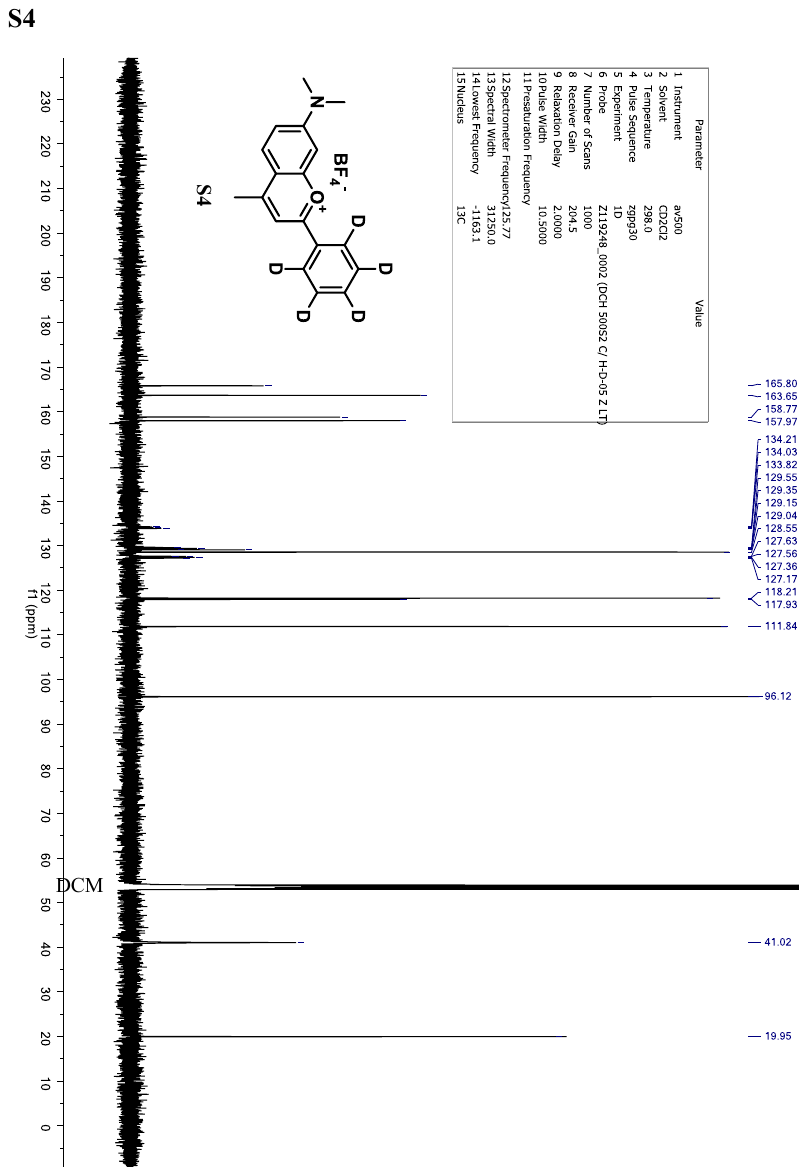}
\caption{$^{13}$CNMR Spectrum of SC-4}
\end{figure}
\begin{figure}
\includegraphics[width=0.7\textwidth]{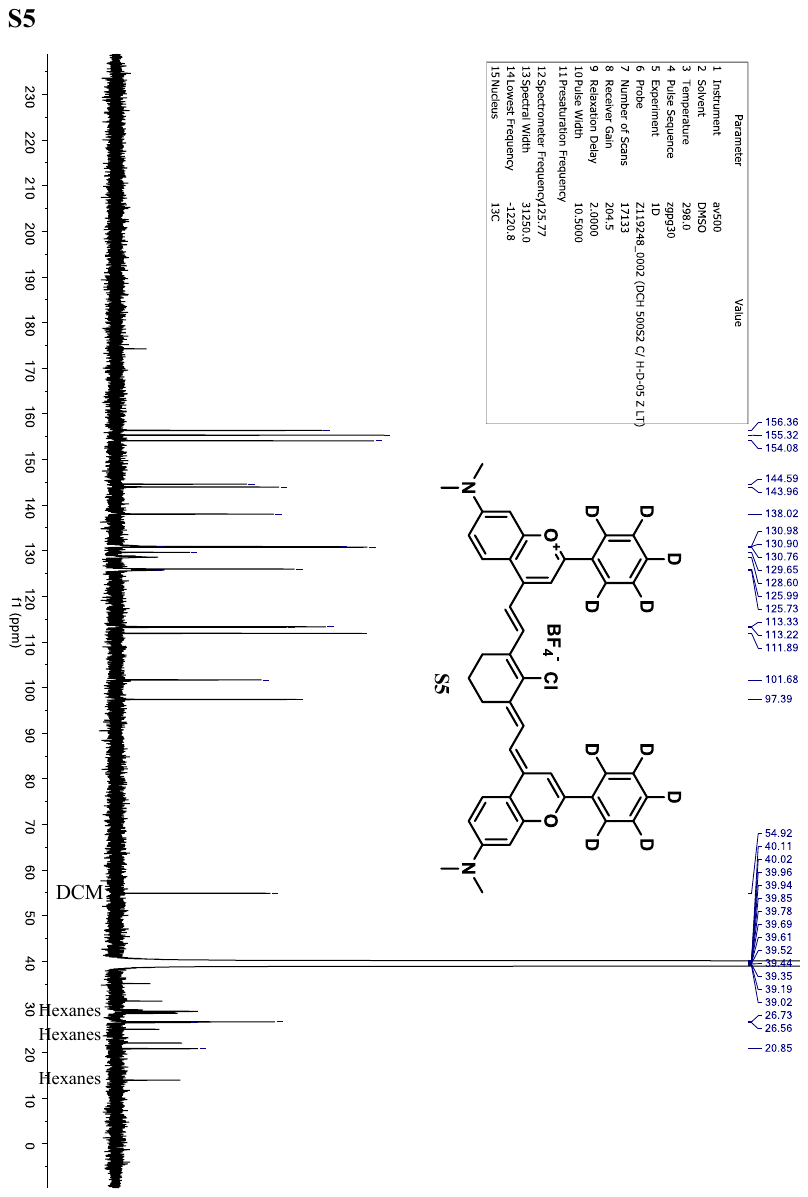}
\caption{$^{13}$CNMR Spectrum of SC-5}
\end{figure}
\end{widetext}

\providecommand{\latin}[1]{#1}
\makeatletter
\providecommand{\doi}
  {\begingroup\let\do\@makeother\dospecials
  \catcode`\{=1 \catcode`\}=2 \doi@aux}
\providecommand{\doi@aux}[1]{\endgroup\texttt{#1}}
\makeatother
\providecommand*\mcitethebibliography{\thebibliography}
\csname @ifundefined\endcsname{endmcitethebibliography}
  {\let\endmcitethebibliography\endthebibliography}{}

\end{document}